\documentclass{aastex63}
\usepackage{xcolor}

\shorttitle{Accretion induced black hole spin up}
\shortauthors{Kr{\'o}l \& Janiuk}
\graphicspath{{./}{figures/}}
\begin{document}
\title{Accretion induced black hole spin up revised by numerical GR MHD simulations\footnote{}}

\correspondingauthor{A. Janiuk}
\email{agnieszka.janiuk@gmail.com}

\author{Dominika {\L}. Kr{\'o}l}
\affiliation{Center for Theoretical Physics, Polish Academy of Sciences \\
Al. Lotnikow 32/46 \\
02-668 Warsaw, Poland}
\affiliation{Astronomical Observatory of the Jagiellonian University \\
Orla 171 \\
30-244, Krak{\'o}w, Poland}
\author{Agnieszka Janiuk}
\affiliation{Center for Theoretical Physics, Polish Academy of Sciences \\
Al. Lotnikow 32/46 \\
02-668 Warsaw, Poland}
\begin{abstract}
  We investigate the accretion induced spin up of the black hole via 
   numerical simulations.
  Our method is based on general-relativistic magneto-hydrodynamics of the slowly-rotating flows in the Kerr metric, where possibly transonic shock fronts may form.
  We account for the changing black hole mass and spin during accretion which enforces dynamical evolution of the space-time metric.
  We first study non-magnetized flows with shocks, and we also include magnetic field
  endowed in the gas. The aim of this study  is to verify whether the high mass black holes may be produced with large spins, even though at birth the collapsars might have contained slowly, or moderately spinning cores. In this way, we put constraints on the content of angular momentum in the 
  collapsing massive stars.
  Our studies are also showing that shock fronts and magnetic fields may halt accretion and limit the black hole spin-up in the exploding supernovae.
\end{abstract}

\keywords{
accretion, accretion disks --- black hole physics --- massive stars --- MHD}

\section{Introduction} 
\label{sec:intro}

Evolution of massive stars ends when the compact remnant is formed in its core, becoming either a neutron star or a black hole. The nature of the remnant mostly depends on the progenitor mass on the Main Sequence \citep{2007Janka, 2018Ott} and is also dependent on its metallicity \citep{2003Nomoto, 2004Eldridge}.
Massive stars heavier than 10 $M_{\odot}$ are frequently born in binary systems, or in multiple stellar configurations \citep{2007Kobulnicky, 2012Sana}.
The evolution of such system is therefore driven by the mass-exchange process, which can proceed either via stable transfer or via an unstable episode called a common-envelope phase (see review by \citet{Langer2012ARA&A}).
On the other hand, some fraction of supernovae might originate from single star progenitors. In this case, the stellar evolution may involve large amount of stellar wind being ejected at the final stages, leading to the self-stripping of stellar envelope 
\citep{2003Heger}. Final mass of the remnant after the core-collapse 
supernova depends on whether the star evolved in a binary system or as a single object, because of an essentially different pre-SN structure revealed by the different stellar evolution models \citep{2020Hirai}.

The measurements of compact remnant masses are possible now independently through the X-ray observations, and gravitational waves detections. In case of X-ray binaries, the known masses of black holes range from $\sim 1.6$ up to 18 $M_{\odot}$ 
\citep{Casares2017}.
In these systems, the first compact object went through a phase of mass transfer, while the progenitor must have been stripped off its Hydrogen envelope \citep{2003Podsiadlowski}.
It is rather unclear though, how the envelope stripping affects the final mass of the compact remnant. As for the spin of such black hole, it is natural to expect that the rotation of the envelope also adds the angular momentum to the stellar core (see however recent study by \citet{2019Fuller} on the angular momentum evolution in stellar cores).

The masses of black holes detected by gravitational wave analysis are systematically higher than these obtained from X-ray observations \citep{2019Perna}. 
The masses of merging components are between 10 and 85 $M_{\odot}$, with the merger products as massive as 142 $M_{\odot}$, for GW190521
\citep{2020Abbott}.
The past history of these most massive black holes is subject to many speculations.
The merging black holes might have been formed in the same binary system and hence have experienced a past mass-exchange episode. The results of population synthesis modeling and comparison with
the data about current merger rates seem to support such scenario \citep{2016Belczynski, 2017Mapelli}. Alternatively, the pre-merger history might contain dynamical encounters in dense stellar clusters \citep{2018Hong, 2018Samsing}, which means that the black holes have been formed independently, possibly from single-star progenitors. 

The information on the black hole spins is important to determine which of the scenarios would be favored in particular objects.
If the spin vectors of merger components are aligned, then most probably the black holes have been formed in a binary system, while the dynamical encounters allow for a more random distribution of the spins \citep{Farr2017Nature}. The preferences
for isotropic spin distribution have to be verified, because the data allow only for determination of an effective spin, which is a combination weighted by the black hole masses (see e.g. \cite{2020Mandel} for the case study of GW190412).
In addition, the large masses of merging black holes should be correlated with their low spins, $a<0.5$, regardless of their formation channel \citep{2016Amaro}.

These new insights into the black hole mass and spin distributions, resulting from the binary interactions or wind-driven stripping, are tackled with the stellar structure and evolution models, and the supernova explosion physics.
Moreover, the paths of black hole formation are affected by the action of magnetic turbulences \citep{2017Obergaulinger}. The finally resulting stellar black hole properties and the fraction of prototypical stellar cores which should result in black hole collapsars depends therefore on a combination of several factors, among which the structure of the progenitor star and the profile of specific angular momentum are probably the most crucial.

 Some information about the rotational velocity in the pre-supernova star may be inferred from its neutrino emission characteristics.
  The recent simulations show that the evolution of standing accretion shock instability (SASI) that strongly depends on both stellar rotation rate and location of the convective zones, might leave imprints on the neutrino Fourier power spectrum \citep{2020PhRvD.101l3013W}. The observational verification o such events is still awaiting.

In this study we analyze the properties of collapsing massive stars, whose envelopes have already been mostly stripped off. We simulate numerically the central parts of a collapsar, where the massive black hole is forming via accretion of mass and angular momentum from the innermost parts of the envelope. 
We perform a parameter study, assuming different levels of rotation in the accreting material, and different initial spins of the black hole that might have resulted from the past stellar evolution before the envelope stripping.
We account for both magnetized and non-magnetized models, and in the latter we probe the formation of transient shocks in the collapsing envelope. With the MHD simulations, we model weakly and strongly magnetized collapsars, where in the latter we adopt the condition of equipartition between thermal and magnetic energy.
All of our simulations are done in full General Relativity and
the coupled evolution of the black hole spin and mass during collapse process is considered with the self-consistent updates of the Kerr metric (see \citet{janiuketal2018}).
The aim of our study is to provide some general constraints on the final mass and spin of the black hole, depending on the collapsar properties and progenitor's history. We also speculate about a possible electromagnetic transient that can occur in some cases, in contrast to the events which leave no luminous counterparts.

In the paper by \cite{2014ApJ...781..119P} 
  the authors study presence of fallback disks around young neutron stars. The angular momentum distribution is taken from the pre-supernova models calculated with public ode MESA, for the stars of masses between 13 and 40 Solar mass.  The specific angular momentum content eventually determines the post-fallback circularization radius. They identify regions in their models parameter space that lead to extended, long-lived disks around BHs, and find that the physical conditions in these disks may be conducive to planet formation.
  In these simulations, the angular momentum profile is based on the reliable assumptions resulting from stellar evolution. On the other hand, these simulations were done in Newtonian gravity, hence all the GR effects, such as the circularization radius dependence on the black hole mass and spin changes, were ignored.

Recently,  the problem of failed supernovae from non-rotating supergiant stars was studied by \cite{2019MNRAS.485L..83Q}. They show that net angular momentum is not necessary condition to form the accretion disk, because of convective motions in the outer parts of the supergiant and that a random angular momentum in supergiant convection zones that exceeds that of the last stable circular orbit of a black hole by a factor of $\sim 10$. The result is obtained using analytic estimates and Cartesian box simulations in the Bussinesq approximation.

In the current work, we follow the approach started in the work of \cite{2006ApJ...641..961L}. The idea described a ballistic motion of particles in the gravitational potential of the black hole, while the authors studied a free fall of the fluid. Superimposed rotation profile was given by an analytic function, $l_{spec}\propto\sin^{2}(\theta)$. Neutrino cooling and contributions to the pressure from degenerate electrons were considered by those authors because the discussion was focused on the resulting GRB power.

A corresponding initial condition set up can be found in \cite{2005astro.ph..2225D}. In their work, the authors conducted a series of GRMHD 2D simulations with a 3D follow-up for chosen models. Initially, the physical system consisted of black hole (with different spin values probed), a Keplerian disk and the background described by Bondi solution. Space-time metric was fixed, and did not evolve under the spin and mass evolution.
  Significant difference in comparison with our simulations is the presence of a Keplerian disk. Also, the authors assumed a smaller ratio between accreting mass and to black hole mass, to justify the fixed metric assumption. The main focus of their work was to discuss the jet lunching process, and they obtained
  a jet present between $0.1 - 1.4$s, with a significant contribution from the coronal wind.

 In this paper we perform the first multidimensional
general relativistic simulations of uniformly rotating, low angular momentum, magnetized flows, where we take into account the Kerr metric evolution due to changing black hole mass and spin. We build our models on the initial conditions of the slowly-rotating quasi-spherical transonic flow and we follow upon the other recent works of our group. However, in \cite{janiuketal2018} and in \cite{Murguia2020}, the magnetic fields were neglected. The latter paper, which actually neglected also the Kerr metric evolution effects, gave the main focus to the comparison of the constraints on the angular momentum resulting from the GR hydro simulations, with the realistic models of the stellar structure and rotation profiles obtained from MESA simulations.

The structure of the article is as follows. In Section \ref{sec:Model} we describe the theoretical framework of our simulations.  In Sec. \ref{sec:Bondiflow} we present the  analytical formulation of transonic spherical accretion onto black hole, in Sect. \ref{sec:rotation} we introduce the slow angular momentum prescription, in Sec. \ref{sec:spacetime} we describe our method for the Kerr spacetime evolution, in Sec. \ref{sec:units} we introduce physical scaling and parameters of our collapsar model, and in Sec. \ref{sec:MagneticField} we describe the initial configuration for magnetic fields used in the MHD simulations. Th results of our numerical analysis are presented in Section \ref{sec:results}, and in particular, the non-magnetized models are presented in Sect. \ref{sec:NonMag}, while the models with magnetic fields are presented in Sec. \ref{sec:Mag}.  Th discussion and conclusions are presented in Section \ref{sec:diss}.

\section{Black hole accretion with low specific angular momentum}
\label{sec:Model}

In the simple collapsar model that we adopt here as an initial condition, we assume that the star has had enough time to reach a quasi-steady state at the end of its stellar evolution. Hence, we adopt a quasi-spherical configuration of density, resulting from the Bondi prescription \citep{Bondi} and endowed with a small angular momentum. During the time dependent simulation, however, we do not add mass to the envelope during the time evolution, so  the  accretion rate is  dynamically  changing.
Moreover, rotation changes the matter distribution, and a disk-like structure forms at the equatorial plane. The microphysics is treated in a simplified way and we use the adiabatic index  of $\gamma =  4/3$,  that  describes  the  stellar  interior, accounting for the contribution from the photon pressure.

\subsection{Initial condition for quasi-spherical flow}
\label{sec:Bondiflow}

The distribution of density as the function of radius comes from
the solution of transonic accretion flow in spherical geometry, and is similar to
\citet{2015MNRAS.447.1565S,2017MNRAS.472.4327S,2019MNRAS.487..755P}.
The initial density profile and the
radial component of the velocity ($u^r$) of the material is determined by the relativistic version of the Bernoulli equation \citep{Michel72,Hawley84}.
In this formalism, the critical point ($r_{\rm s}$),
where the flow becomes supersonic, is set as a parameter.
Here we take the value of $r_{s}=80 r_{g}$.

The fluid is considered a polytrope with a pressure $P=K\rho^\gamma$, where $\rho$ is the density,  $\gamma=4/3$ is the adiabatic index, and $K$ is the constant specific entropy, in this case taken to be that of a relativistic gas. 

Once the critical point, $r_{s}$, is set, the velocity at that point is:
\begin{equation}
  (u^r_{\rm s})^2= \frac{GM}{ 2r_{\rm s}},
  \label{eq:us}
\end{equation}
 where $r$ is the radial coordinate, $M$ is the mass of the BH and the sound speed is:
 \begin{equation}
 \label{eq:cs}
 c_{\rm s}^2=\frac{\gamma\frac{P_{\rm s}}{\rho_{\rm s}}}{1+\frac{\gamma}{\gamma-1}\frac{P_{\rm s}}{\rho_{\rm s}}}.
 \end{equation}

 The constant specific entropy can be obtained using the sound speed:
\begin{equation}
K=\frac{c_{\rm s}^2}{\rho^{\gamma-1}\gamma}
\end{equation}
where the density is set by the mass accretion rate $\dot{M}$: 
\begin{equation}
\rho=\frac{\dot{M}}{r^2u^r}
\end{equation}

The radial velocity profile is obtained by numerically solving the equation \citep{Shapiro}:
\begin{equation}
  \bigg{(} 1+\frac{\gamma}{\gamma-1}\frac{P}{\rho}\bigg{)} ^2 \bigg{(}  1-\frac{2GM}{r}+(u^r)^2 \bigg{)} = \rm{constant}
  \label{eq:bernoulli}
\end{equation}
 and 
the proper velocity of the fluid at infinity is zero, while it equals to the speed of light at the horizon.

In construction of the numerical solution, we integrate the Eq. \ref{eq:bernoulli}
upwards and downwards of the critical point, and the normalized accretion rate $\dot M=1$ is used (note that our code works in dimensionless units of $G=c=1$).
To convert the quantities into physical units, we adopt the parameterization of density in the
accreting gas as in \cite{janiuketal2018}.

\subsection{Initial conditions for small angular momentum configuration}
\label{sec:rotation}

The  accreting material is 
endowed with small angular momentum scaled to the one at the circularization radius of
$r_{\rm circ}$, being the ISCO radius (equal to 6 $r_{g}$ for a non-rotating black hole). 
The angular momentum 
is further scaled with 
the fraction of the angular momentum at the ISCO, $S$, and scaled with polar angle to have its maximum value on the equatorial plane, at $\theta=\pi/2$: 
\begin{equation}
l=S l_{\rm isco} \sin^2{\theta} 
\end{equation}
where $l$ is the specific angular momentum,  defined as $l=u_{\phi}=g_{\phi\mu}u^{\mu}$.
Here $l_{\rm isco}$ is the specific angular momentum at the ISCO of the black hole, and $\theta$ is the polar coordinate. 
In order to get the angular velocity at each point, we use the auxiliary energy $\epsilon_{\rm isco}$ and auxiliary angular momentum at the ISCO ($l_{\rm isco}$):
\begin{equation}
\epsilon_{\rm isco}=\frac{1-2/r_{\rm isco}+a/r_{\rm isco}^{3/2}}{\sqrt{1-3/r_{\rm isco}+2a/r_{\rm isco}^{3/2}}}
\end{equation}

\begin{equation}
l_{\rm isco}=\frac{r_{\rm isco}^{1/2}-2a/r_{\rm isco}+a^2/r_{\rm isco}^{3/2}}{\sqrt{1-3/r_{\rm isco}+2a/r_{\rm isco}^{3/2}}}
\end{equation}

where $a$ is the spin of the BH, and the radius of the ISCO ($r_{\rm isco}$) changes self-consistently with changing spin. In these units, black hole mass is $M=1$.

We can then construct the angular velocity in the Boyer-Lindquist coordinates as:
\begin{equation}
u^\phi=-g^{t\phi}\epsilon_{\rm isco}+g^{\phi\phi}l_{\rm isco}
\end{equation}
where the components of the metric are for a Kerr BH $g^{t \phi}=-2ar/(\Sigma \Delta)$ and $g^{\phi \phi}=(\Delta-a^2\sin^2\theta)/(\Sigma \Delta \sin^2\theta)$, and $\Sigma=r^2+a^2\cos^2\theta$, $\Delta=r^2-2r+a^2$.

Normalization of angular momentum content in the collapsar's envelope is
done with $S$, which is a free parameter of our model.
Note that $S=0$ reduces to the Bondi spherical accretion. 
                 
\subsection{Time dependent General Relativistic MHD code}
\label{sec:code}

We use the general relativistic magneto-hydrodynamic code, \textit{HARM} \citep{Gammie_2003, Noble_et_all_2006}.
In this version of the code, we use the method developed in \citep{janiuketal2018} to account for the change of Kerr metric due to the dynamically increasing black hole mass and its spin.

The \textit{HARM} code is a finite volume, shock capturing scheme that solves 
the hyperbolic system of the partial differential equations of GR MHD.
The numerical scheme is using 
the plasma energy-momentum tensor, $T^{\mu\nu}$, with contributions from gas and electromagnetic field
\begin{eqnarray}
{T_{\left(m\right)}}^{\mu\nu}= \rho h u^\mu u^\nu + p g^{\mu\nu} \\
{T_{\left(em\right)}}^{\mu\nu}=b^\kappa b_\kappa u^\mu u^\nu+\frac{1}{2} b^\kappa b_\kappa g^{\mu\nu} - b^\mu b^\nu\\
T^{\mu\nu}={T_{\left(m\right)}}^{\mu\nu}+{T_{\left(em\right)}}^{\mu\nu}
\end{eqnarray}
where $u^{\mu}$ is the four-velocity of gas, $u$ denotes internal energy density,  $b^{\mu}$ is magnetic four-vector, and
$h$ is the fluid specific enthalpy.
The continuity and momentum conservation equations read:
\begin{equation}
\
(\rho u^{\mu})_{;\mu} = 0;
\hspace{1cm}
T^{\mu}_{\nu;\mu} = 0,
\end{equation}
They are brought in conservative form, by implementing a Harten, Lax, van Leer (HLL) solver to calculate numerically the corresponding fluxes. 

In terms of the Boyer-Lindquist coordinates, $\left(r,\theta,\phi\right)$, the black hole is located at $0<r \leq r_{\rm h}$, where $r_{\rm h} = \left(1+\sqrt{1-a^{2}}\right) r_{\rm g}$ is the horizon radius of a rotating black hole with mass $M$ and angular momentum $J$ in geometrized units, $r_{\rm g}=G M /c^2$, and $a$ is the dimensionless Kerr parameter, $a=J/(Mc), 0 \leq a \leq 1$.

The {\it HARM} code doesn't perform the integration in the Boyer-Lindquist coordinates, but instead in the so called Modified Kerr-Schild ones: $t,x^{(1)},x^{(2)},\phi$ \citep{Noble_et_all_2006}. The transformation between the coordinate systems is given by:
$r= R_0 + \exp\left[{x^{(1)}}\right]$,
and
$\theta = \frac{\pi}{2} \left(1 + x^{(2)}\right) + \frac{1 - h}{2} \sin\left[\pi\left(1 +x^{(2)}\right)\right]$,
where $R_0$ is the innermost radial distance of the grid, $0 \leq x^{(2)} \leq 1$, and $h$ is a parameter that determines the concentration of points at the mid-plane. In our models we use $h=0.3$ (notice that for $h=1$ and a uniform grid on $x^{(2)}$ we obtain an equally spaced grid on $\theta$, while for $h=1$ the points concentrate on the mid plane).
The exponential transformation in the $r$-direction leads to higher resolution in the innermost regions.
The inner and outer radii of our grid are located at $0.98 r_{h}$ and $10^{3} r_{g}$, respectively. 
Our grid resolution is $256 \times 256$.

\subsection{Kerr spacetime evolution}
\label{sec:spacetime}

The changing black hole spin and mass affect the spacetime metric. In our simulation, the spacetime is described by a sequence of quasi-stationary Kerr solutions, where the metric is
changed discretely between the consecutive time steps, according to mass and spin increments, $\Delta M = (M^{i}/M^0-1)$, and $\Delta a = (\dot{J}/M^{i} - a^{i-1}/M^{i}\dot{E})\Delta t$, where $M^{i}$ denotes the current black hole mass at time $t>0$, $M^0$ denotes initial black hole mass at $t=0$, and $\dot{J}$ and $\dot{E}$ are the flux of angular momentum and energy flux transmitted through the black hole horizon at a given time (cf. \cite{janiuketal2018}).
The six non-trivial components of the Kerr metric, namely 
$g_{tt}$, $g_{tr}$, $g_{t\phi}$, $g_{rr}$, $g_{r\phi}$, $g_{\phi\phi}$, are
updated at every time step and they get new values.

This procedure is developed to take into account the change of black hole spin and mass, and the metric update, but does not consider the effects of self-gravity of diluted envelope and the effect of its changing mass on the metric. We argue that the self-force of the envelope would have negligible impact on our simulations results, however we plan to address this issue in more detail and make quantitative comparisons in the future work.

We notice that the external material does have an impact on the overall Kerr metric evolution, however in the present simulations we neglect this effect.
  The adopted density of matter in the accreting cloud is large (see Section \ref{sec:units} below). Nevertheless, the large size of the computational domain
  ensures that cloud is still not as compact as the central black hole, so the black hole still dominates the gravitational potential in this system.

 The effect of the collapsing cloud on the metric evolution is driven by the self force, which is induced on the matter element located at every grid point by the mass contained inside the sphere of that radius. It effectively acts as an additional perturbative term, apart from the increasing mass of the black hole.
 This perturbative term is changing with time during the collapse
(cf. \citet{ishika2020_appol}).
 We verified that this term is large only at the beginning of collapse, when the cloud mass is largest, and very quickly decreases with time.

\subsection{The physical simulations setup}
\label{sec:units}

The code HARM operates in dimensionless units $G=c=M=1$, so the units do not appear explicitly in the physical equations solved neither in the metric coefficients.
Because we want to model a particular case of collapsing star, we convert our simulation results to physical scale, which allows to obtain the proper intuition about astrophysical objects of our interest.
Following \citet{Janiuk2013}, we use the value of the total (volume integrated) initial mass of the accreting matter to scale the density in the spatial domain 
and we use the unit conversion system as:
\begin{eqnarray*}
L_{\rm unit}=r_g=\frac{G M}{c^2}=1.48\cdot 10^5 \frac{M}{M_\odot} \rm{cm} \\
T_{\rm unit}=\frac{r_g}{c}=4.9\cdot 10^{-6} \frac{M}{M_\odot} \rm{s} \\
\end{eqnarray*}
for the spatial and time units, respectively.
The density scale is related to the spatial unit as $D_{\rm unit}=M_{\rm scale}/L_{\rm unit}^{3}$.

In this work, the collapsing stars are modeled with the following global parameters: initial mass of the core black hole $M^{0}_{\rm BH}$, the cloud mass enclosed within the simulation volume, $M_{\rm cloud}$, and the initial black hole dimensionless spin, $A_{0}$.
The fiducial parameters are $M^{0}_{\rm BH}=3 \,M_{\odot}$, and $A_{0}=0.3-0.85$, while the cloud mass is 25 $M_{\odot}$,
as resulting from its size in geometrical units, and adopted density scaling.
The accretion rate onto the black hole, measured as the mass flux transported
through the horizon, is varying with time. Its value in expressed
in physical units of $M_{\odot}$s$^{-1}$, as converted from the density scaling and time unit.

\subsection{Magnetic field configuration for MHD models}
\label{sec:MagneticField}

We assume that the initial accreting cloud is embedded in a poloidal magnetic field,
prescribed with the vector potential of 
\begin{equation}
\label{eq:mag}
A_{\varphi} \propto (1-\cos(\theta))
\end{equation}
where the normalization comes from the ratio between the gas pressure, given by $p_{\rm gas}=(\gamma-1) u$, and magnetic pressure, given by
$p_{\rm mag}={1\over 2} b_{\mu}b^{\nu}$.
Our code parameter is initial gas to magnetic pressure ratio, $\beta_{0}\equiv p_{\rm gas} / p_{\rm mag}$.
As a consequence, the magnetized flows considered here are not in equilibrium, and the angular momentum is transported also through the rotationally supported disk in the equatorial plane, thus affecting the accretion.

We examine here models with initial $\beta_{0}=1.0$ and $\beta_{0}=100$ and we also probe different values of $A_{0}$, equal to 0.3, 0.5, and 0.85 for the initial spin of the black hole.
In addition, every set of magnetized collapsar models, includes different rotation parameters, $S$. 
The $\beta$ parameter is normalized to the gas pressure at the inner boundary of the grid which position depends on  $A_0$ (due to the scaling of the innermost stable circular orbit).

\section{Results}
\label{sec:results}

\begin{deluxetable}{lccccccr}

\tablecaption{Summary of the models \label{tab:in}}

\tablehead{
  \colhead{Model} 
  & \multicolumn{1}{p{1.5cm}}{\centering BH spin\\ $A_{0}$} 
  & \multicolumn{1}{p{1.5cm}}{\centering $S$\\ $l/l_{\rm crit}$ } 
  & \multicolumn{1}{p{1.5cm}}{\centering $\beta_{0}$\\  }
  & \multicolumn{1}{p{1.5cm}}{\centering $t_{\rm f}$\\ s }
  & \multicolumn{1}{p{1.5cm}}{\centering $M_{\rm BH}^{\rm f}$ } 
  & \multicolumn{1}{p{1.5cm}}{\centering $A_{\rm f}$ \\ }
  &  \multicolumn{1}{p{1.5cm}}{\centering $\dot M^{max}$\\ $M_{\odot}$ s$^{-1}$ }
}
\startdata
      HS-T-04   & 0.85 & 0.4 & $\infty$ & 2.1 & 15.45 &0.15 &  $45.1$ \\
      HS-LM-04  & 0.85  & 0.4 & 100 & 1.4 & 17.35 & 0.13 &  $49.6$ \\
      HS-HM-04  & 0.85  & 0.4 & 1.0 & 1.3 & 17.30 & 0.13 &  $49.6$ \\
      LS-T-04   & 0.5 & 0.4 & $\infty$ & 1.9 &16.09 & 0.16 &  $46.4$\\
      LS-LM-04 & 0.5  & 0.4 & 100 & 1.4 &18.07 & 0.13 &  $51.0$ \\
      LS-HM-04 & 0.5  & 0.4 & 1.0 & 1.4 &18.07 & 0.13 &  $51.1$ \\
      NS-T-04  & 0.3 & 0.4 & $\infty$ & 2.0 &16.22 & 0.17 &  $46.7$\\
      NS-LM-04 & 0.3  & 0.4 & 100 & 1.4 & 18.20 & 0.14 &   $51.4$ \\
      NS-HM-04 & 0.3  & 0.4 & 1.0 & 1.4 & 18.20 & 0.14 &   $51.4$ \\
      \hline
    HS-T-10  & 0.85 & 1.0 & $\infty$ & 4.5 &14.30 & 0.40 &  $41.1$ \\
      HS-LM-10  & 0.85  & 1.0 & 100 & 1.4 & 15.90 & 0.33 &  $47.7$ \\
      HS-HM-10  & 0.85  & 1.0 & 100 & 1.4 & 15.90 & 0.34 &  $43.6$ \\
      LS-T-10   & 0.5 & 1.0 & $\infty$ & 2.4 &14.95 & 0.43 &  $42.6$\\
      LS-LM-10   & 0.5  & 1.0 & 100 &1.3 & 17.65 & 0.34 &  $49.7$ \\
      LS-HM-10   & 0.5  & 1.0 & 1.0 &1.5 & 16.90 & 0.35 &  $46.2$ \\
      NS-T-10  & 0.3 & 1.0 & $\infty$ & 2.8 &12.22 & 0.57 &  $34.1$\\
      NS-LM-10  & 0.3  & 1.0 & 100 & 1.2 &17.60 & 0.36 &   $49.7$\\
      NS-HM-10  & 0.3  & 1.0 & 1.0 & 1.2 &17.14 & 0.37 &   $45.8$\\
      \hline
      HS-T-14   & 0.85 & 1.4 & $\infty$ & 1.5 &11.90 & 0.65 &  $42.0$ \\
      HS-LM-14  & 0.85  & 1.4 & 1.0 &1.4 & 15.70 & 0.50 &  $44.1$\\
      HS-HM-14  & 0.85  & 1.4 & 1.0 &1.5 & 15.20 & 0.51 &  $43.2$\\
      LS-T-14  & 0.5 & 1.4 & $\infty$ & 1.4 &12.43 & 0.71 &  $48.3$\\
      LS-LM-14  & 0.5  & 1.4 & 1.0 & 1.3 & 16.55 & 0.53 &  $45.7$ \\
      LS-HM-14  & 0.5  & 1.4 & 1.0 & 1.3 & 16.05 & 0.52 &  $42.7$ \\
      NS-T-14   & 0.3 & 1.4 & $\infty$ & 1.4 &12.60 & 0.74 &  $79.3$\\
      NS-LM-14   & 0.3  & 1.4 & 1.0 & 1.2 & 16.50 & 0.55 &   $45.9$\\  
      NS-HM-14   & 0.3  & 1.4 & 1.0 & 1.2 & 16.20 & 0.57 &   $43.9$\\      
\enddata
\tablecomments{The HS models refer to a highly spinning black hole at birth, the LS models represent moderately spinning black hole, and NS models are for slowly-spinning black hole at birth.
The angular momentum magnitude, $S$, the spin of the initial black hole, are given as the initial state parameters. In magnetized models, denoted with $HM$, the plasma $\beta$ was adopted to 1.0, $LM$  models correspond to $\beta=100$, while in thermal models, denoted with $T$, the magnetic field was neglected, $\beta=\infty$. The initial mass of the accreting cloud was set to 25 $M_{\odot}$, while the final mass reached at time $t_{\rm f}$ was close to zero. The last three columns give resulting final black hole mass, its final spin, and  time-averaged mass accretion rate.}
  \label{tab:models}
\end{deluxetable}

\subsection{General structure of the flow}
\label{sec:general}

The overall evolution follows a similar pattern for all hydrodynamical models that have been investigated. We always start from a spherically symmetric distribution of density, that is quickly broken by the imposed azimuthal velocity component that is maximal at the equator. In consequence, the distribution of density becomes flatter towards equator.
For the radial velocity of the flow, we always start from transonic Bondi solution and the sonic radius is initially located at 80 $r_{g}$.
In most cases, the angular momentum imposed on the flow is changing drastically the topology of solutions, and 
the sonic surface shape becomes aspherical (hence we use the term 'sonic surface' rather than the 'point'). Also, in some cases multiple transonic solutions can be found, depending on the model, and an inner sonic point is found closer to the black hole, where a transient shock develops. Inside this region, the 'bubble', or 'mini-disk' is formed. The mini-disk is rotationally supported, so in general its size increases for a higher angular momentum content. It is present for
a relatively long simulation time, and
its innermost, densest part
is enclosed with $\sim 20 r_{g}$.
Its temporal behavior and quantitative properties are however sensitive to the global parameters of the model, and presence or absence of magnetic fields.

The angular momentum distribution at initial state is given by the parameter $S$ and specific function of the polar angle. The black hole rotation is 
initially parameterized by the dimensionless Kerr parameter $A_{0}$, and then the black hole spin changes during the simulation. It may either decrease or increase, and the final spin is measured when the simulation ends (the end time of most simulations is fixed at $20000 M$ \footnote{ gravitational times, $t_{\rm g} = GM/c^{3}$, in the units of BH mass. For the black hole of 3 Solar masses the physical time of our simulation would be $\sim 1.5$ s.}).

\begin{figure*}
 \begin{tabular}{ccc}
  \includegraphics[width=0.3\textwidth]{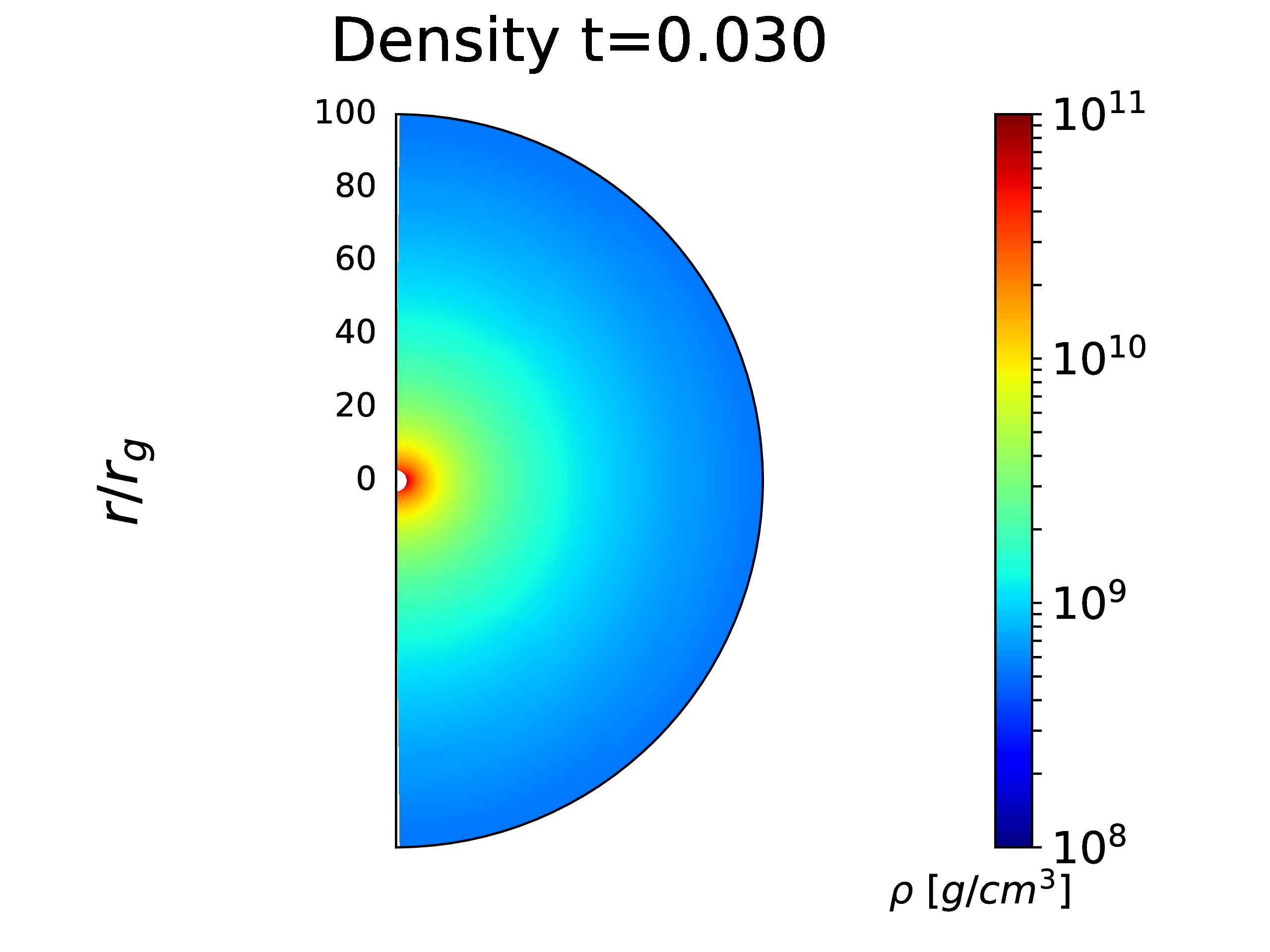} &
    \includegraphics[width=0.3\textwidth]{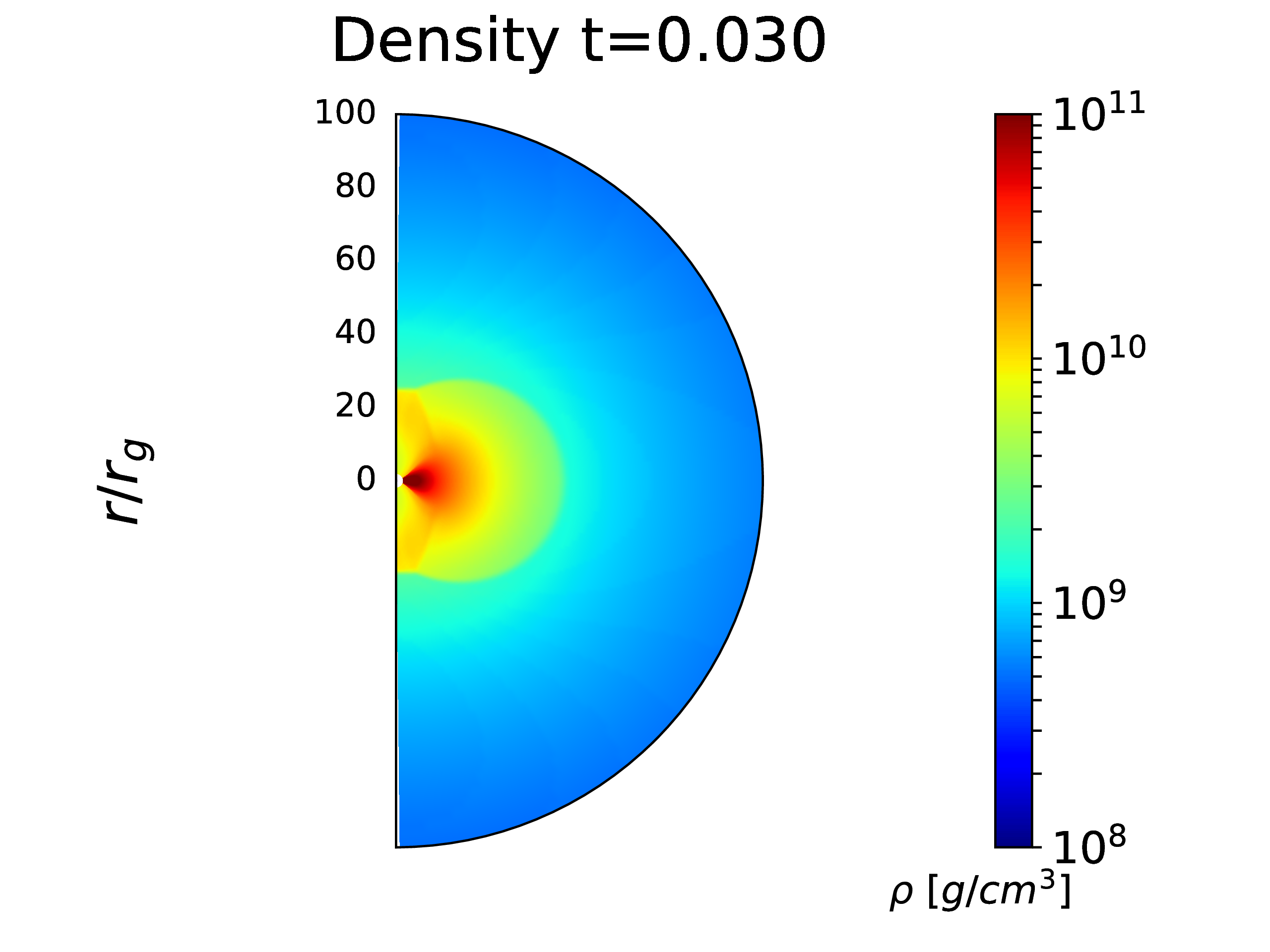} &
    \includegraphics[width=0.3\textwidth]{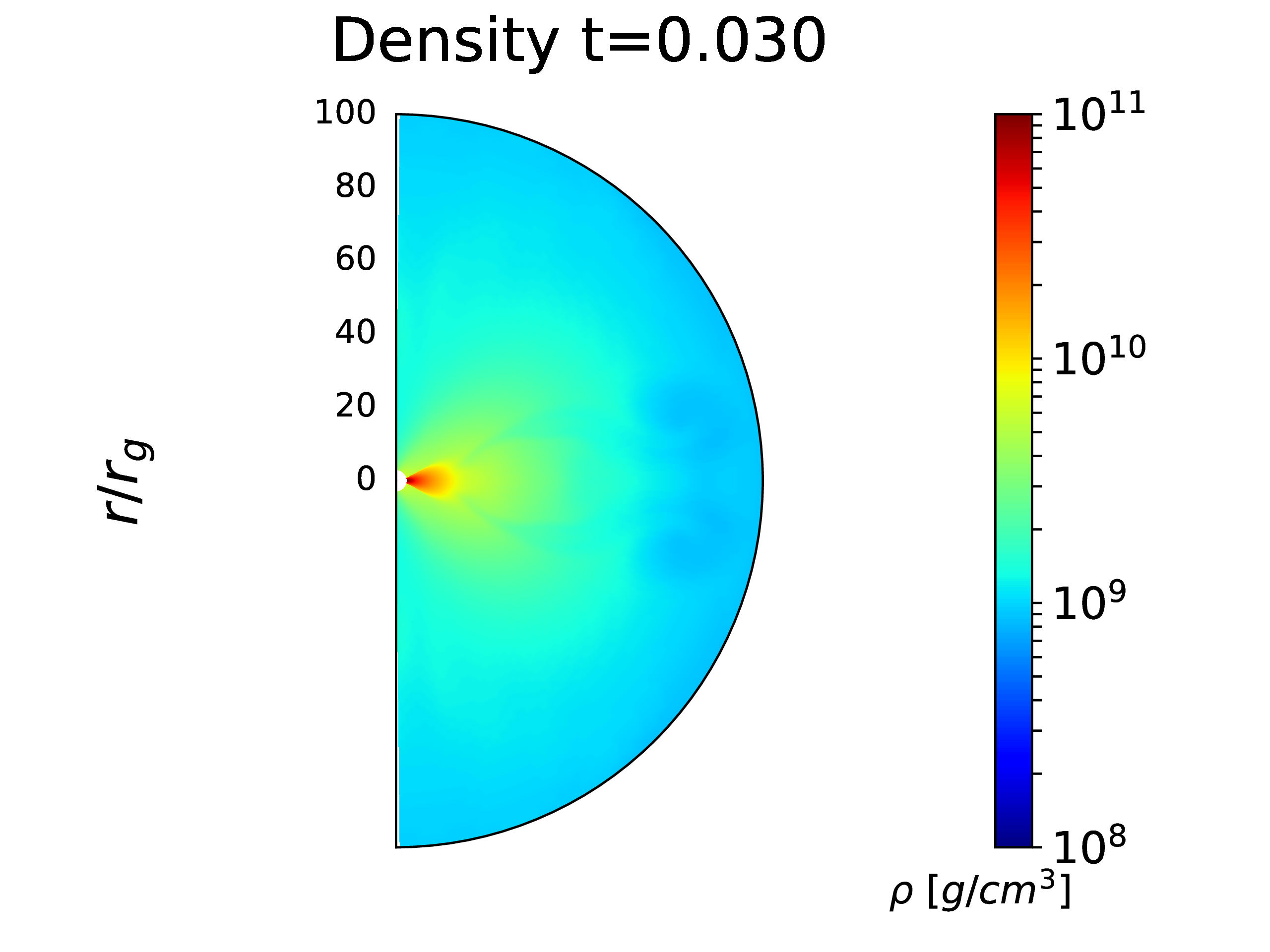}
    \end{tabular}
    \caption{Density distribution in the innermost regions of the accretion flow, within 100 $r_{\rm g}$. Snapshots show models with $A_0=0.3$ and rotation parameter of $S=0.4$ (NS-T-04), $S=1.0$ (NS-T-10), and $S=1.4$ (NS-T-14), from left to right, as taken at the beginning of these simulations. Time snapshots were taken at time $t=0.03$ s. }
     \label{fig:Profiles_initial}
     \end{figure*}

In Figure \ref{fig:Profiles_initial} we show the initial distributions of density in the cloud depending on the rotation parameter.
The sub-critical models do not contain enough angular momentum in the envelope to form a disk bubble at the equatorial plane. Hence, the distribution of density in the gas is spherically symmetric and resembles the Bondi flow. Material from both polar and equatorial regions contributes to the growth of black hole mass, as it cannot be kept by rotation on the orbit and will flow under the horizon. 
We note however, that even if there is a small amount of angular momentum in the gas, there will be dissipation
of energy at the equator, because of shocks and compression 
(cf. \citet{Murguia2020}).
For critical rotation in the envelope, a rotationally supported torus may form at the equator, and material can accrete only through the polar regions.

Below we present results for two sets of models: non-magnetized and magnetized.
The summary of adopted model parameters and the results are given in Table \ref{tab:models}. 

\subsection{Simulations with non-magnetized accretion flows}
\label{sec:NonMag}

\begin{figure*}
 \begin{tabular}{ccc}
  \includegraphics[width=0.3\textwidth]{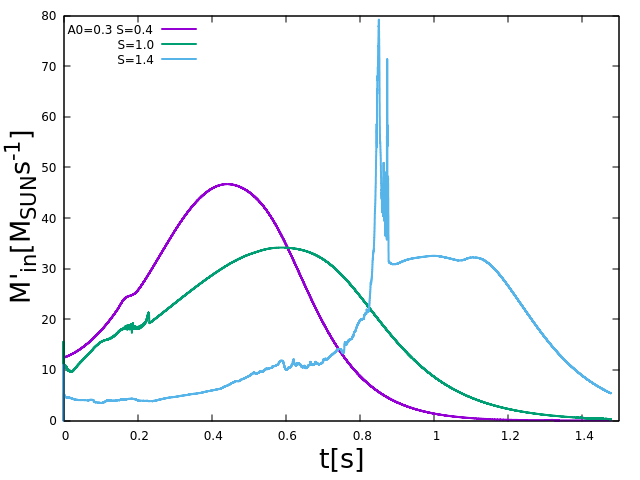} &
    \includegraphics[width=0.3\textwidth]{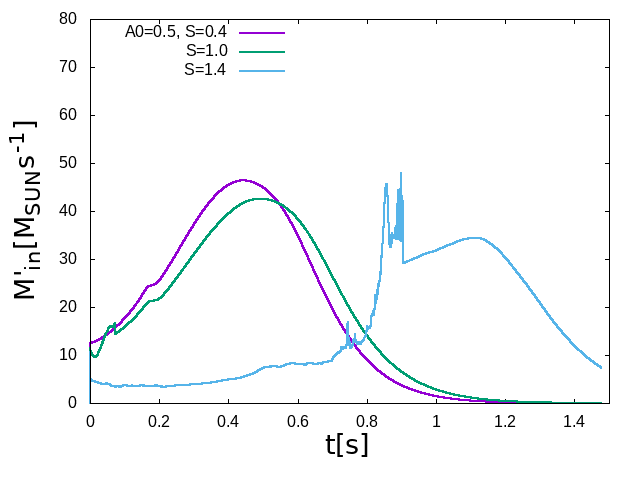} &
    \includegraphics[width=0.3\textwidth]{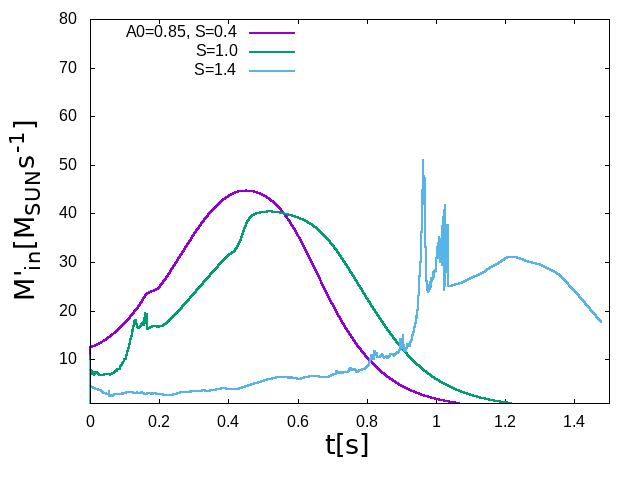}
    \end{tabular}
    \caption{Time dependence of the mass flux through the inner boundary for three different initial black hole spin. Left panel -- $A_0=0.3$, middle panel -- $A_0=0.5$, right panel -- $A_0=0.85$. 
      Cyan lines denote the models with $S=1.4$
      while the violet lines denote $S=0.4$ and green lines $S=1.0$.
            }
     \label{fig:mdotin}
     \end{figure*}

In Figure \ref{fig:mdotin} we present the time evolution of the main important quantity, the mass accretion rate onto black hole, for all the 9 models without magnetic field. 
The models shown differ with respect to rotation magnitude in the accretion flow, and with the initial black hole spin. Our fiducial models were run until the end time of $t=1.5 - 4.5 \textrm{s}$.

We observe that the accretion rate onto black hole is small initially, but then rises, and the steepest rise is always found in the subcritical rotation models ($S=0.4$). The critical and supercritical rotation of the flow results in a somewhat slower rise of the accretion rate, and smaller value at the peak. This is because large part of the material is kept by rotation in the centrifugally-supported mini-disk, while only the flow accreting from the polar regions contributes to the mass accretion rate through the inner radius, $\dot M_{in}$. Moreover, for critical rotation some fluctuations of the accretion rate appear with low amplitudes, at the initial times of the simulation. They appear for all simulations with $S=1.0$, but at different time, depending on the initial value of the spin parameter. For supercritical rotation, $\dot{M}$ at the beginning is steady and low in comparison to other cases. Small amplitude fluctuations develop later in time (around $\sim 0.6s$ to $\sim 0.8s$). Also, for $S=1.4$ models, the accretion rate oscillations with very large amplitude appear transiently, before the $\dot{M}$ becomes roughly constant for a period of time, and then starts decreasing. 
In this case moment of appearance of this feature is weakly sensitive to the value of initial black hole spin.
The largest amplitude of oscillations is obtained for the model with small initial spin, $A_{0}=0.3$.

\begin{figure*}
 \begin{tabular}{ccc}
  \includegraphics[width=0.3\textwidth]{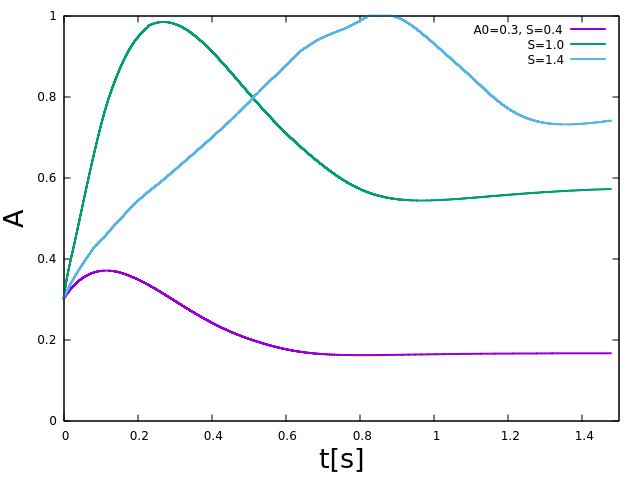} &
    \includegraphics[width=0.3\textwidth]{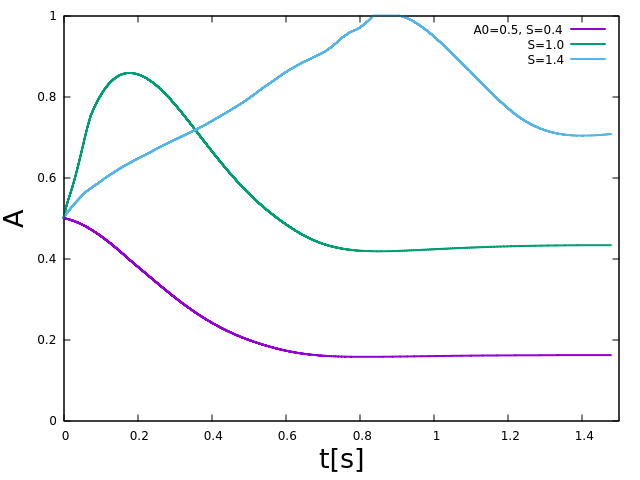} &
    \includegraphics[width=0.3\textwidth]{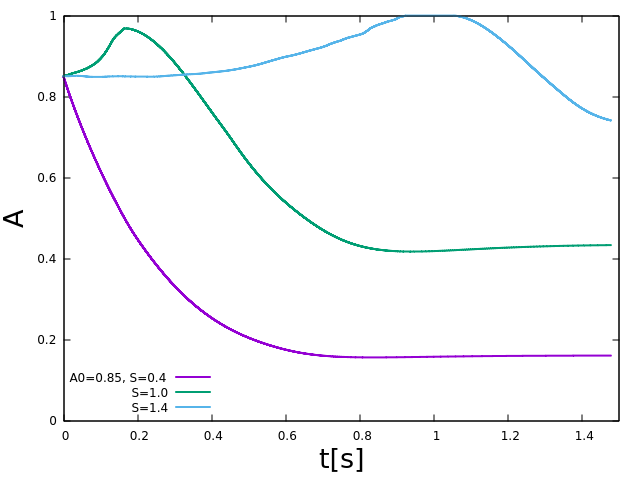}
    \end{tabular}
    \caption{Time dependence of the black hole spin for three different initial black hole spins. Left panel -- $A_0=0.3$, middle panel -- $A_0=0.5$, right panel -- $A_0=0.85$. 
      Cyan lines denote the models with $S=1.4$
      while the violet lines denote $S=0.4$ and green lines $S=1.0$.     }
      \label{fig:BHspin}
     \end{figure*}

Evolution of the black hole spin is presented in Fig. \ref{fig:BHspin}. For subcritical rotation only in the case of $A_0=0.3$ we observe increase of the black hole spin at any stage of the simulation, when it grows to the value of $\sim 0.37$. Then it starts to decrease and finally saturates at the value of $A \sim 0.17$. In case of other initial spin parameters, the value of bh spin starts to decrease from the very beginning of the simulation and  saturates at a similar value.
In case of critical rotation of the envelope there is no simple correspondence between the initial and maximal value of spin parameters. The highest maximal spin is achieved  in case of $A_0=0.3$  ($A_{max}\sim 0.98$), intermediate initial spin $A_0\sim 0.5$ gives the lowest value of maximal spin ($A_{max}\sim 0.87$), and the highest initial spin, $A_0\sim 0.85$, gives maximal spin value during evolution at the level of $A_{max}\sim 0.92$. Final spin value is similar for every critical case with values from $\sim 0.4$ to $\sim 0.57$.
In the case of supercritical rotation models it is worth to compare spin evolution to the accretion rate evolution described above. For all $A_{0}$ values, the spin is growing since the beginning of the simulation and achieves the maximal value $\sim 1$ when the high amplitude oscillation of accretion rate appears. Spin maintains its maximal value during those oscillations. Then it starts to drop to the final value which is around $\sim 0.7$ for all supercritical models. Exact values can be found in the Table \ref{tab:models}.    

We conclude that the main factor which determines evolution of the spin and accretion rate during the simulation is the angular momentum content in the envelope, i.e., the value of $S$ parameter. Models are further depending on the $A_0$ value, and it seems to shift the moment in time at which a characteristic feature arises, and also changes its duration and the amplitude. Nevertheless, the occurrence of those characteristic features is preserved.

\begin{figure*}
 \begin{tabular}{ccc}
  \includegraphics[width=0.3\textwidth]{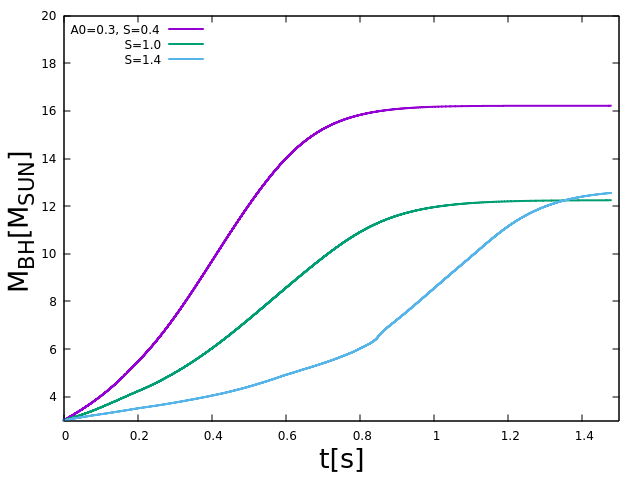} &
    \includegraphics[width=0.3\textwidth]{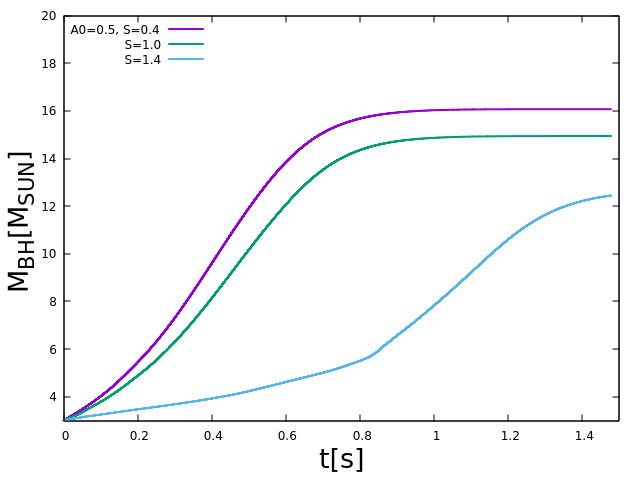} &
    \includegraphics[width=0.3\textwidth]{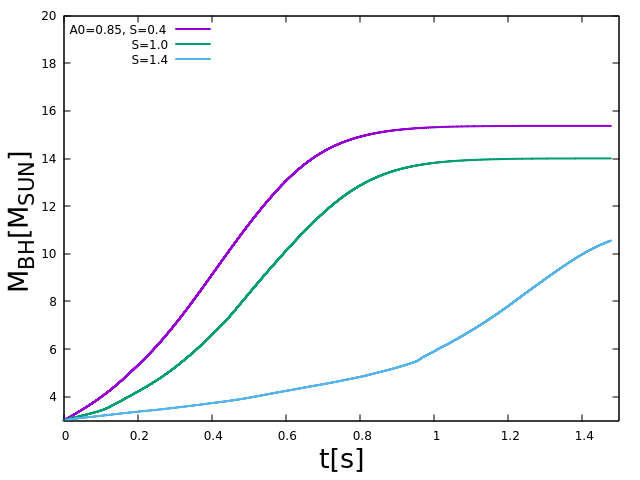}
    \end{tabular}
    
    \caption{Time dependence of the black hole mass for three different initial black hole spins. Left panel -- $A_0=0.3$, middle panel -- $A_0=0.5$, right panel -- $A_0=0.85$. 
      Cyan lines denote the models with $S=1.4$
      while the violet lines denote $S=0.4$ and green lines $S=1.0$.   
    }
      \label{fig:BHmass}
     \end{figure*}

In Fig. \ref{fig:BHmass} we present the evolution of the black hole mass during simulations. For every initial spin value the highest black hole mass is achieved for $S=0.4$. Initial spin $A_0=0.3$ is the only case when  the supercritial rotation is more efficient for mass gain than critical. For the initial part of the simulation in every case the black hole mass increases linearly. In case of supercritical rotation the slope of $M_{BH}(t)$ changes, which corresponds to the beginning of high amplitude oscillations of accretion rate.

The two dimensional structure of accretion flow in our simulations is as follows.

\begin{figure*}
\begin{tabular}{ccc}
 \hspace{-10mm}\includegraphics[width=0.33\textwidth]{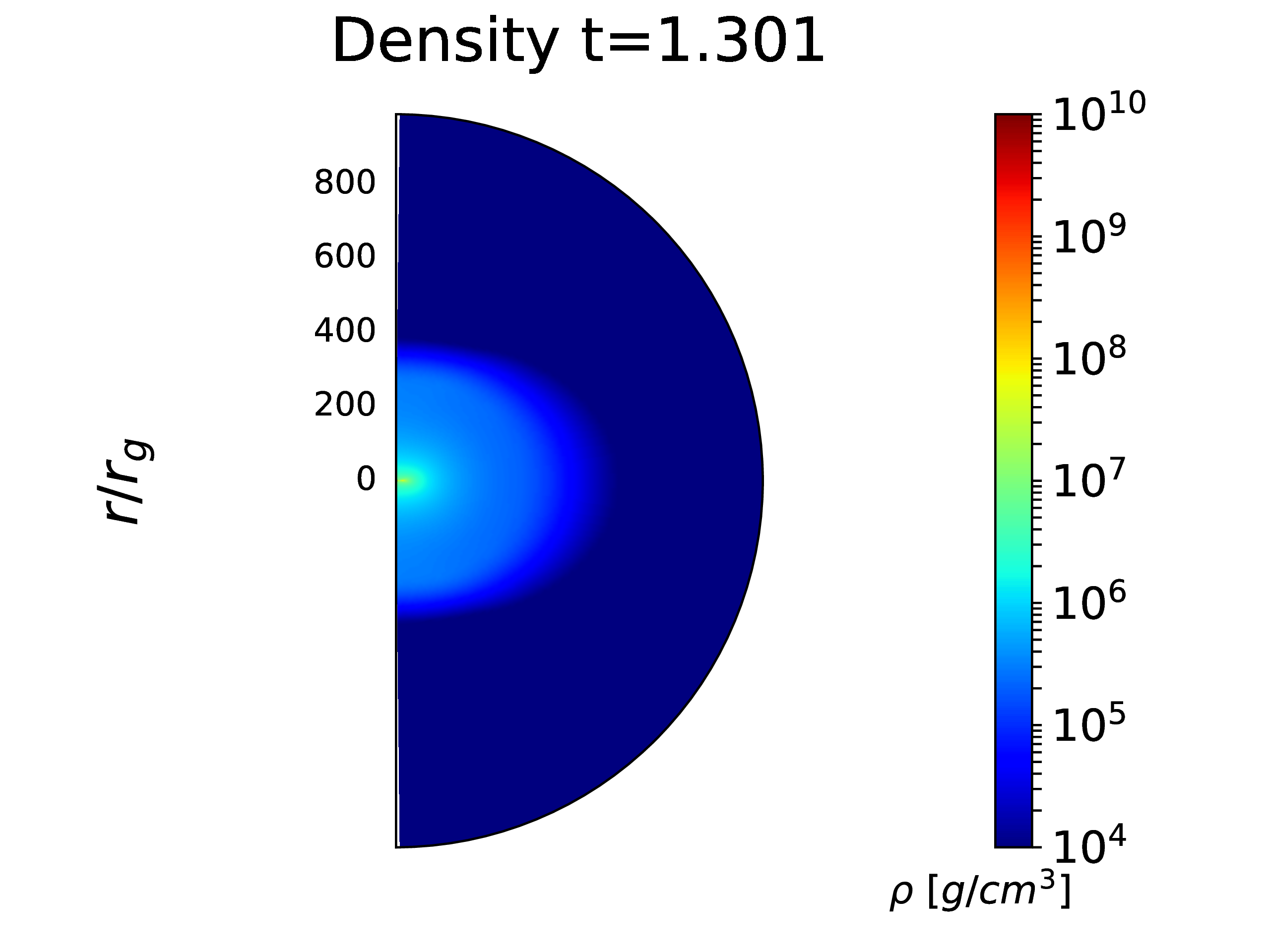} &
\includegraphics[width=0.33\textwidth]{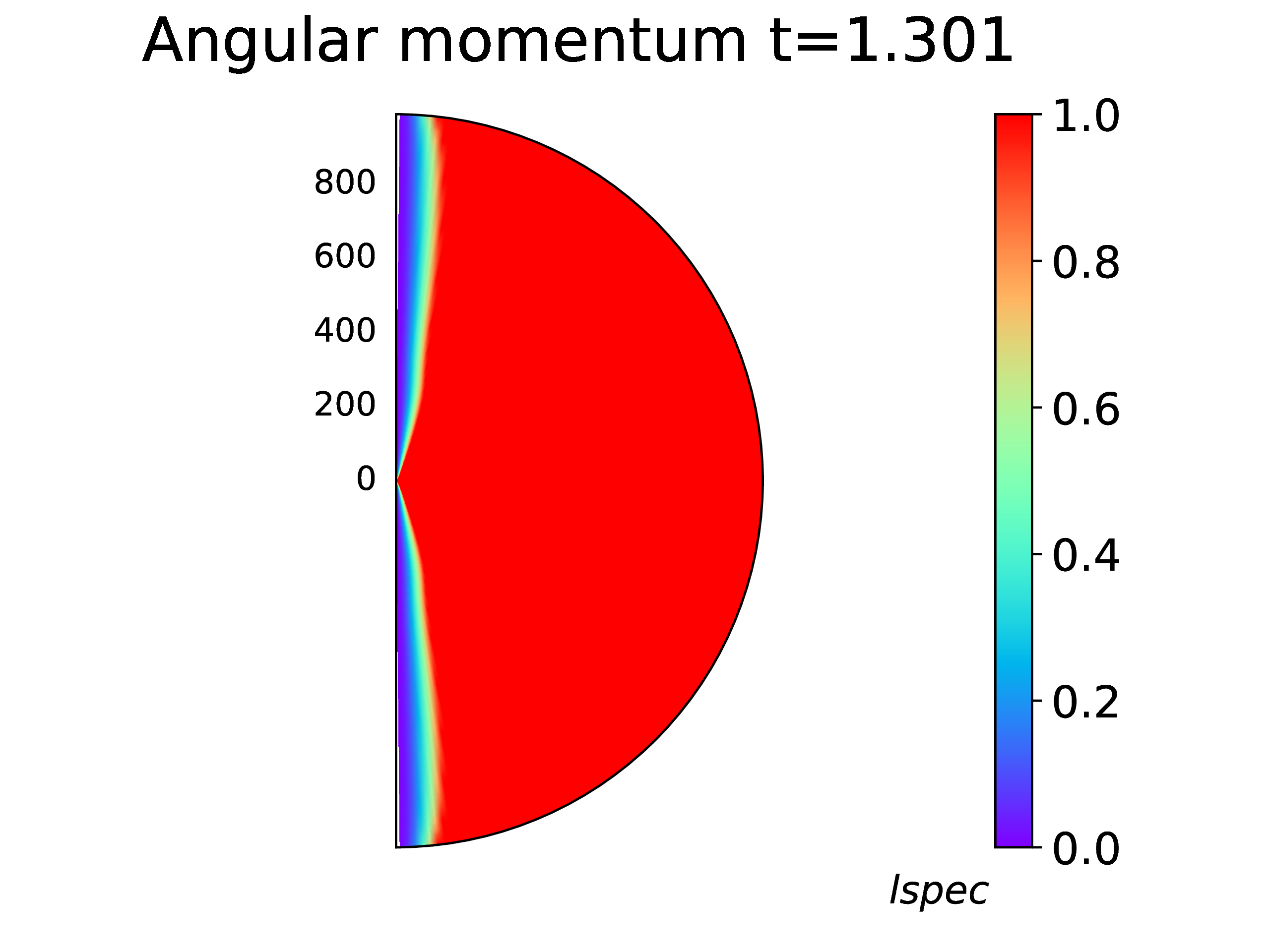} &
\includegraphics[width=0.33\textwidth]{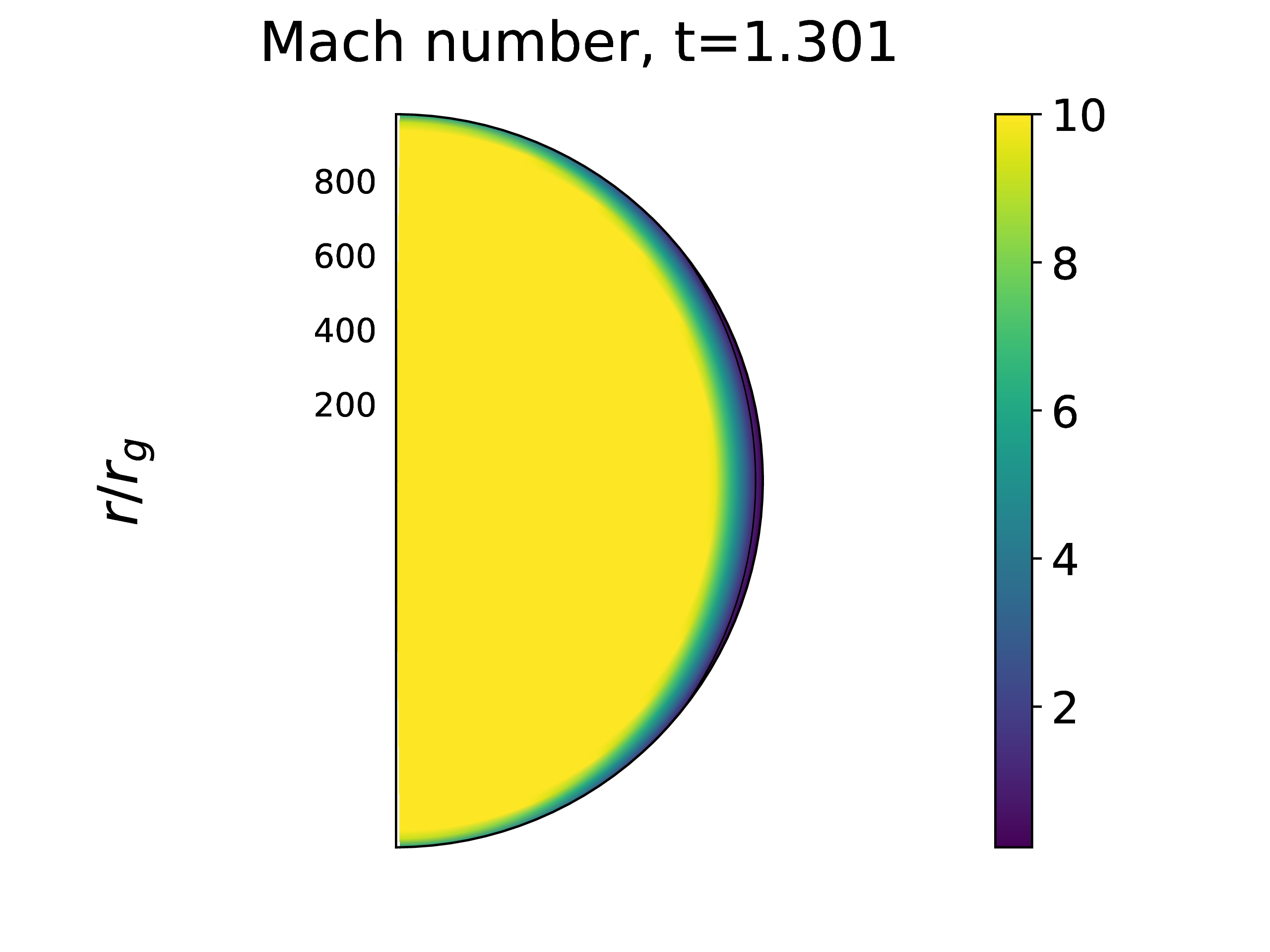} 
\\
 \hspace{-10mm}\includegraphics[width=0.33\textwidth]{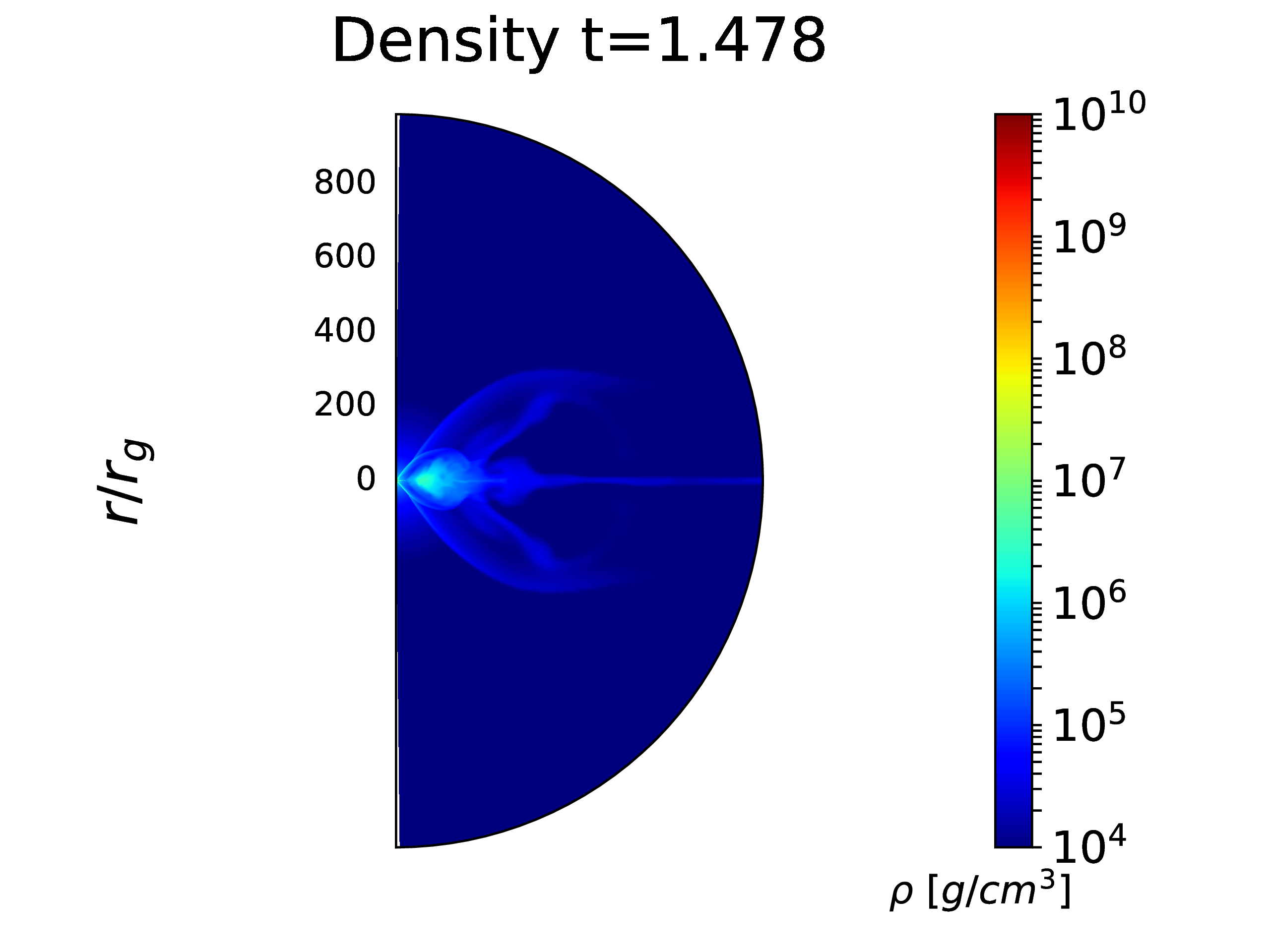} &
\includegraphics[width=0.33\textwidth]{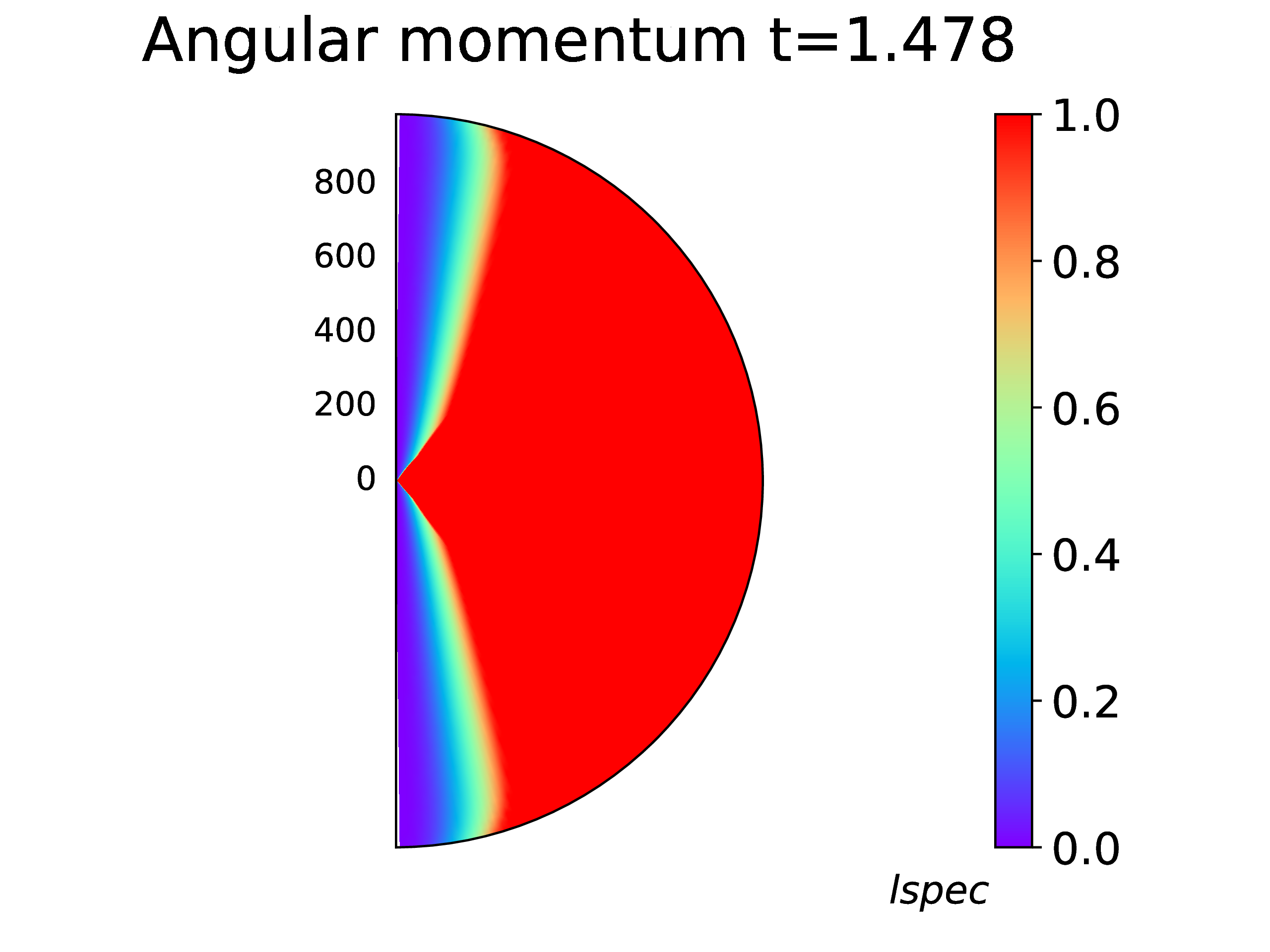} &
\includegraphics[width=0.33\textwidth]{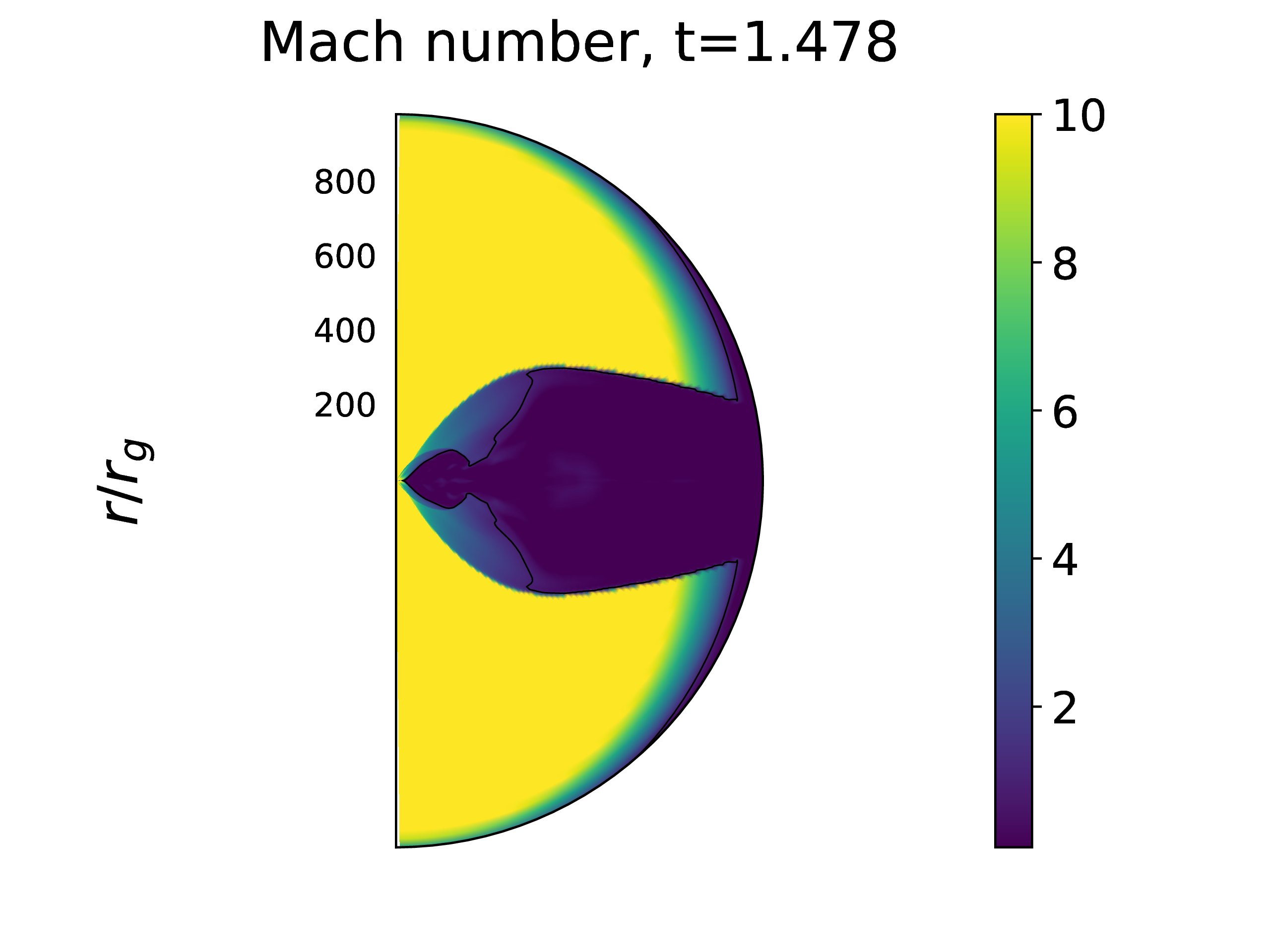} 
\end{tabular}
\caption{The results, from left to right, for the \textit{Density, Specific Angular Momentum, and Mach Number} distributions, for one of the models with neglected magnetic fields. The initial black hole spin was $A_{0}=0.3$, and rotation parameter was $S=0.4$ (NS-T-04 model, non-magnetized).
  The color maps are taken in the end of the simulation.
  Note that in the middle panels, the spatial scale of the plot is zoomed in to 100 $r_{\rm g}$.
  In addition contour of $M=1$ is shown with a black line.
}    \label{fig:models_lowA0_endsim}
\end{figure*}

In Figure \ref{fig:models_lowA0_endsim} we show the 2-D distributions of density, angular momentum, and Mach number at the end of simulation of the most slowly rotating flow with $S=0.4$. The model shows the case of initial black hole spin $A_{0}=0.3$ (model $NS-T-04$). Rotation in the envelope was below critical, so the structure of the flow was close to spherical, for most of the time. Somewhat flattened, Bondi-like structure accreted onto black hole uniformly, and the envelope became almost empty by the time $t \sim 1.5$ s. Mass of the black hole reached about $16 M_{\odot}$, while its spin has dropped down to about $A=0.17$, because of accretion of very low angular momentum material (see above).
An interesting feature (a 'butterfly-shape') developed in the equatorial plane of the envelope at the end of the collapse, after time $t\sim 1.35$ s, which we show in the Figure \ref{fig:models_lowA0_endsim}.

The fiducial model (NS-T-04) was ran until time $t=1.4$. In addition, we ran a longer simulation to verify that the 'butterfly-shape' feature develops into a disk-like structure at the equatorial plane, which is sustained for up to $2 \textrm{s}$.
The duration of this feature additionally depends on the $A_0$ value.

In the equatorial plane, a remnant material with relatively high angular momentum reached the innermost regions of the collapsar and is kept by rotation. This mini-bubble with subsonic accretion speed survived at the equator for a long time, preventing the cloud from getting completely evacuated. The supersonic matter overpasses the bubble at higher latitudes, and sinks under the black hole horizon.

Evolution of the density structure in collapsing star corresponds to features visible in $\dot{M}$ behavior. As mentioned above they are determined mostly by the $S$-parameter value. Simulations for the same rotation magnitude $S$  and different initial spin $A_0$ give similar results, and characteristic moments are just shifted in time.
\begin{figure*}
	\begin{tabular}{ccc}
		\hspace{-10mm}\includegraphics[width=0.35\textwidth]{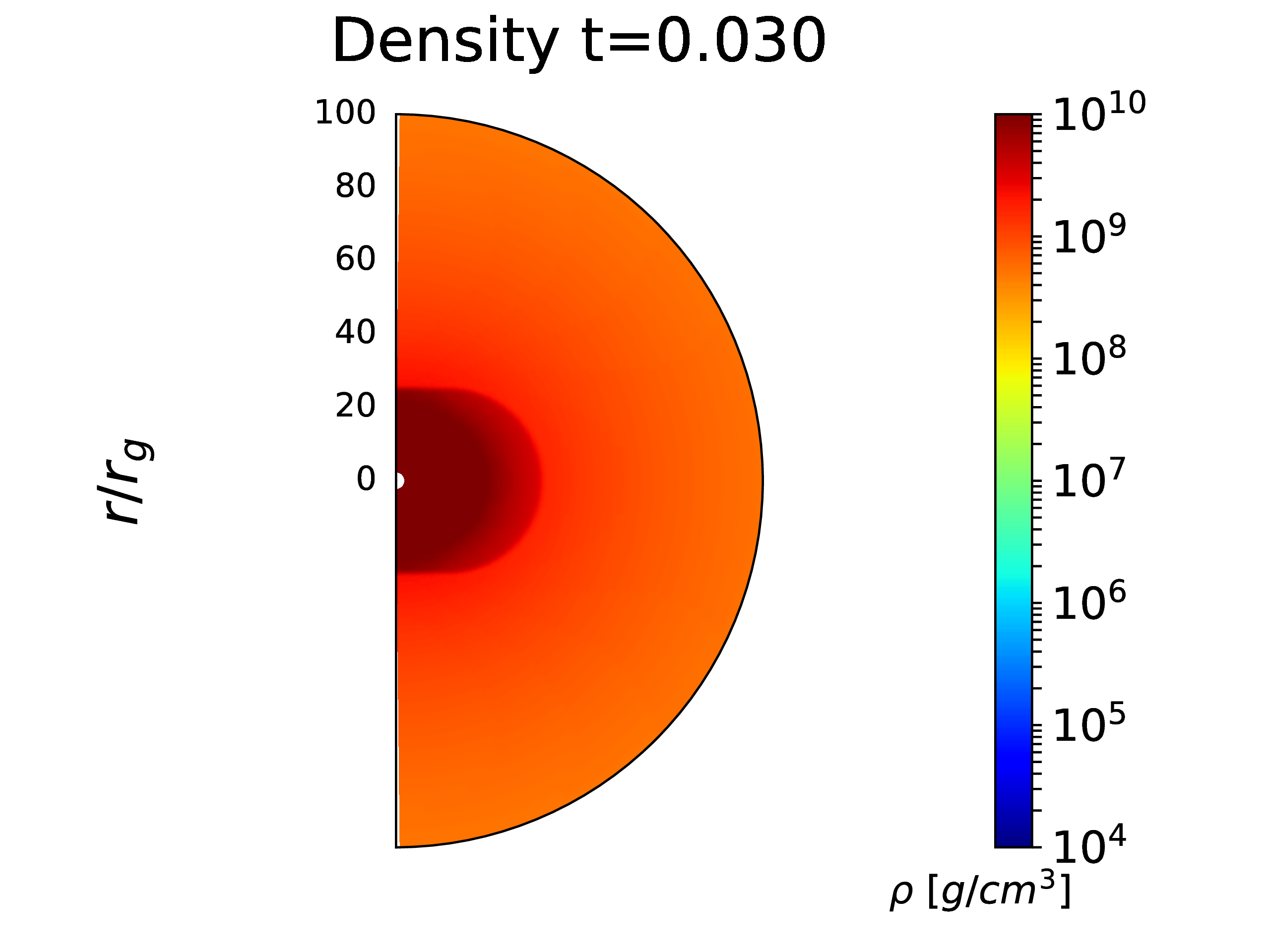} &
		\includegraphics[width=0.35\textwidth]{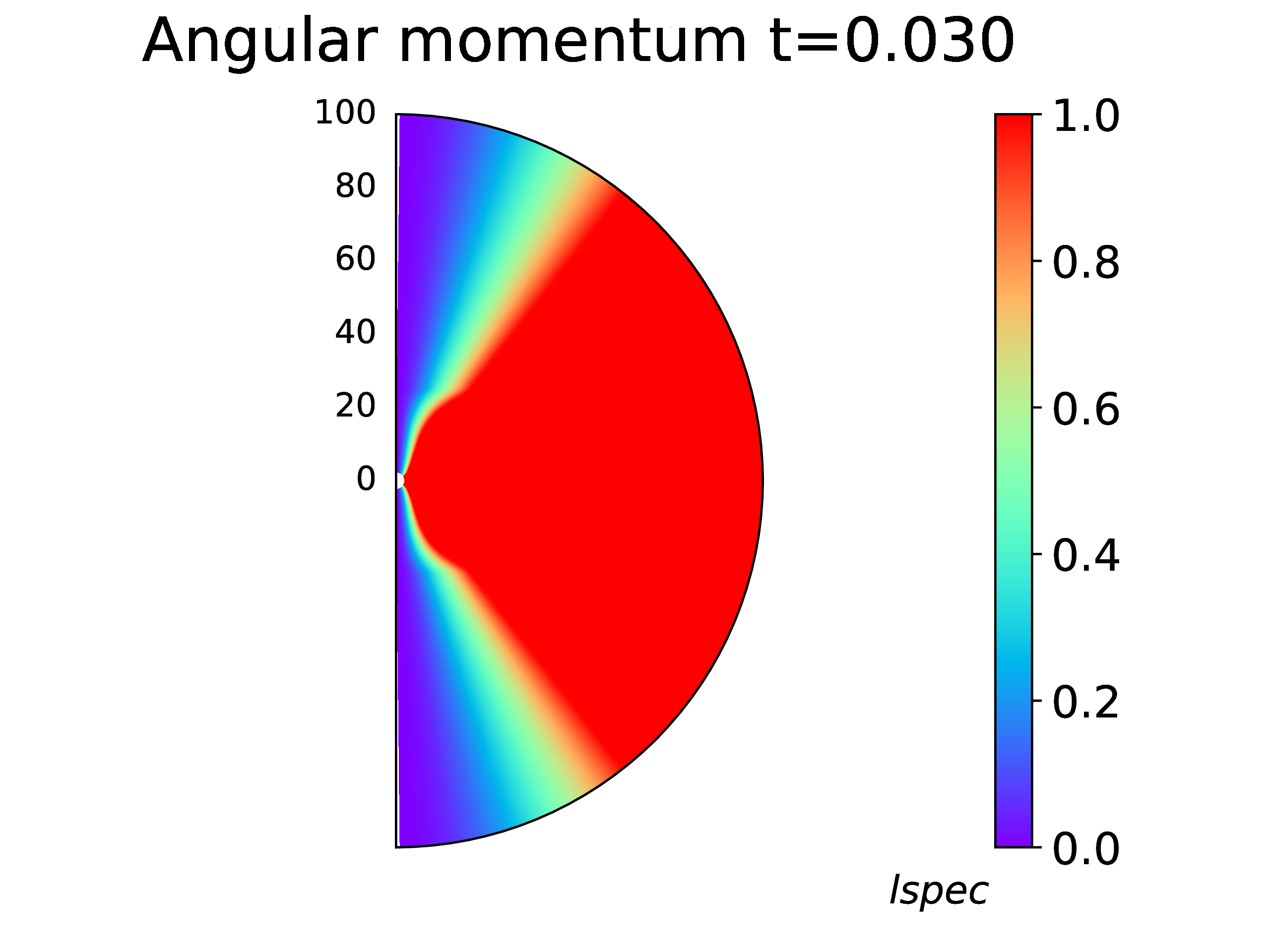} &
		\includegraphics[width=0.35\textwidth]{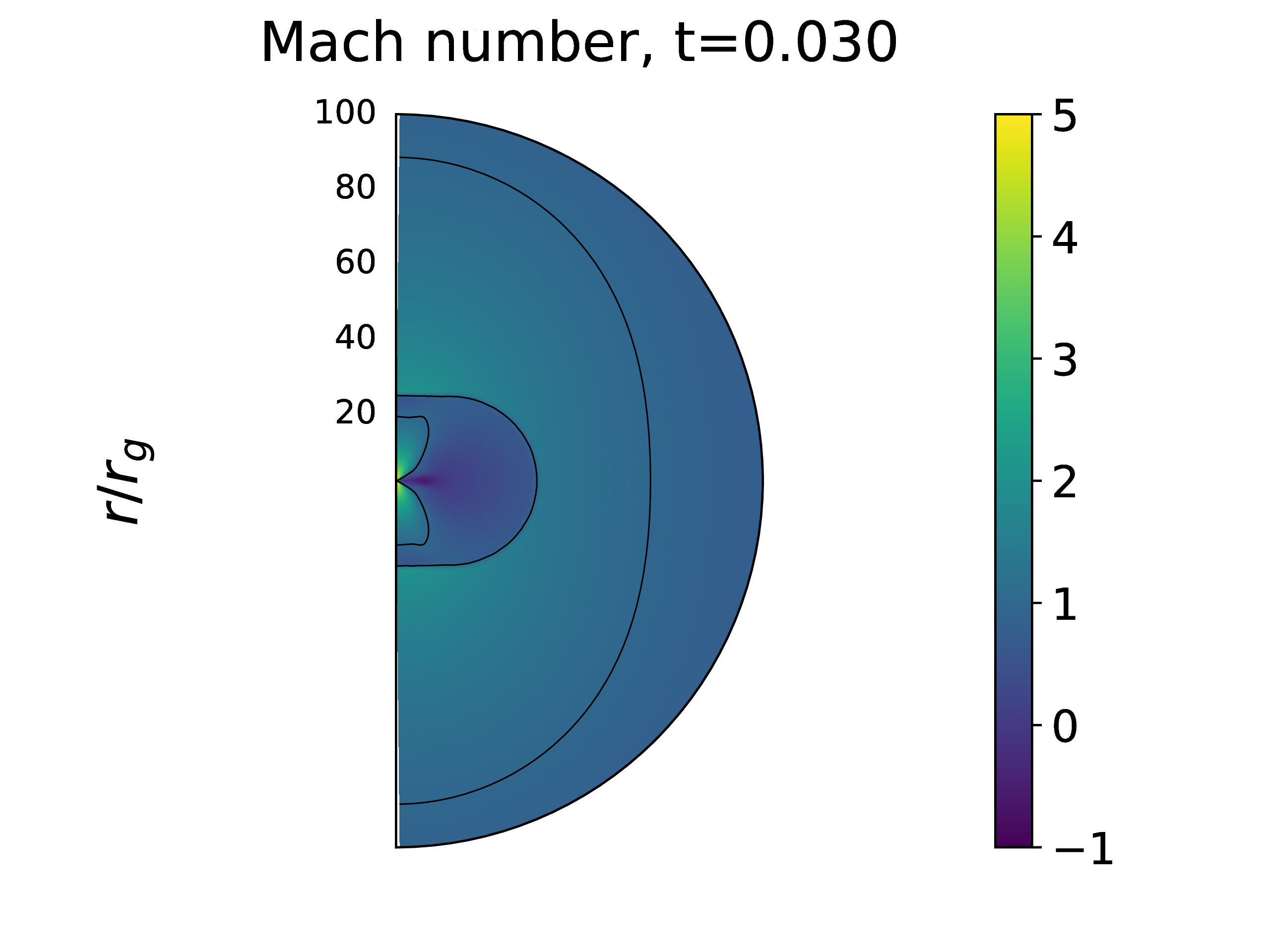}
		\\
		\hspace{-10mm}\includegraphics[width=0.35\textwidth]{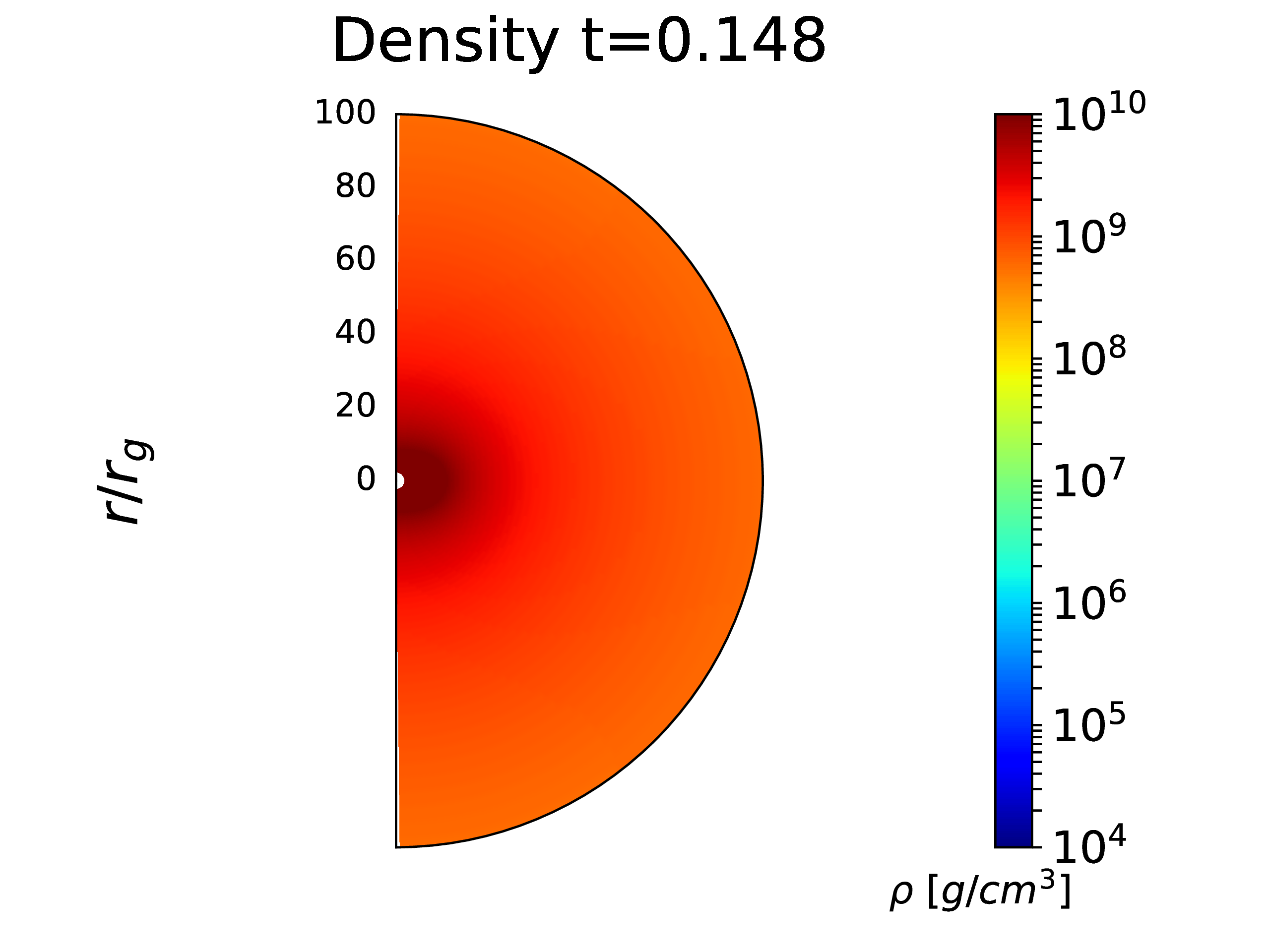}  &
		\includegraphics[width=0.35\textwidth]{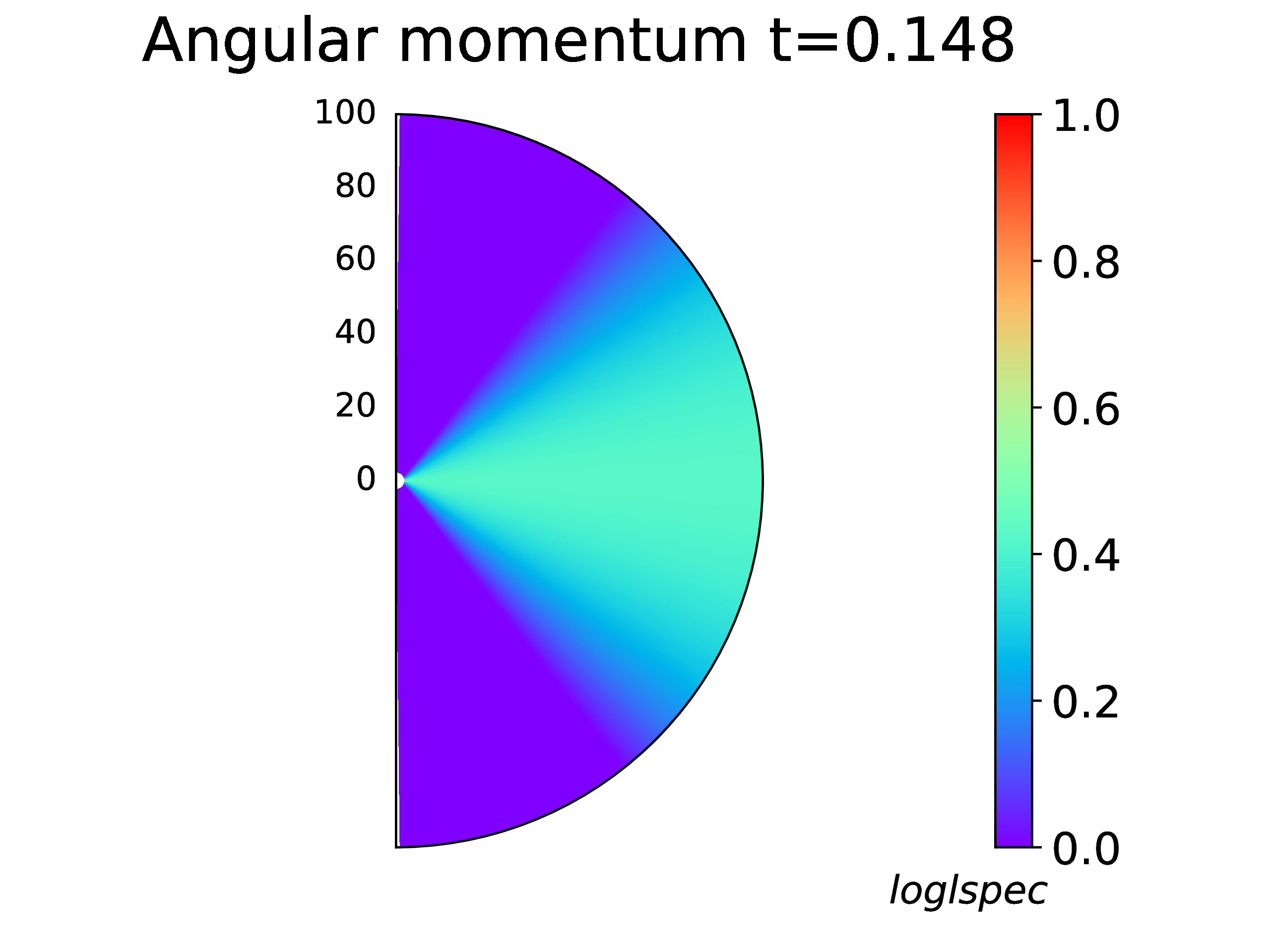}   &
		\includegraphics[width=0.35\textwidth]{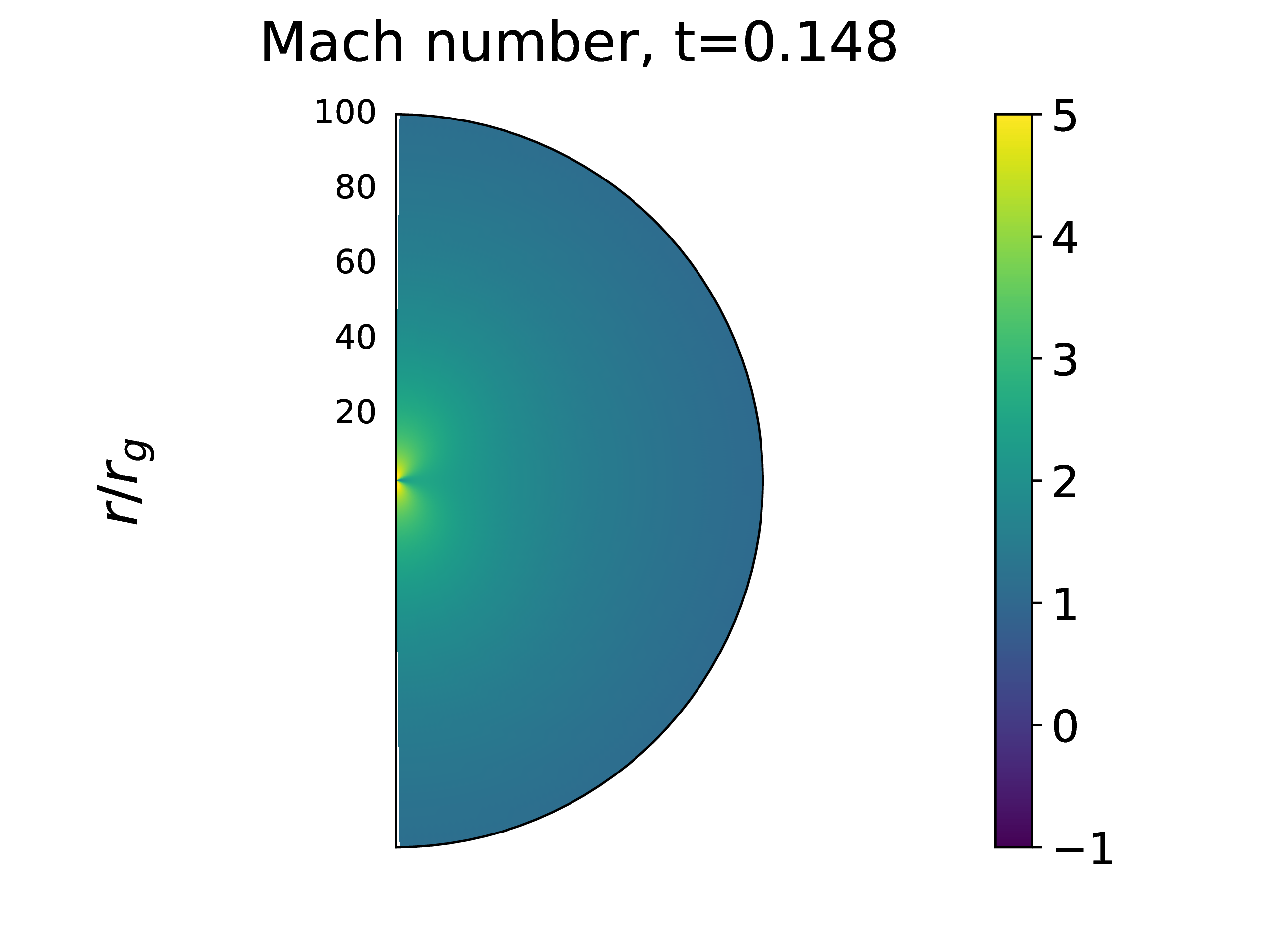}
	\end{tabular}
	\caption{The results, from left to right, for the \textit{Density, Specific Angular Momentum, and Mach Number} distributions, for the model with neglected magnetic fields and initial black hole spin of $A_{0}=0.85$ and the rotation parameter $S=1.0$ (model HS-T-10).
		The color maps show  profiles at the beginning of the simulation.
	}
	\label{fig:a085s10_1}
\end{figure*}

\begin{figure*}
\begin{tabular}{ccc}
 \hspace{-10mm}\includegraphics[width=0.35\textwidth]{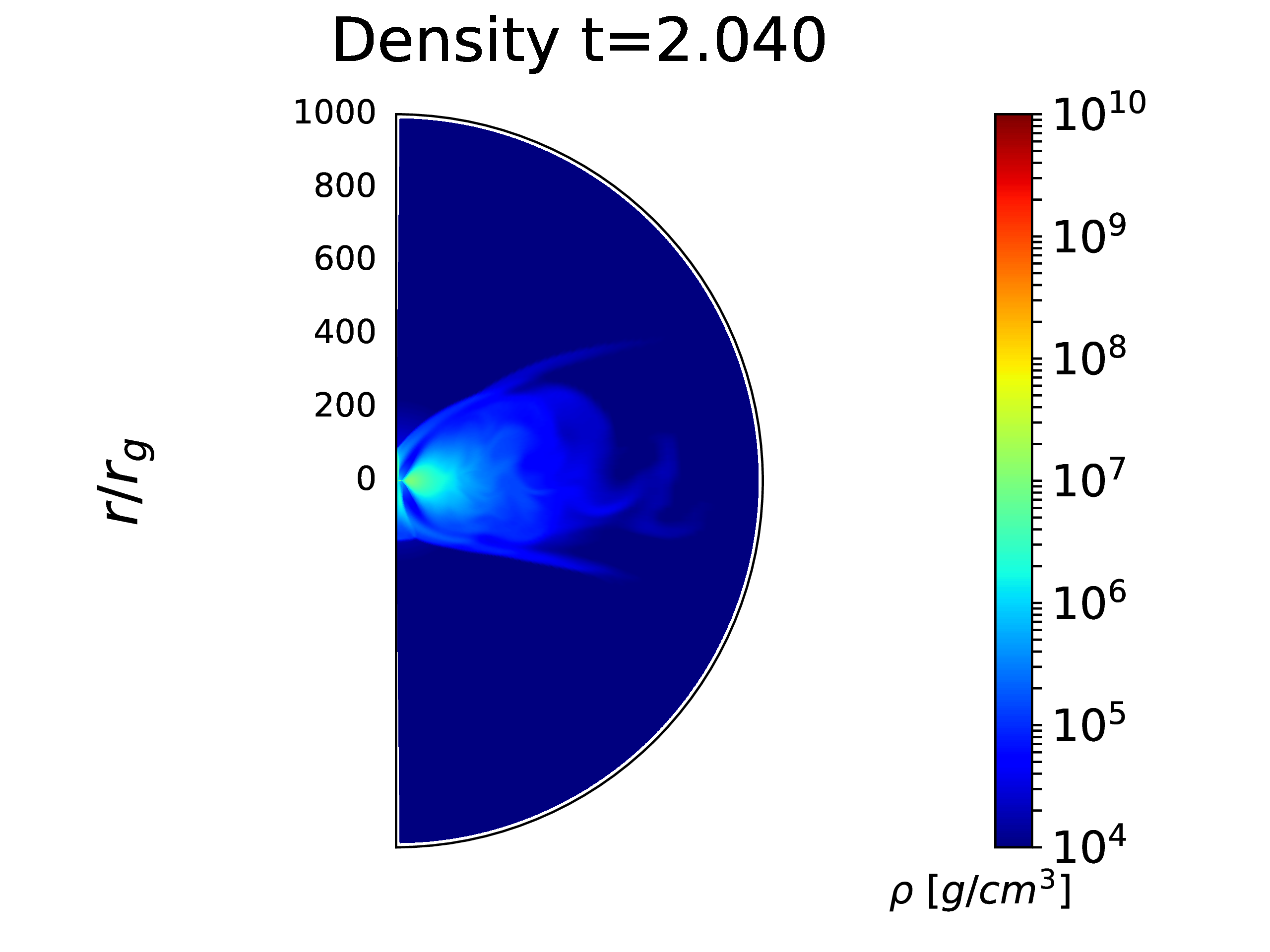} &
\includegraphics[width=0.35\textwidth]{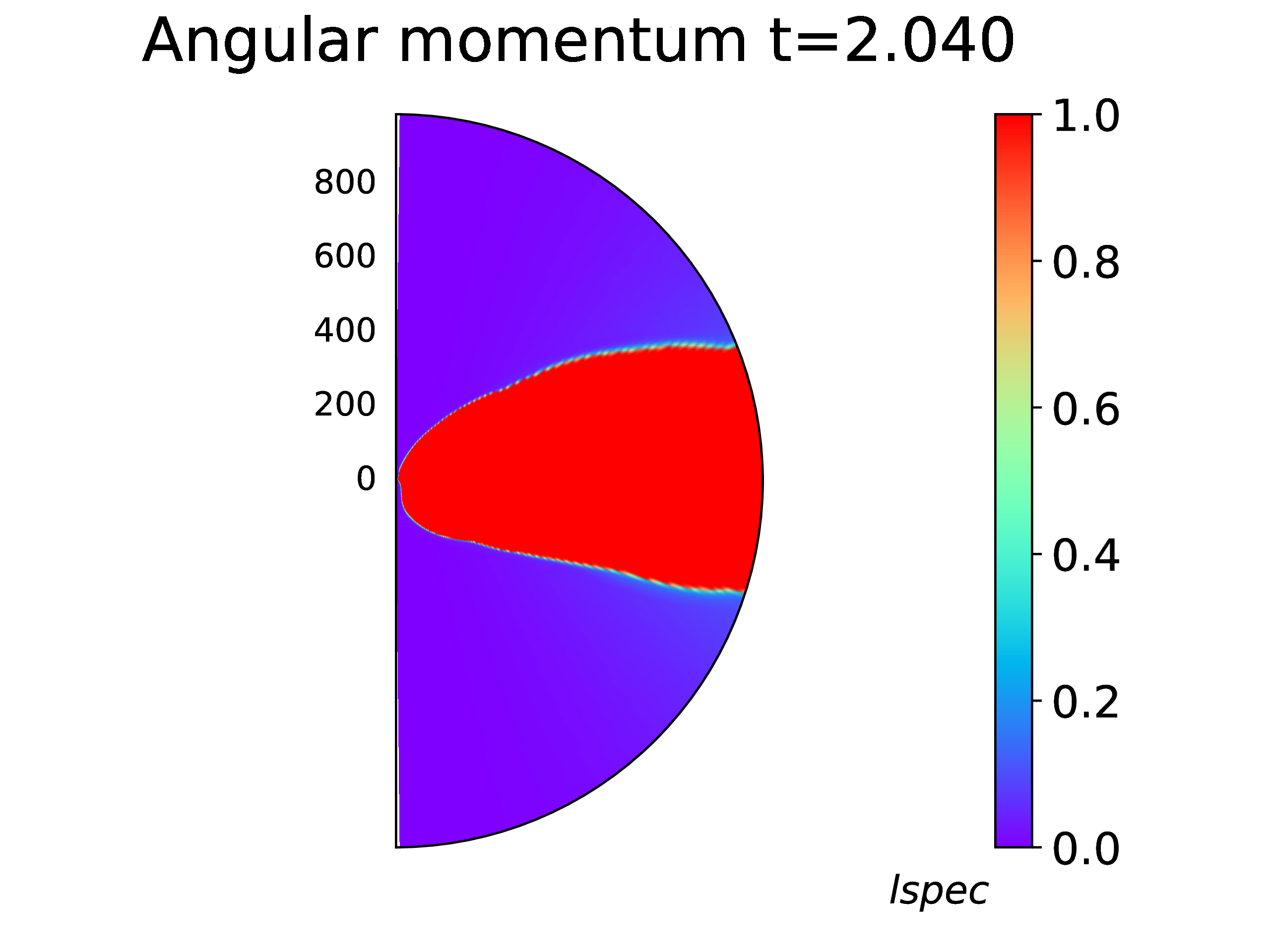} &
\includegraphics[width=0.35\textwidth]{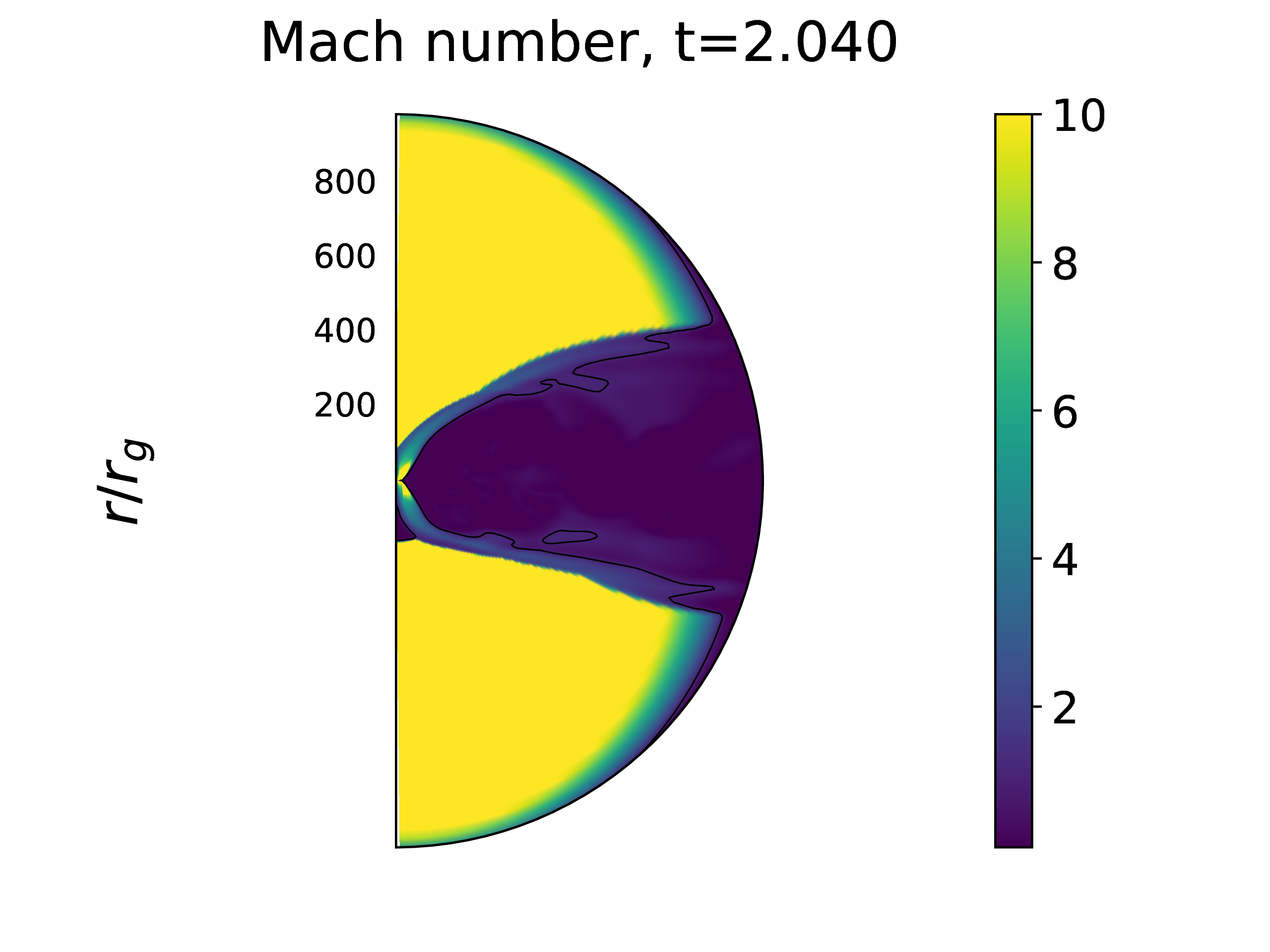}
\\
  \hspace{-10mm}\includegraphics[width=0.35\textwidth]{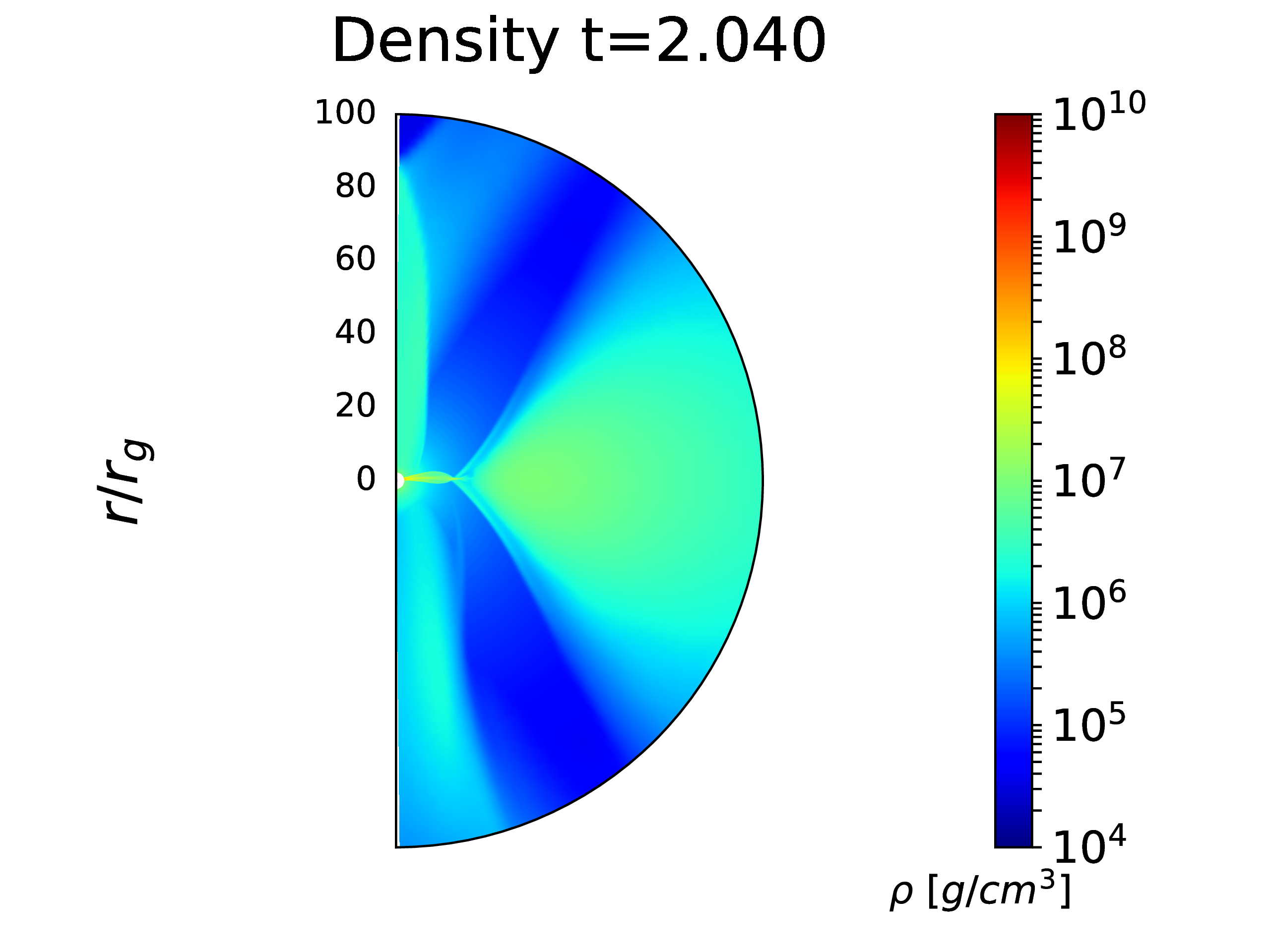}  &
\includegraphics[width=0.35\textwidth]{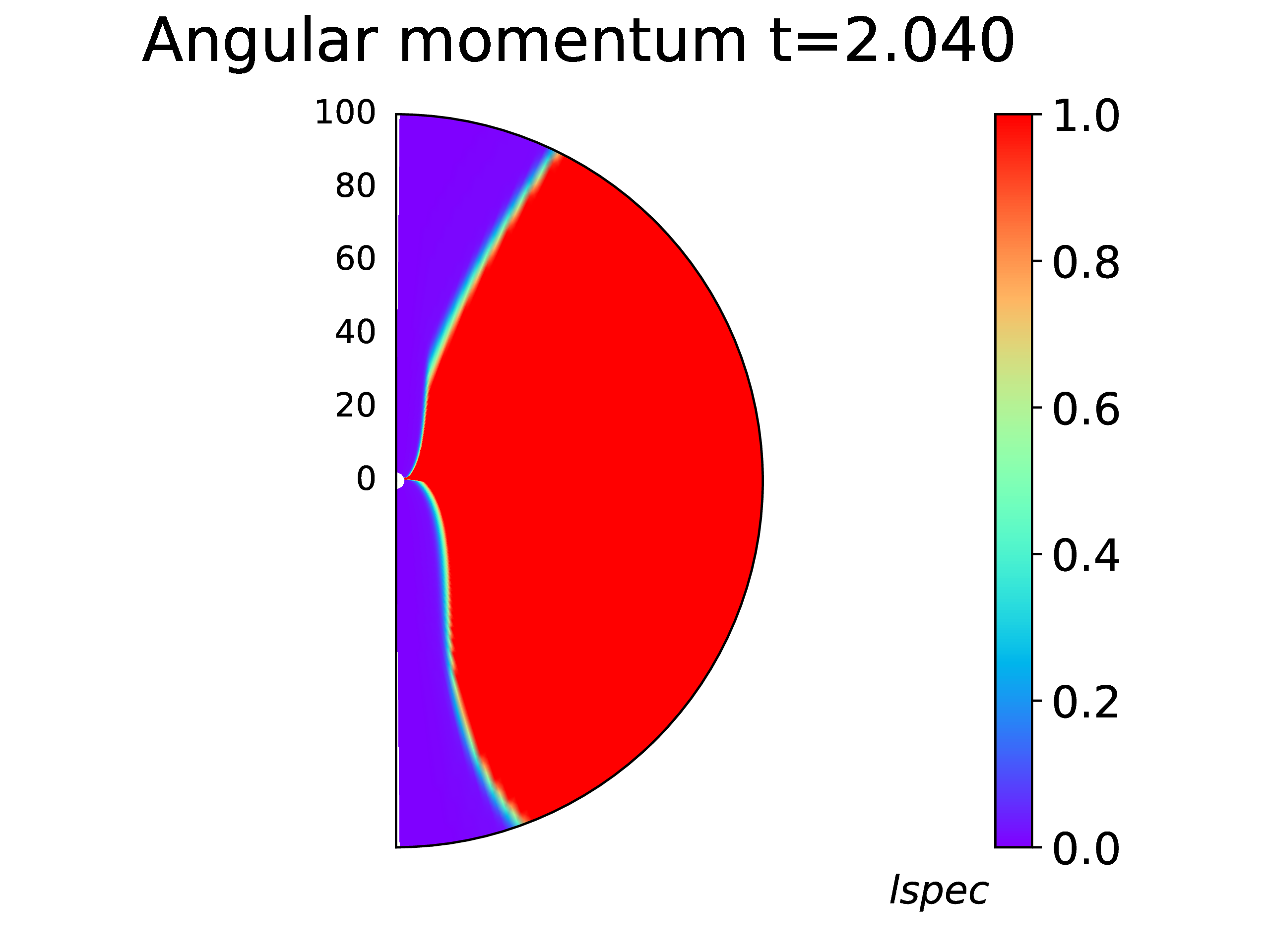}   &
\includegraphics[width=0.35\textwidth]{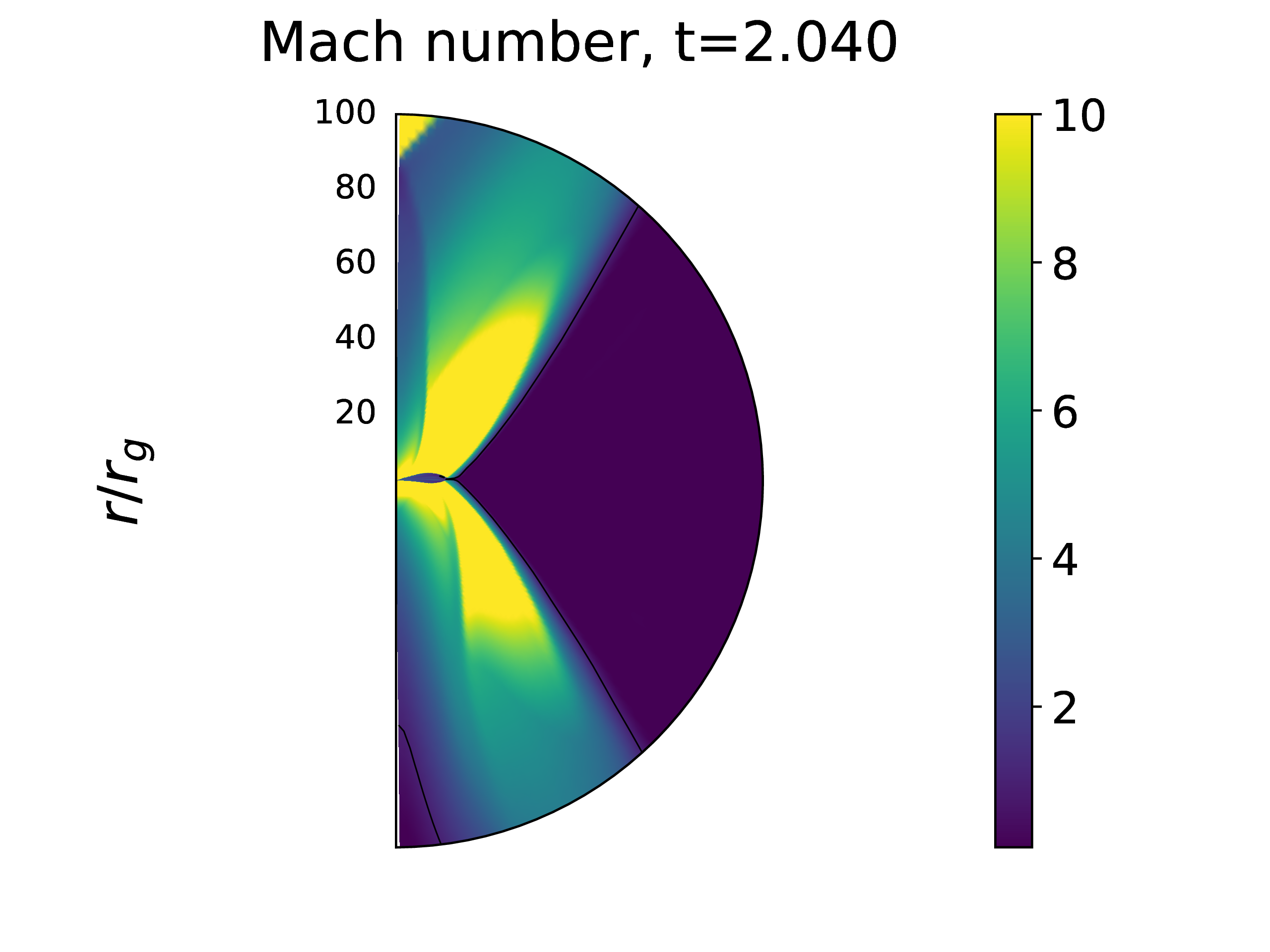}
  \end{tabular}
\caption{The results, from left to right, for the \textit{Density, Specific Angular Momentum, and Mach Number} distributions, for the model with neglected magnetic fields and initial black hole spin of $A_{0}=0.85$ and the rotation parameter $S=1.0$ (model HS-T-10).
  The color maps show disk structure which appears in this simulation and is sustained for $\sim 4s.$ Contour of $M=1$ is marked with a black line.
  (Note different spatial scale of density and angular momentum profiles in the first and second row.)}
     \label{fig:a085s10_2}
\end{figure*}

 As an example of simulation with critical initial angular momentum, $S=1.0$, in the envelope let us discuss the case with the highest initial spin value, $A_0=0.85$.
  At the beginning of the calculations we observe accumulation of matter in the innermost part of the cloud, forming also  multiple sonic surfaces in the flow.
  It  corresponds to slower growth of the $\dot{M}$ than in case of sub-critical rotation.
  Subsequently, matter from denser inner regions gets accreted onto the black hole (reflected in the growth of the accretion rate), and multiple sonic surfaces no longer exist.
  The only one, spherically symmetric sonic surface moves outwards.
  Density and angular momentum profiles corresponding to this initial stage of the simulation are shown in Fig. \ref{fig:a085s10_1}.
  Later, the shape of the cloud is mostly spherical. At the final times, around
  $t=1.5 \textrm{s}$ the density distribution flattened towards the equator, creating a mini-disk structure. The additional run of this simulation, carried for longer time,
  revealed the formation of the elongated disk structure. In this model, the disk was present for longest time of $\sim 3\textrm{s}$.
  In case of different values of initial spins, such structure did not sustain longer than for $\sim 1 \textrm{s}$.
  In the Fig. \ref{fig:a085s10_2} we show the elongated
  disk structure which was formed and sustained.

\begin{figure*}
\begin{tabular}{ccc}
 \hspace{-10mm}\includegraphics[width=0.35\textwidth]{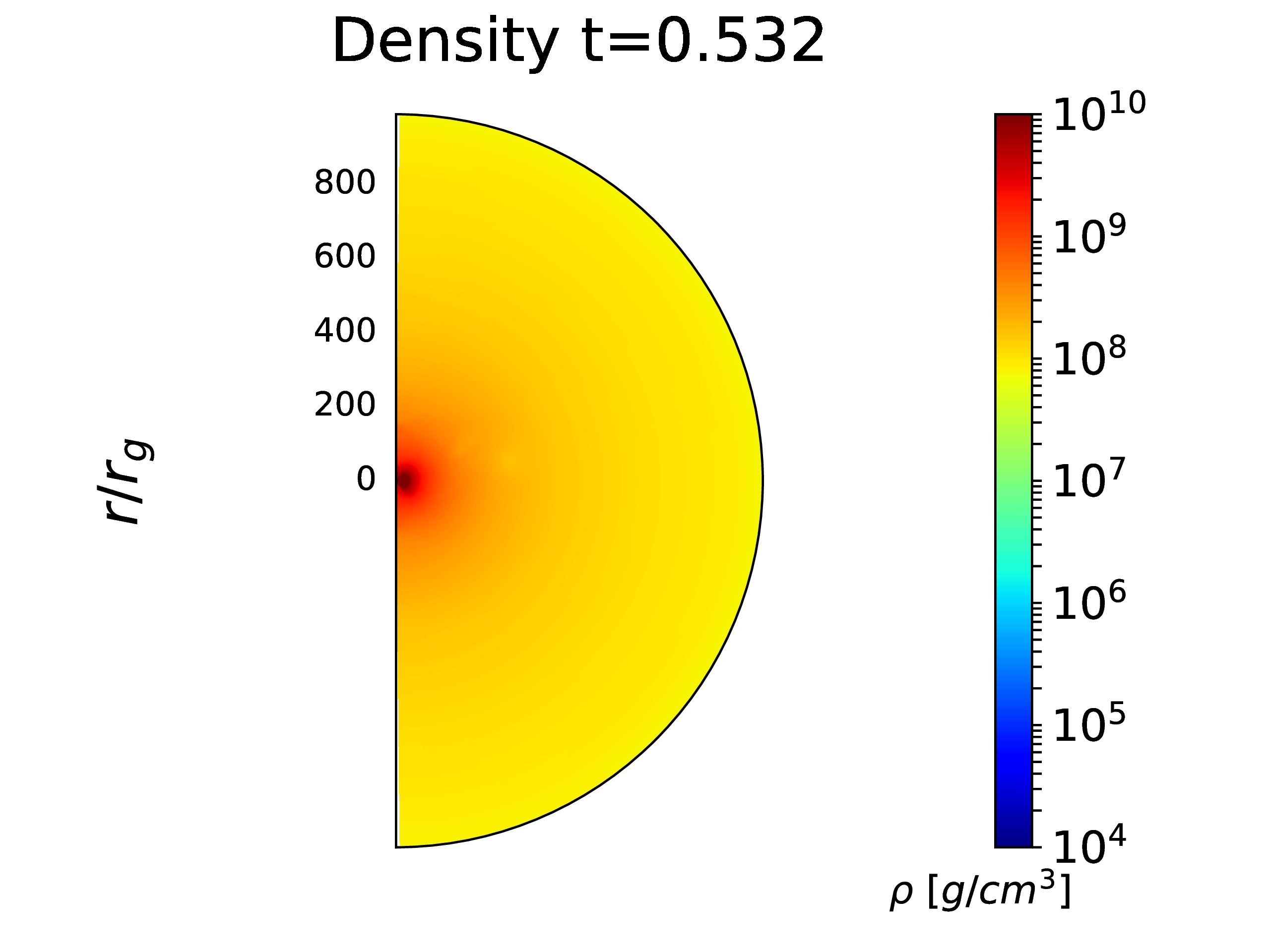} &
\includegraphics[width=0.35\textwidth]{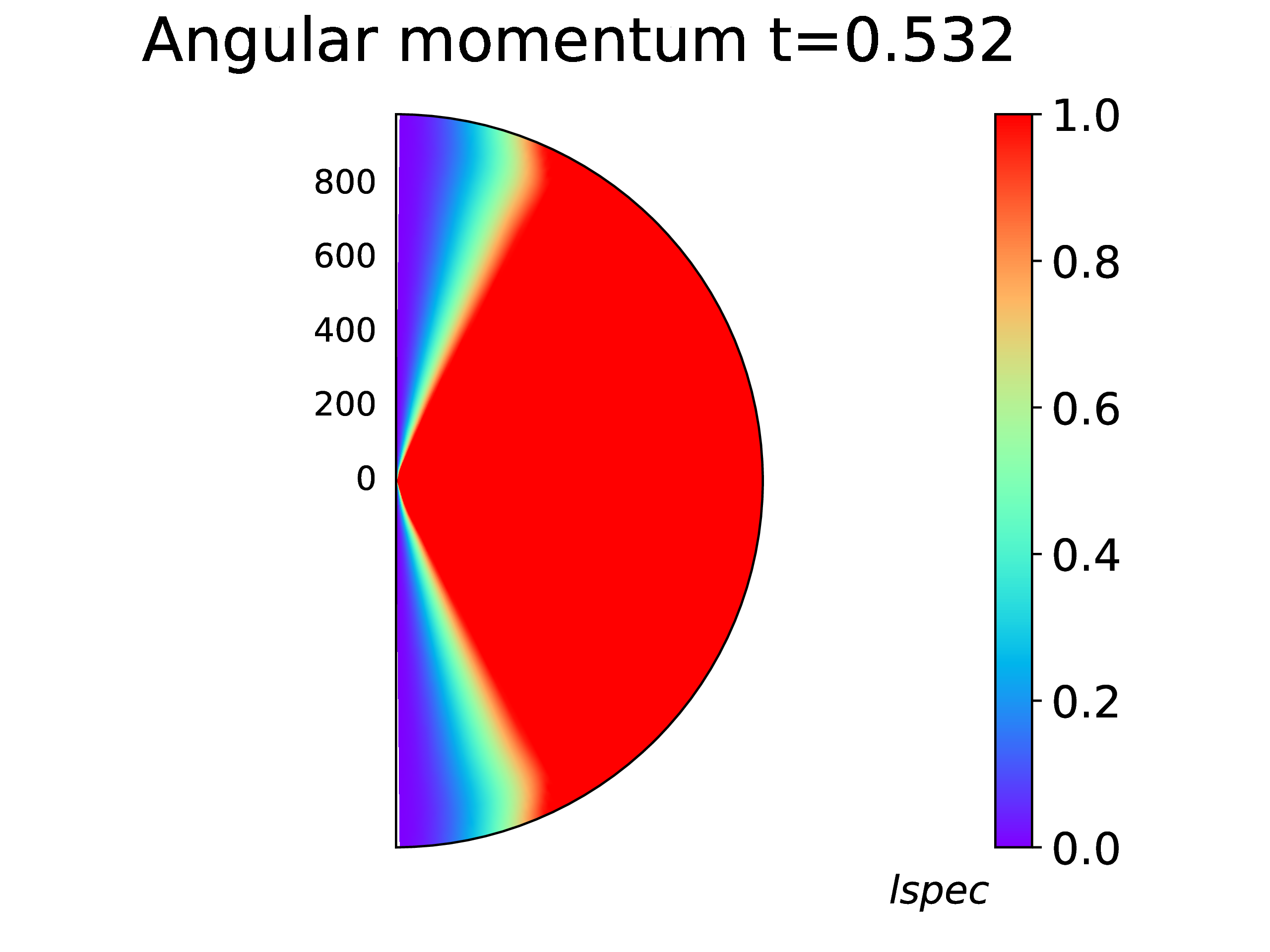} &
\includegraphics[width=0.35\textwidth]{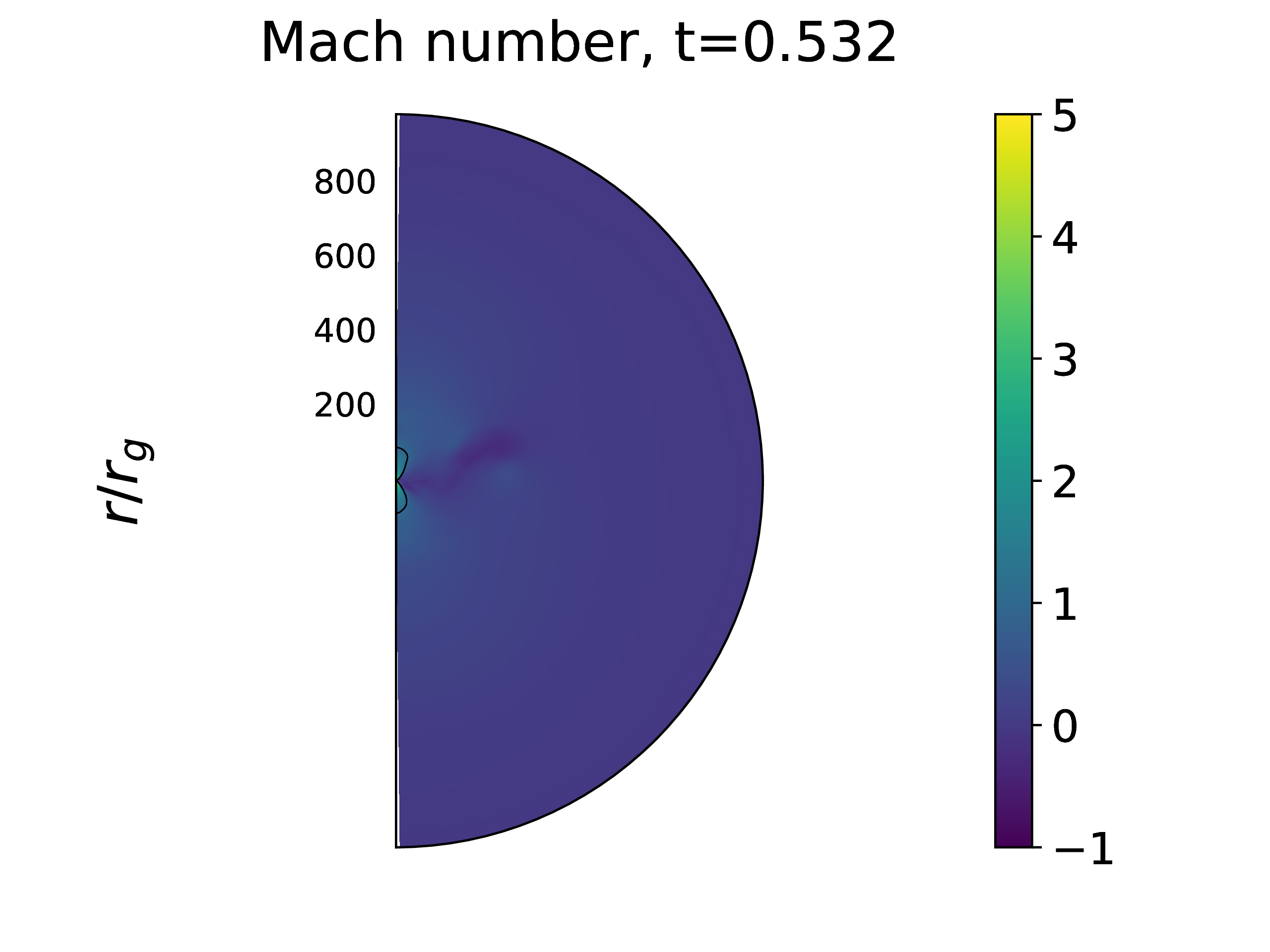} 
\\
 \hspace{-10mm}\includegraphics[width=0.35\textwidth]{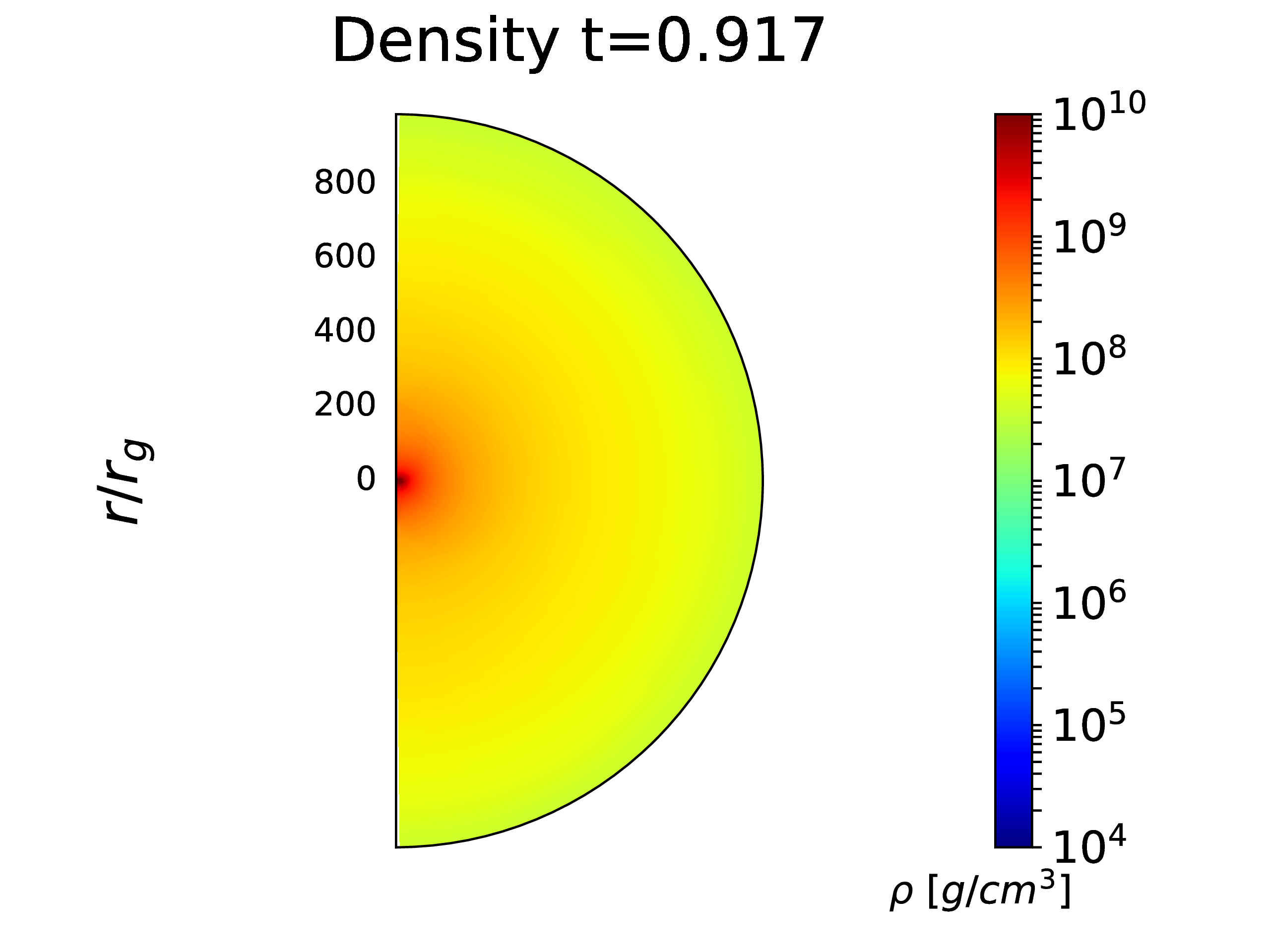}  &
\includegraphics[width=0.35\textwidth]{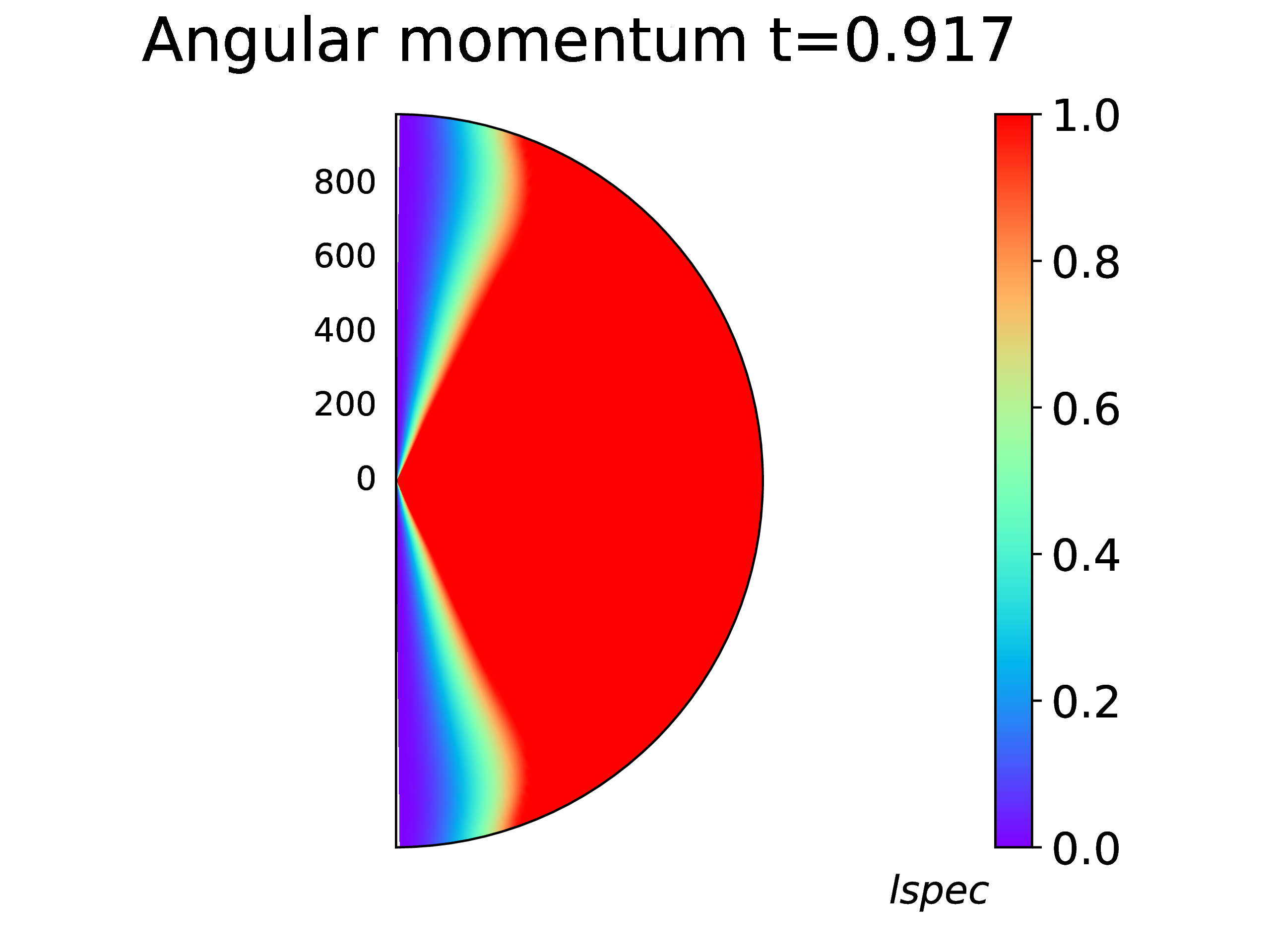}  &
\includegraphics[width=0.35\textwidth]{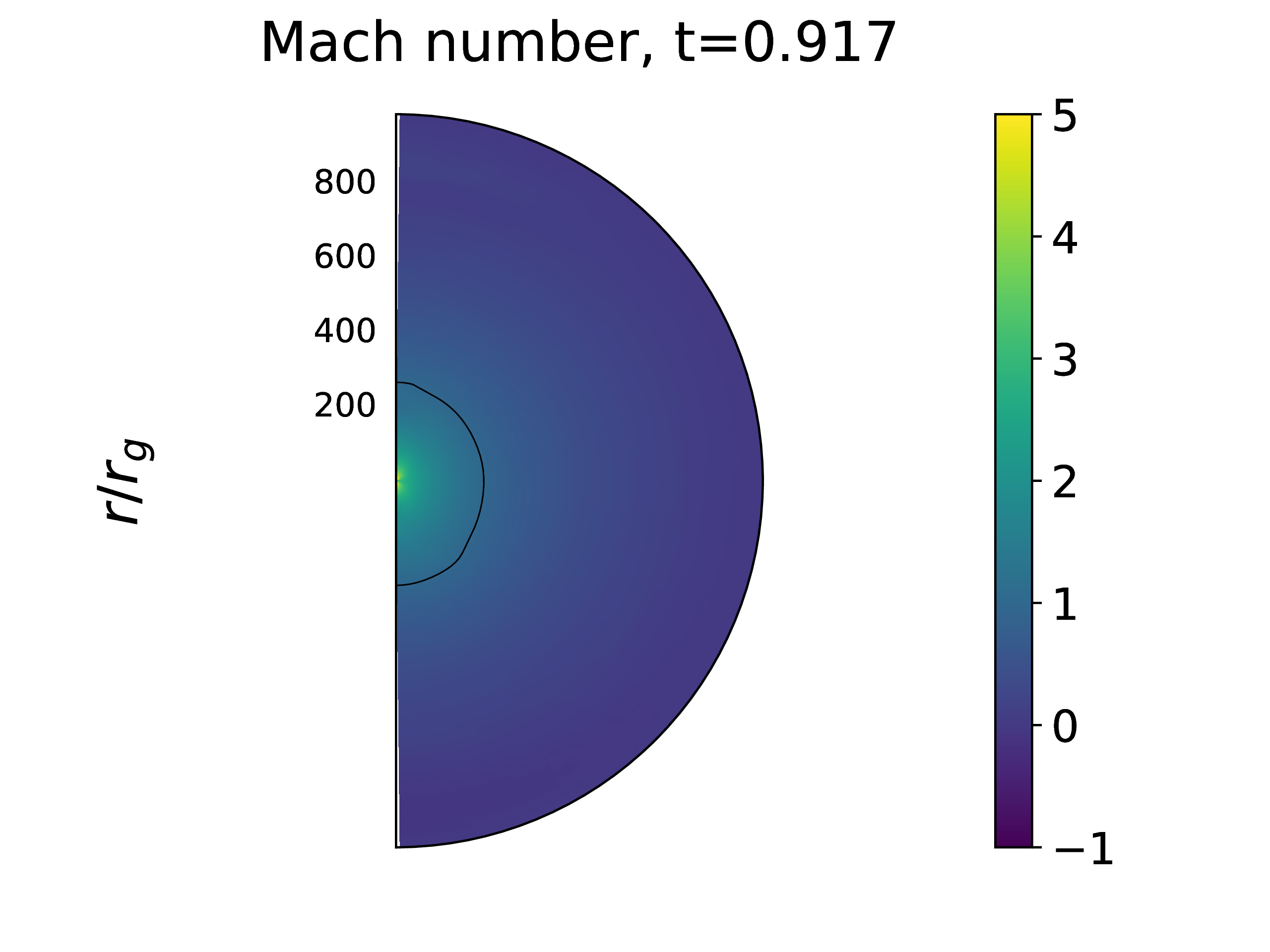}  
\\
 \hspace{-10mm}\includegraphics[width=0.35\textwidth]{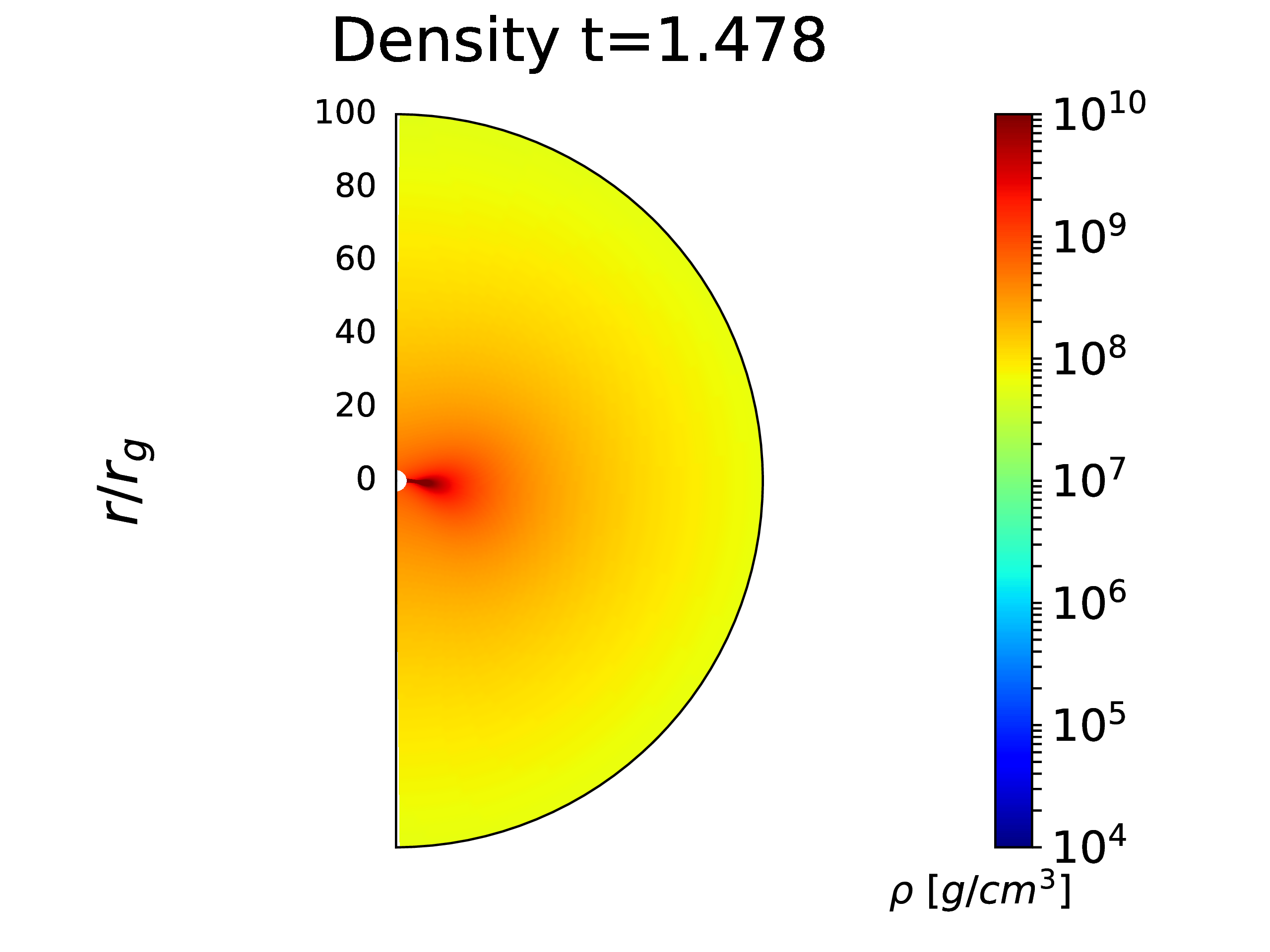} &
 \includegraphics[width=0.35\textwidth]{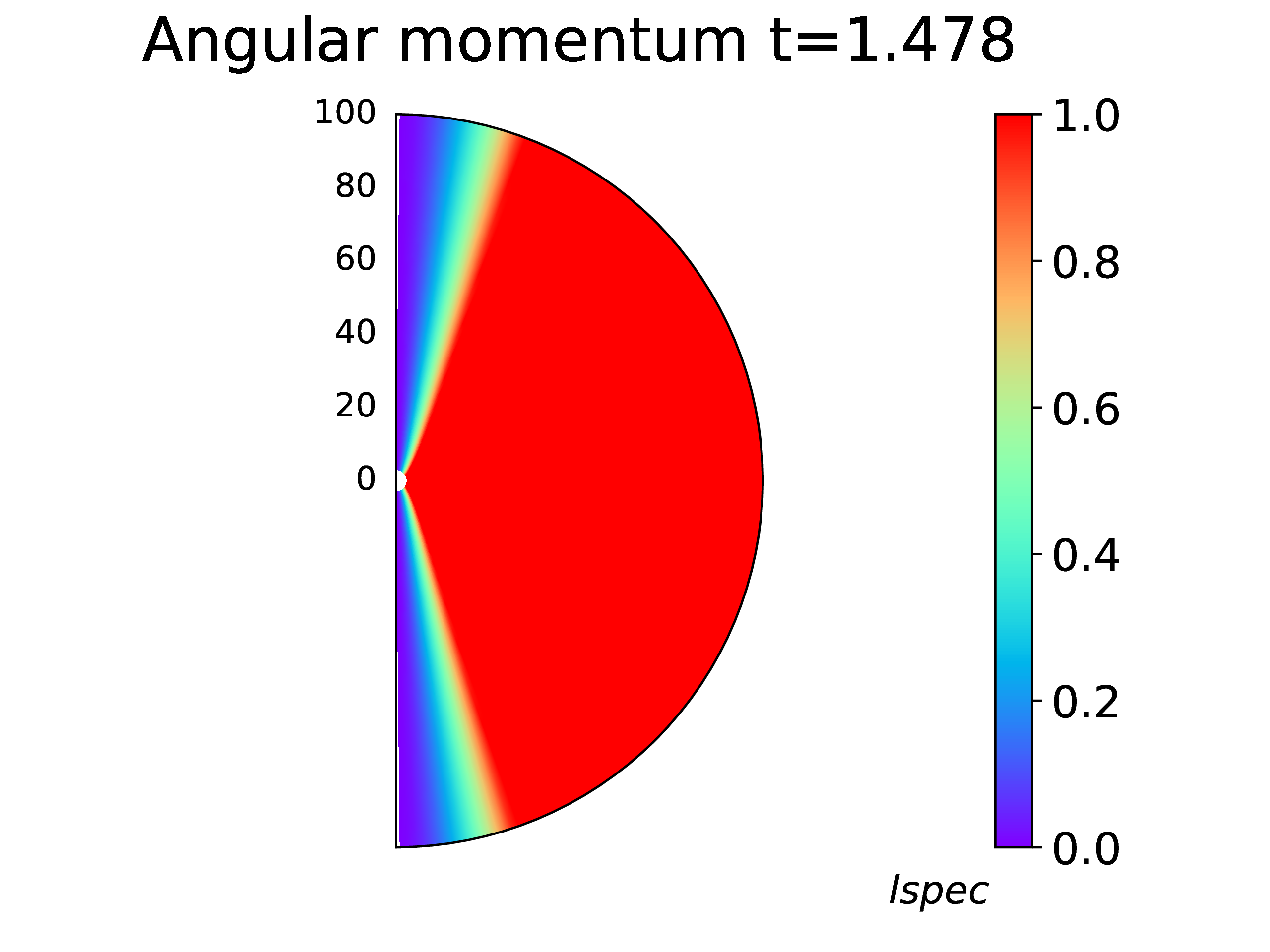} &
\includegraphics[width=0.35\textwidth]{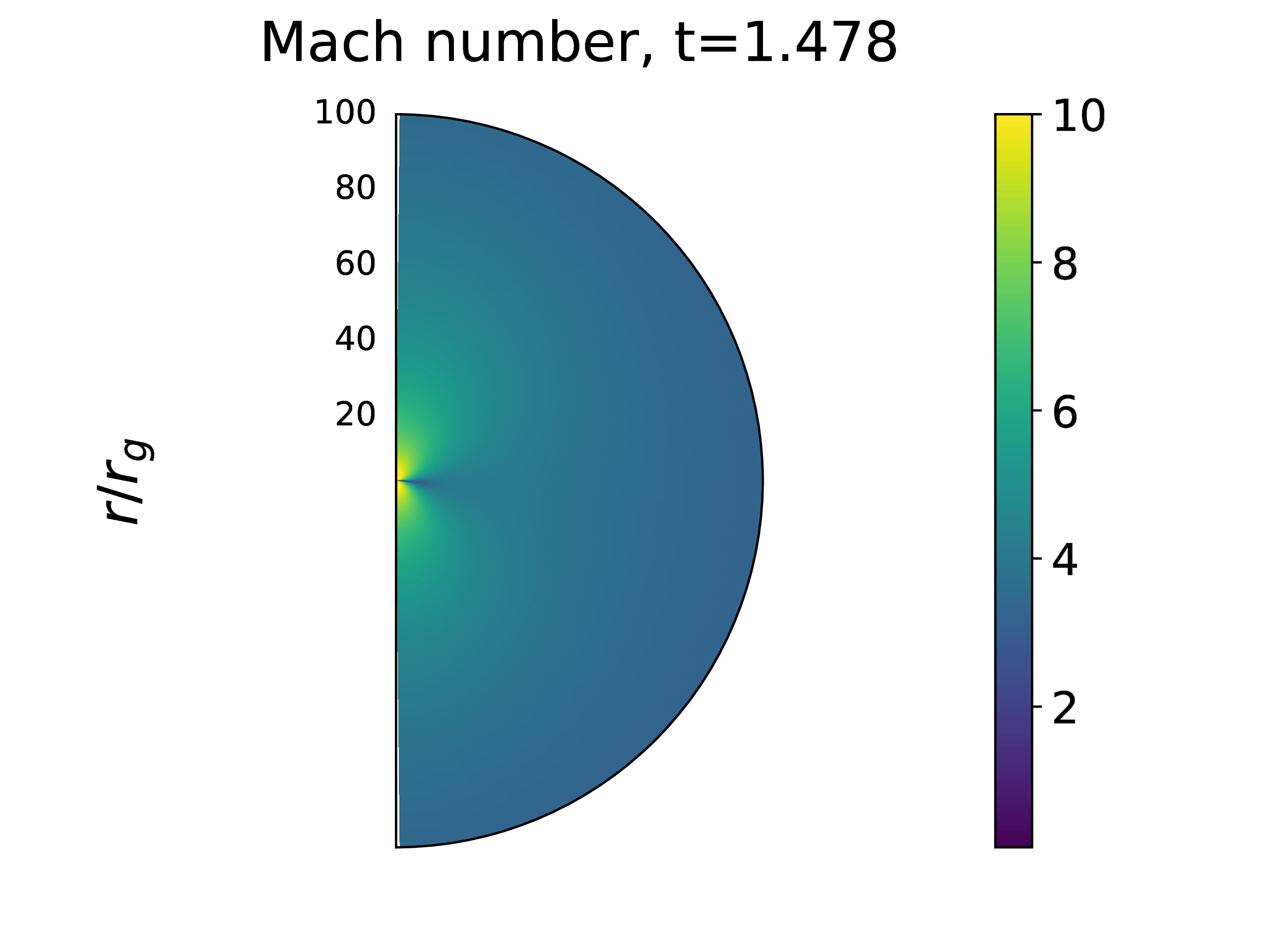} 
\end{tabular}
\caption{The results, from left to right, for the \textit{Density, Specific Angular Momentum, and Mach Number} distributions, for the model with neglected magnetic fields and initial black hole spin of $A_{0}=0.3$. The rotation parameter was $S=1.4$ (model NS-T-14), there was no magnetic field.
 The color maps are taken at two intermediate times $t=0.53$, $t=0.92$, and at $t=1.48$, which is at the end of simulation.
 In addition, contour of $M=1$ is shown with a black line. (Note different spatial scales of density and angular momentum profiles in the third row.)
  }
     \label{fig:models_lowA0_s14sim}
\end{figure*}

In Fig. \ref{fig:models_lowA0_s14sim} we present selected profiles of density, specific angular momentum and radial Mach number, $M_{r}$, distributions, for model with $A_{0}=0.3$ and supercritical rotation, $S=1.4$.  
Selected snapshots correspond to different stages of evolution of accretion rate which is reflected in the the flow hydrodynamical structure. First row of Fig. \ref{fig:models_lowA0_s14sim} shows the snapshots taken at $t=0.532$ s. It corresponds to a steady $\dot {M}$ value. Sonic surface is very close to the black hole at the equator and forms an "eight-shape" structure, elongated in the polar direction. Close to the equatorial plane drop of $M_{r}$ value is visible.
Middle row shows a snapshot at $t=0.917$ s. It corresponds to the high amplitude oscillations in the accretion rate. Density profiles around that time flicker. Sonic surface becomes spherical and with time it starts to move to the outer boundary of the mesh.  The last row shows the final structure of the flow at the end of the simulation.
In super-critical simulations with $S=1.4$, we do see only a mini-disk structure at the innermost region of the flow, up to $r\sim 10 r_g$ at  time $t= 1.5 \textrm{s}$. At larger scales, the density profile is spherically symmetric.
Now the angular momentum is largest close to the black hole, while the density of entire cloud has dropped significantly in the whole envelope, leaving only a small centrifugally supported dense remnant disk at the equator below $\sim 20 r_{g}$ (note different spatial scales of density and angular momentum maps in the last row in Fig\ref{fig:models_lowA0_s14sim} on which a small scale disk structure is visible).

\begin{figure}
   \centering
   
   \includegraphics[width=0.45\textwidth]{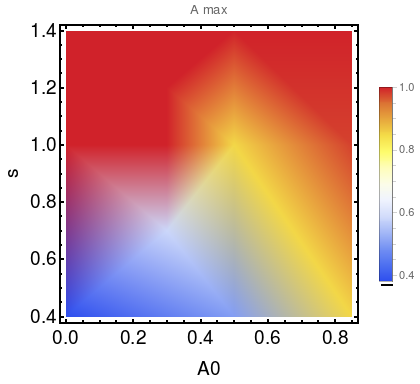}
   \includegraphics[width=0.45\textwidth]{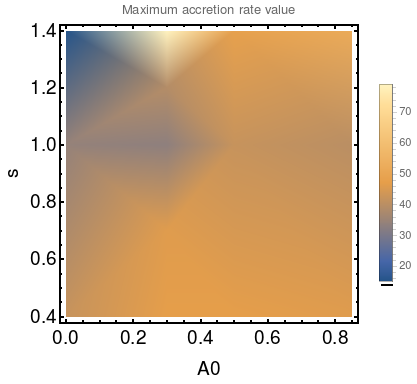}

   \includegraphics[width=0.45\textwidth]{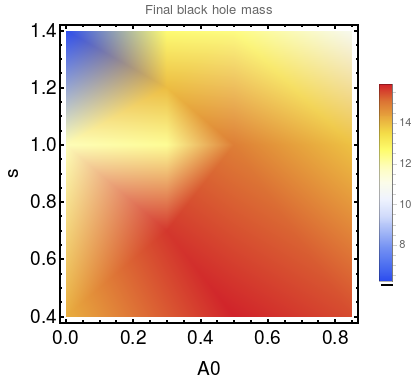}
   \includegraphics[width=0.45\textwidth]{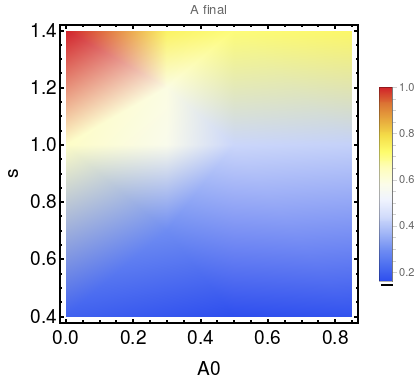}

    \caption{Parameter space studied in our simulations, and results for maximal spin during the simulation (upper left panel), maximal accretion rate (upper right panel) and final values of mass (lower left panel) and black hole spin (lower right panel). The maps show results with non-magnetized flows}
     \label{fig:maps_hydro}
\end{figure}

To summarize the parameter space of our models, in Figure \ref{fig:maps_hydro} we present color-maps showing the results of final and maximal global quantities obtained for our grid of models, plotted with the parameter-space contours.  The resulting black hole mass and final spin, obtained at the end of the simulation, is depicted with a color scale (bottom left and right panels, respectively). We also show the maximal spin and maximal accretion rate values reached during the simulation (upper left and right panels, respectively).

\subsubsection{Shocks}
\label{sec:shocks}

At critical rotation, the shock discontinuity forms and pressure builds up to slow down the incoming material near the equator. For even higher rotation parameter, $S=1.4$, the accretion flow forms a geometrically thin structure at the equatorial region. Here, a transient shock structure may form to redistribute the angular momentum, but the shock is still subsonic for $S<2$. Shock for the case $A_0=0.3$ and s=1.4 appears only for $t=0.857 \textrm{s}$ at $r=16.37r_{g}$.

We investigated the Mach number profiles in order to check for shocks existence. Our algorithm for shock finding was simple, i.e. we were looking for the radii where the flow changes from supersonic to subsonic. Additionally we checked if this change in the flow is connected with a steep change in the thermodynamic properties of the flow.  We rejected the apparent shocks that appear with a supersonic velocity value only in one, the outermost, grid cell, and are numerical artifacts.   We investigated the shock presence in the equatorial plane and we found that for almost every combination of $A_0$ and $S$ parameters, a shock appeared at some point. The only exception  was model with $A_0=0.3$ and $S=1.0$ in which a shock did not appear at any stage of the simulation. For $A_0=0.3$ and $S=1.4$ a shock appears only in one moment, at time $0.857$, so that it last less than $0.03\textrm{s}$. For $A_0=0.3$ and $S=0.4 $ shocks start to appear at the final stages of the simulation and they are placed at the radii around $100 - 500 r_{g}$.

In the case of $A_0=0.5$, for $S=0.4$ shock appears at  $r\sim 100 r_g$ in the end of the simulation, for $S=1.0$ it can be found at a similar time but closer to the black hole, at $r\sim 20 r_g$. In addition, a transient shock is present at the beginning of the simulations below $50 r_g$. In the case of super-critical rotation, the shock was found below $50 r_g$ at intermediate times.

In the case of high initial black hole spin situation is similar. For $S=0.4$ shock appears at the end of the simulation at high $r$, in case of critical rotation shock appears at the very beginning and at the end of the simulations and for super-critical rotation  it appears for a moment at intermediate time.

\begin{figure}
   \centering
    \hspace{-10mm}\includegraphics[width=1.0\textwidth]{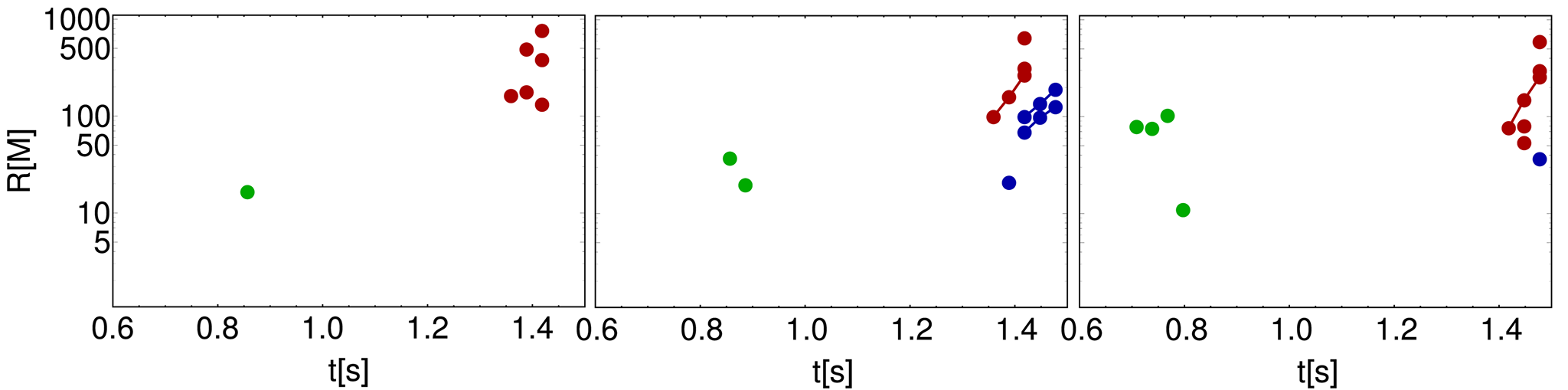}
    \caption{Shocks positions as a function of time. Red dots---$S=0.4$, blue dots--- $S=1.0$, green dots--- $S=1.4$. Panels from left to right, present models with various initial spin of the black hole: $A_0=0.3$, $A_0=0.5$, $A_0=0.85$. Note that plots present time from $0.8  \textrm{s}$ to the end of the simulation, so shocks which appear as initial condition relaxation are not visible. They do appear for critical initial rotation for $A_0=0.85$ and $A_0=0.5$ before $0.1\textrm{s}$ at $\sim 20r_g$ and $\sim 50 r_g$. 
    }
     \label{fig:shock}
\end{figure}

Shock positions evolving in time are shown in Fig. \ref{fig:shock}. In simulations with $A_0=0.5$ $S=0.4$, $A_0=0.5$ $S=1.0$ and $A_0=0.85$ $S=0.4$ (models LS-T-04, LS-T-10, and HS-T-04, respectively) the shocks which appear at later stages are standing over some time and then expand to the outer boundary of the grid. We calculated their velocities, which are given in Table \ref{tab:2}. We compare them to the escape velocity, corresponding with position of the shock. Because this ratio is always below 20\%, we conclude hat such shocks are too slow to be able to dissipate enough energy and disrupt the outer envelope.

\begin{deluxetable}{lccr}

\tablecaption{Summary of the models}

\tablehead{
   \multicolumn{1}{p{1.5cm}}{\centering  $A_{0}$ \\} 
  & \multicolumn{1}{p{1.5cm}}{\centering $S$\\ $l/l_{\rm crit}$ } 
  & \multicolumn{1}{p{1.5cm}}{\centering $v$\\ $c$ }
    & \multicolumn{1}{p{1.5cm}}{\centering $v_e$\\ $c$ }
}
\startdata
      0.5   & 0.4  & 0.041 & 0.26\\
      0.5   & 1.0  & 0.014 & 0.34\\    
      0.5   & 1.0  & 0.022 & 0.28\\
      0.85   & 0.4  & 0.045& 0.28 \\      
\enddata
\tablecomments{Velocities of shock fronts and escape velocities for different models.}
  \label{tab:2}
\end{deluxetable}

\subsection{Simulations with magnetized accretion flows}
\label{sec:Mag}

\begin{figure*}[h]
\begin{tabular}{ccc}
\hspace{-10mm}
\includegraphics[width=0.33\textwidth]{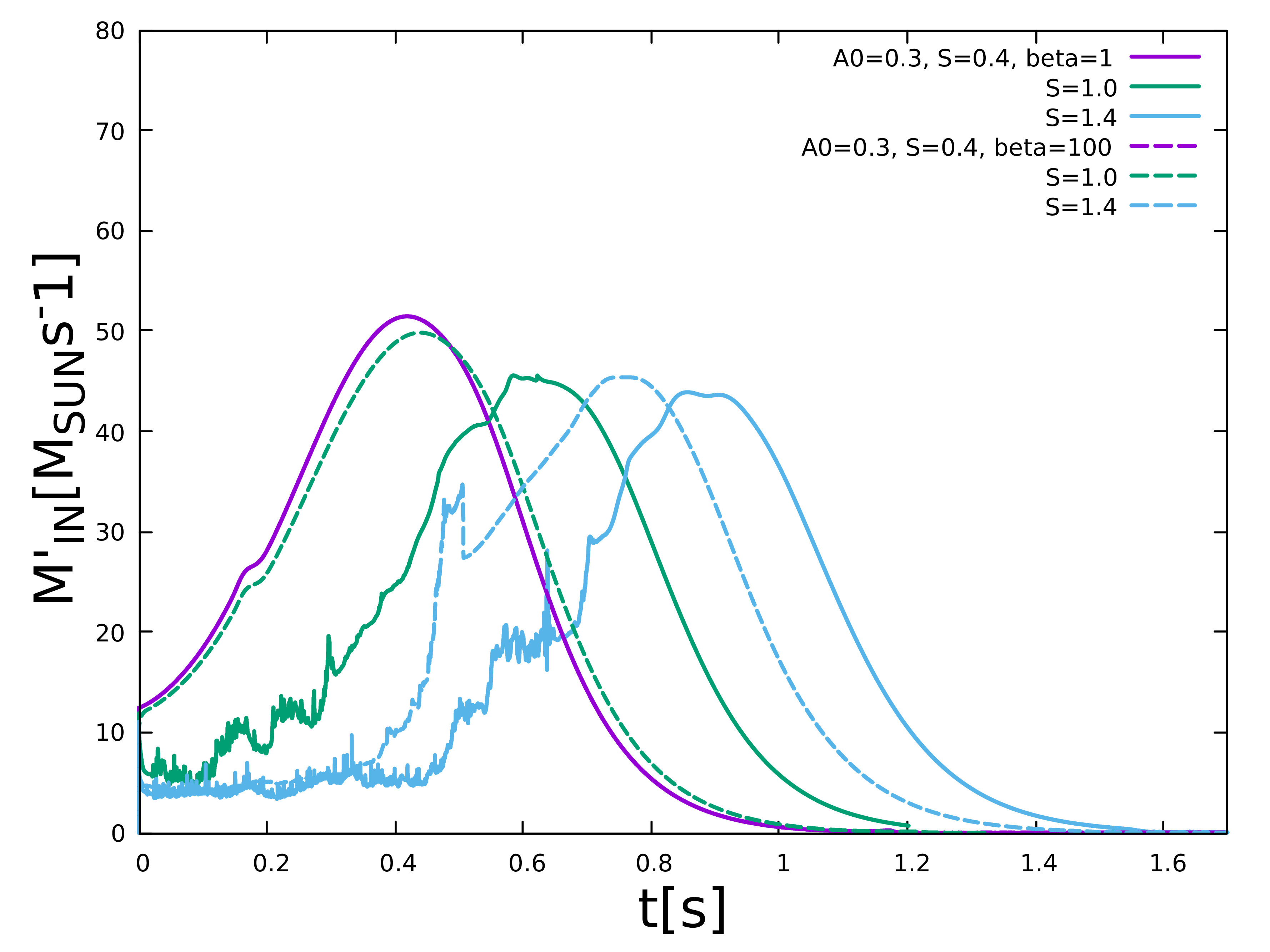}
  \includegraphics[width=0.33\textwidth]{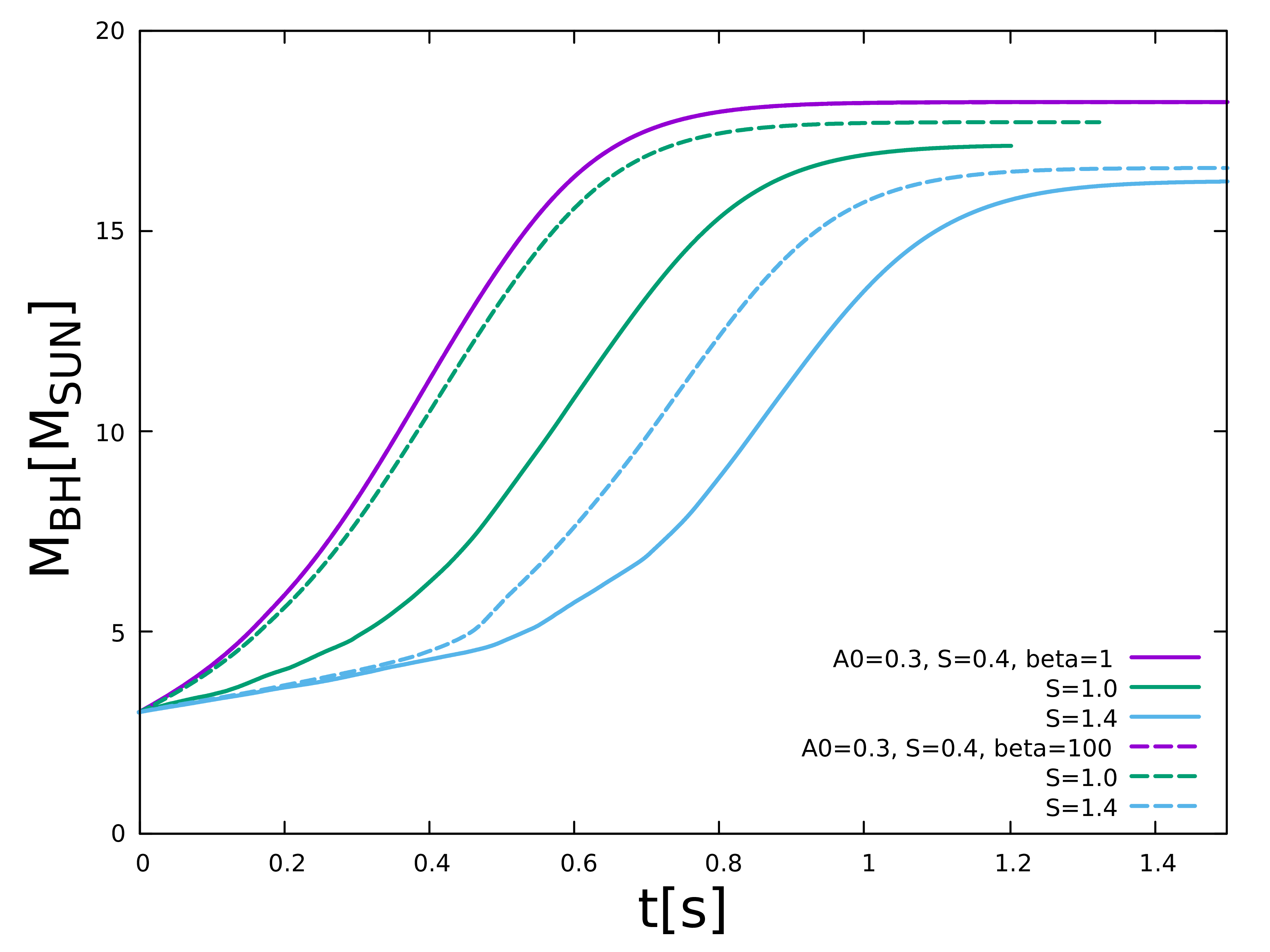}
  \includegraphics[width=0.33\textwidth]{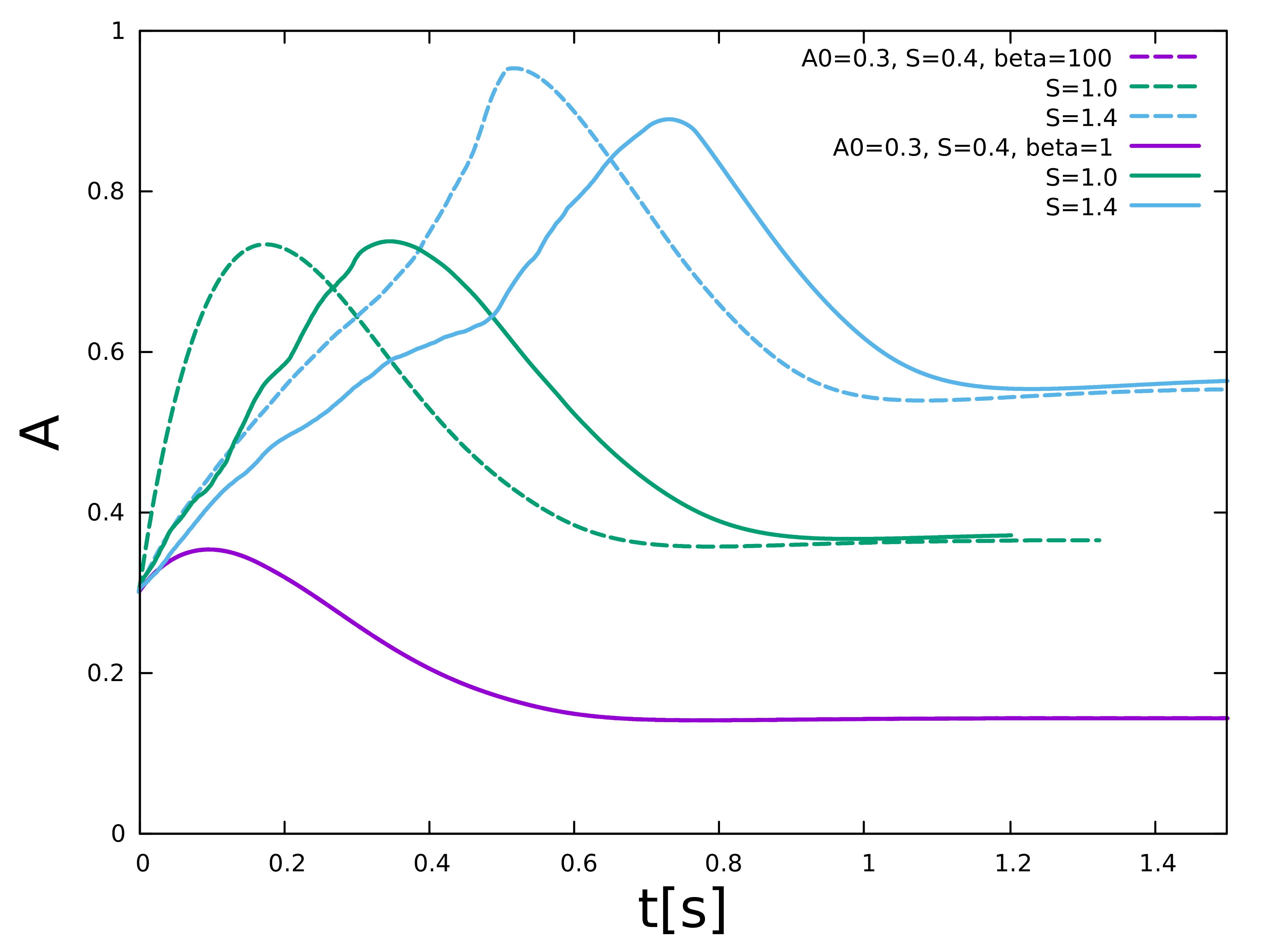}
\end{tabular}
  \caption{Time dependence of the mass accretion rate onto black hole (left panel), black hole mass (middle panel), and its spin evolution (right panel) 
  for the magnetized collapse case with  $\beta=100$ (dotted line) and $\beta=1$ (solid lines). The initial black hole spin was $A_{0}=0.3$.
      Cyan lines denote the models with $S=1.4$
      while the violet lines denote $S=0.4$ and green lines are for $S=1.0$.  }
      \label{fig:evol_magA03}
\end{figure*}

\begin{figure*}[h]
 \begin{tabular}{ccc}
\hspace{-10mm}
\includegraphics[width=0.33\textwidth]{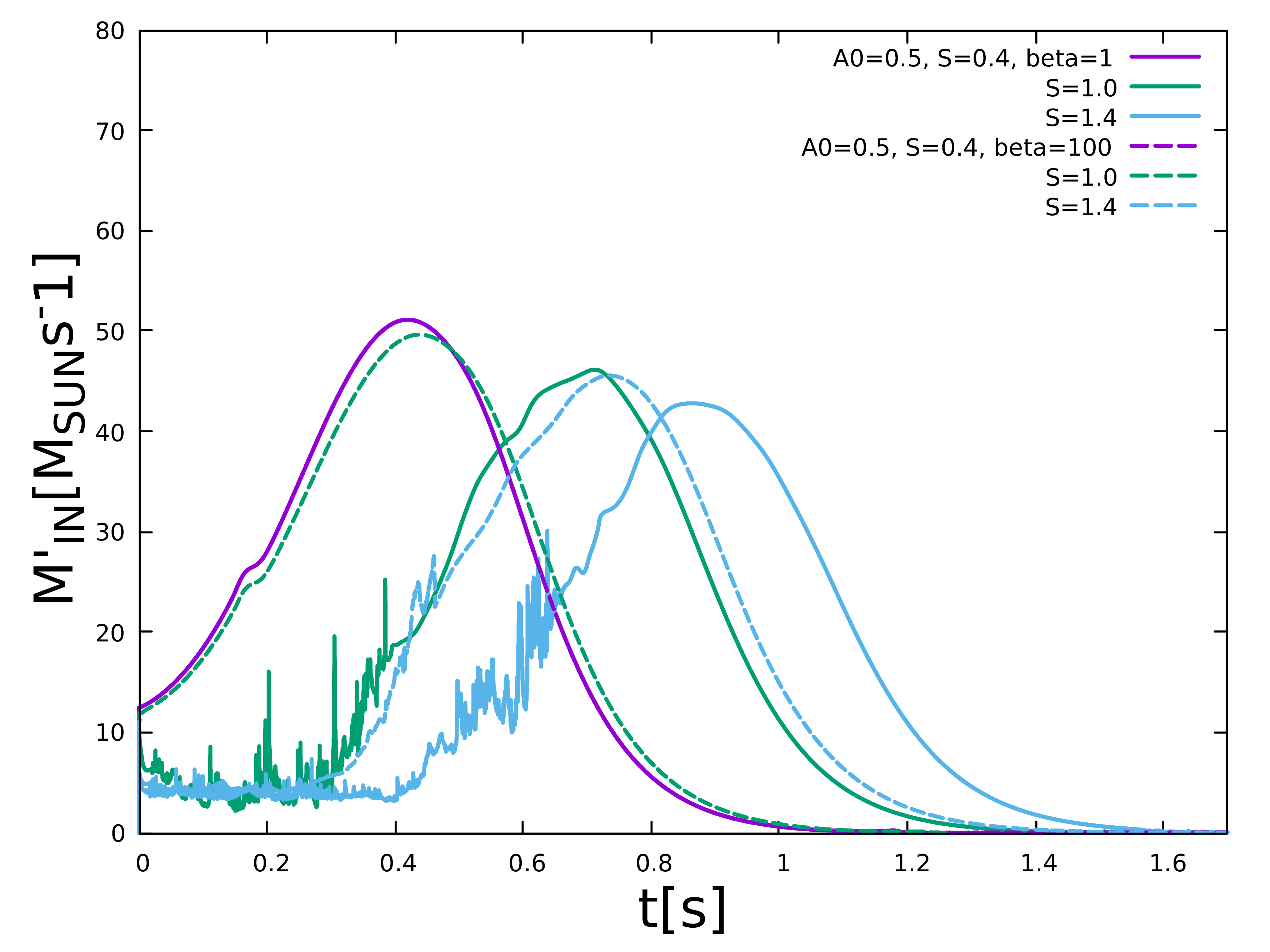} &
  \includegraphics[width=0.33\textwidth]{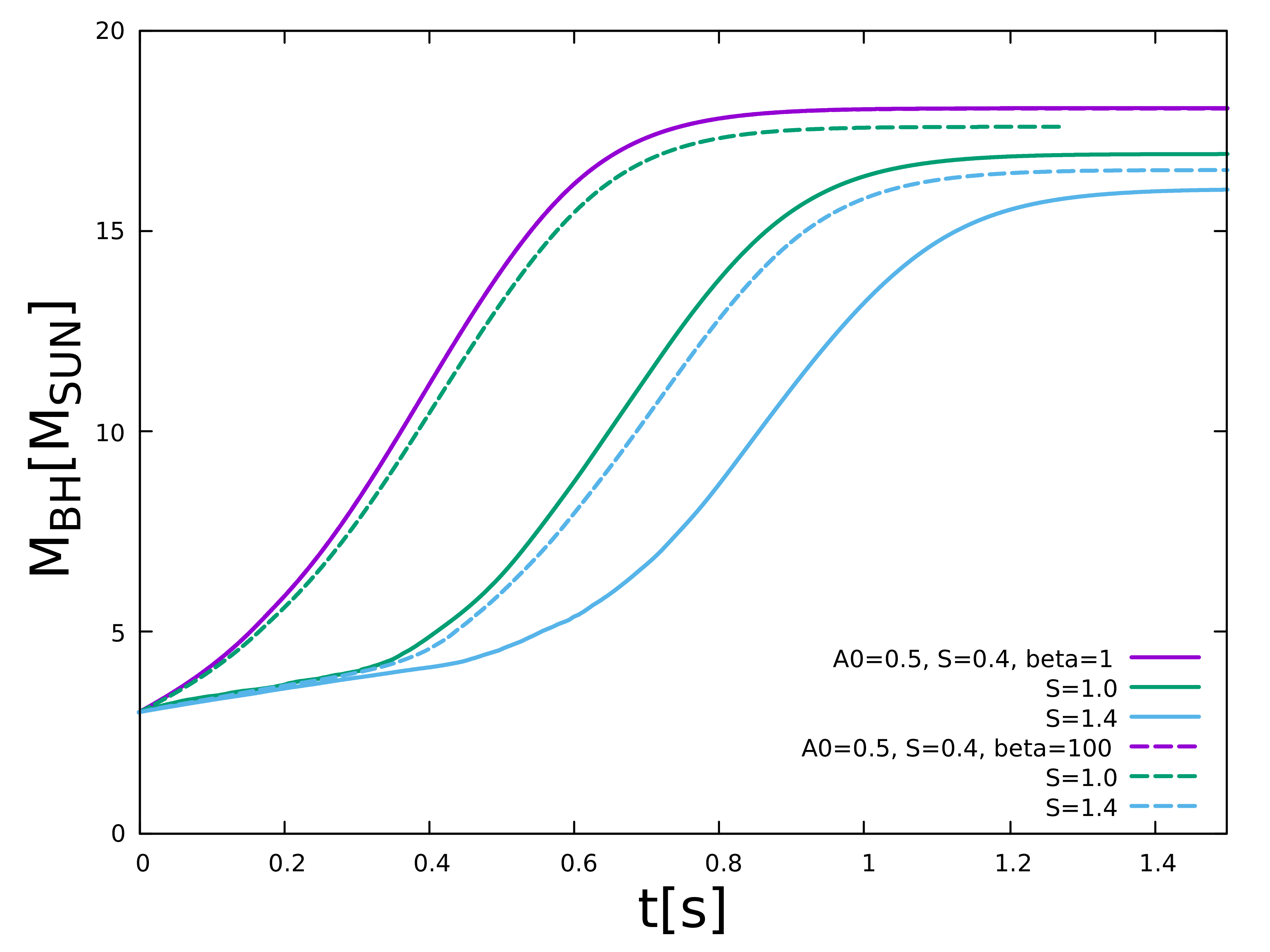} &
 \includegraphics[width=0.33\textwidth]{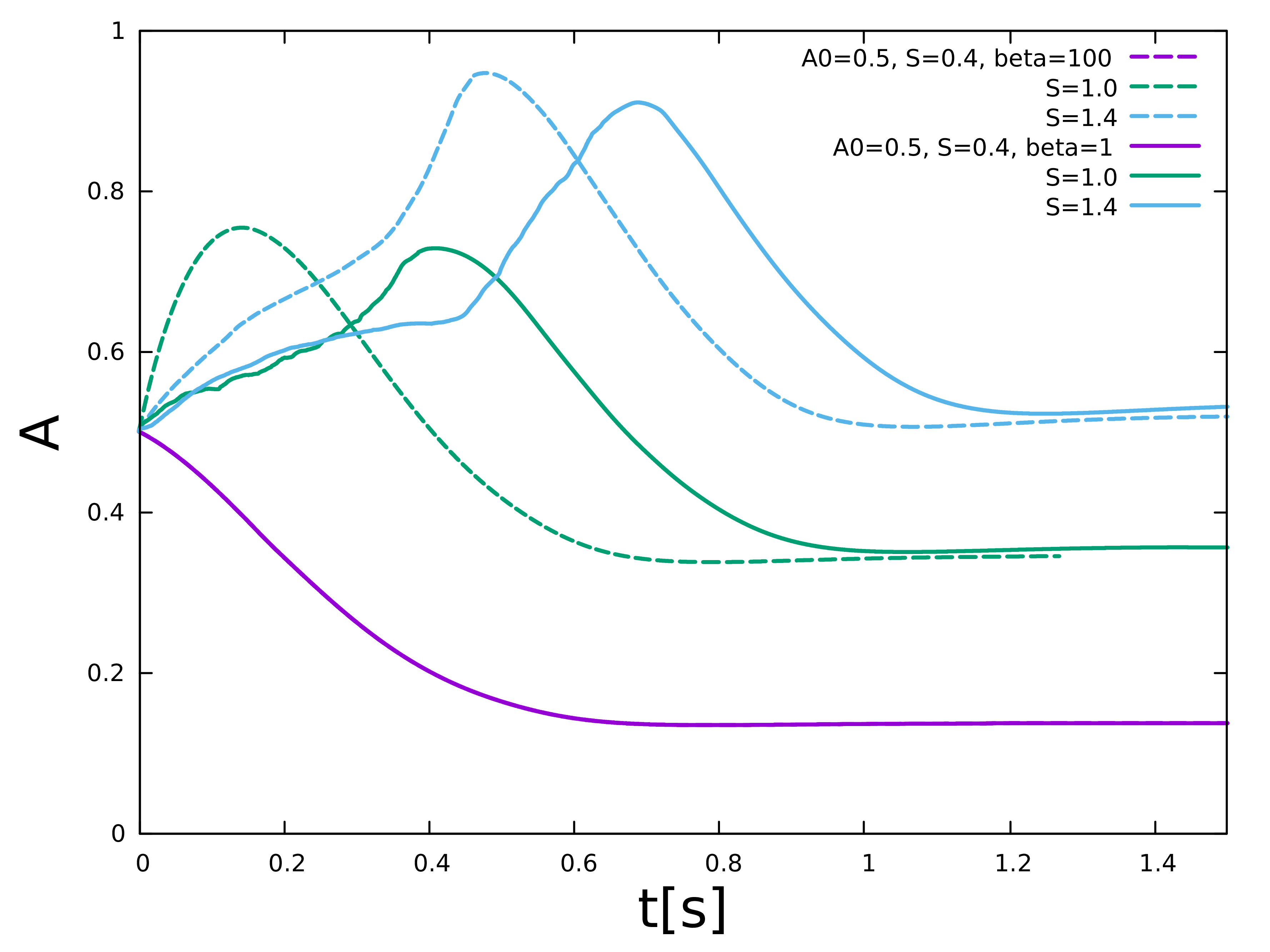}
    \end{tabular}
    \caption{Time dependence of the mass accretion rate onto black hole (left panel), black hole mass (middle panel), and its spin evolution (right panel),   for the magnetized collapse case with  $\beta=100$ (dotted lines) and $\beta=1$ (solid lines). The initial black hole spin was $A_{0}=0.5$.
      Cyan lines denote the models with $S=1.4$
      while the violet lines denote $S=0.4$ and green lines are for $S=1.0$. 
    }
      \label{fig:evol_magA05}
     \end{figure*}

\begin{figure*}[h]
 \begin{tabular}{ccc}
  \hspace{-10mm} \includegraphics[width=0.33\textwidth]{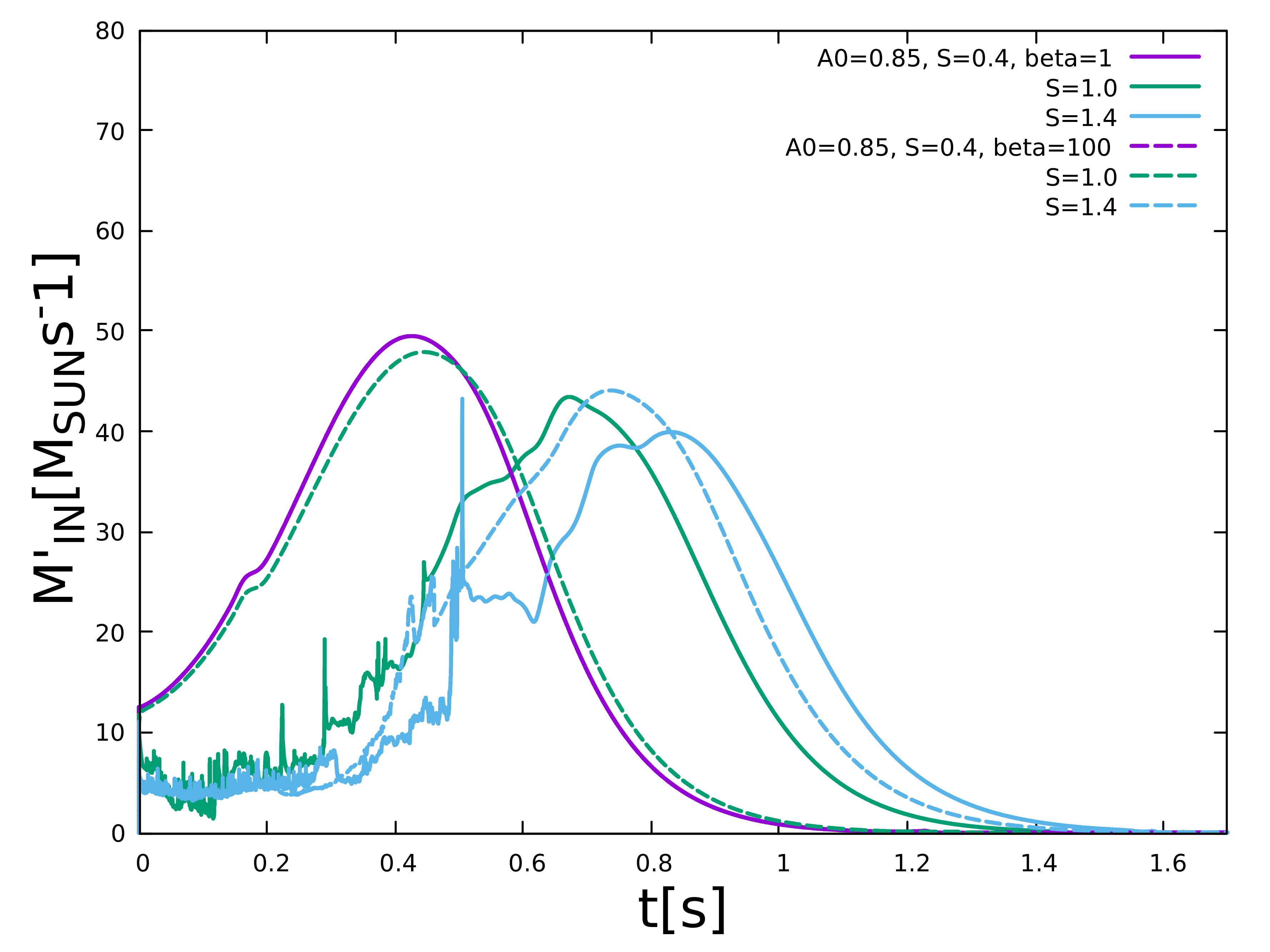} &
  \includegraphics[width=0.33\textwidth]{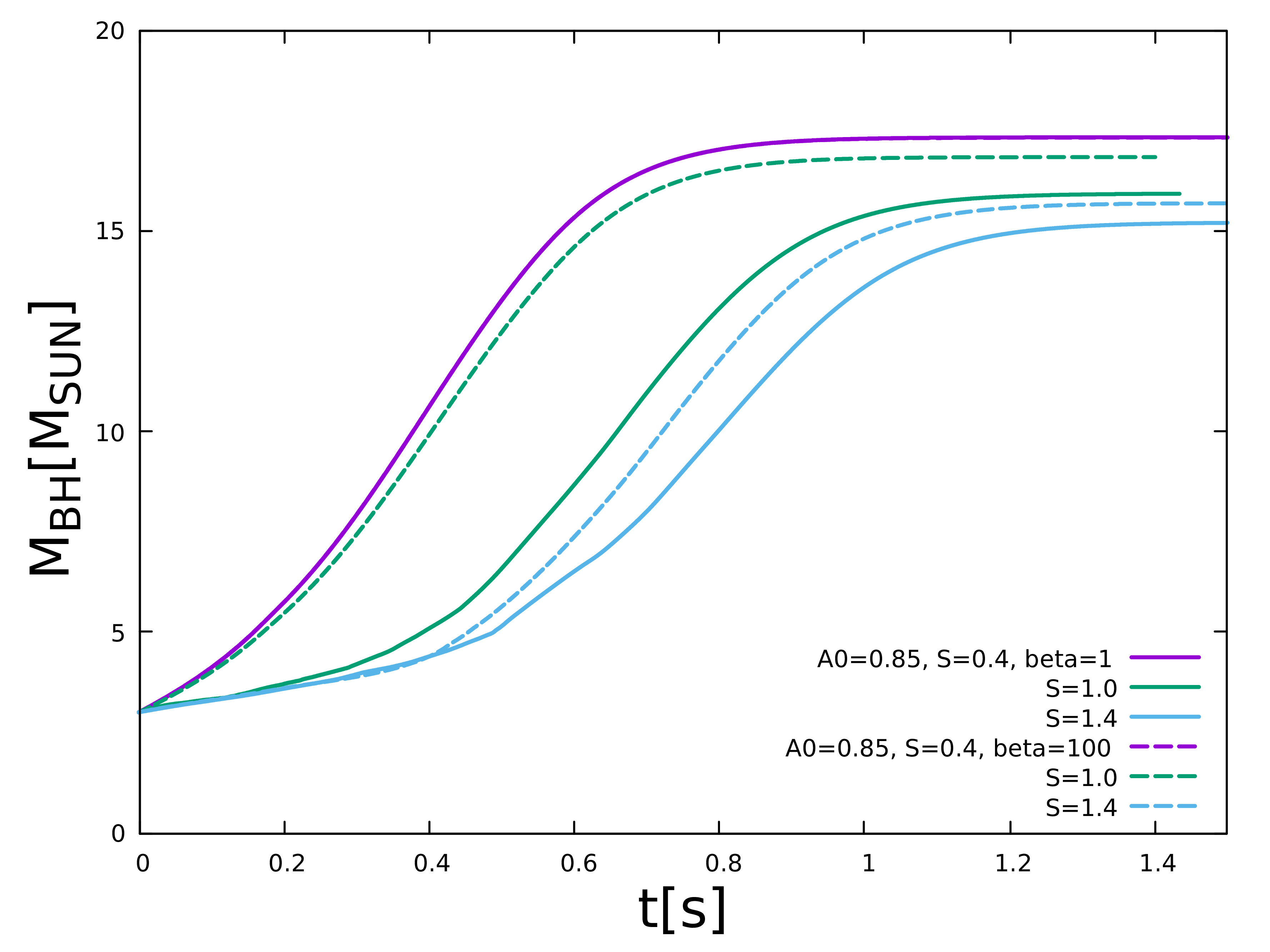} &
 \includegraphics[width=0.33\textwidth]{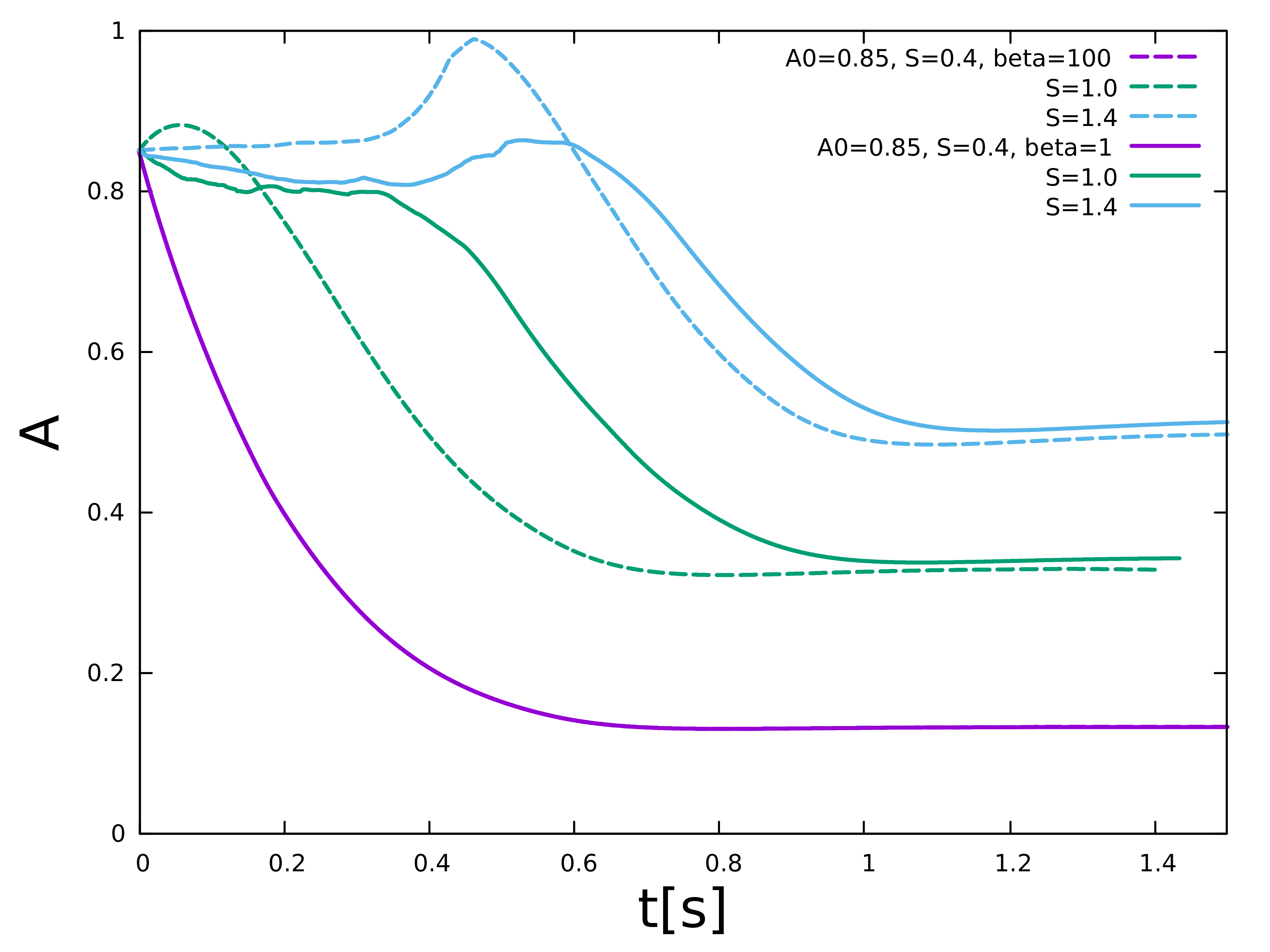}
    \end{tabular}
    \caption{Time dependence of the mass accretion rate onto black hole (left panel), black hole mass (middle panel), and its spin evolution (right panel), for the magnetized collapse case with  $\beta=100$ (dotted line) and $\beta=1$ (solid lines). The initial black hole spin was $A_{0}=0.85$.
      Cyan lines denote the models with $S=1.4$
      while the violet lines denote $S=0.4$ and green lines are for $S=1.0$.  
    }
      \label{fig:evol_magA085}
     \end{figure*}

In this Section, we present the simulations of magnetized stellar envelopes, where the low-angular momentum flow is embedded in a poloidal magnetic field.
Specific pattern of the evolution depends on the $\beta$ value, but also on the combination of $A_0$ and the $S$ parameters. We performed two sets of simulations for the same combinations of $A_0$ and $S$  as for the case  without magnetic field discussed above. 
Tested values of $\beta$ were $1$ and $100$.  We present evolution of the global parameters for both $\beta$ values in Figures \ref{fig:evol_magA03},
\ref{fig:evol_magA05}, and \ref{fig:evol_magA085}. 
The evolution of black hole mass and spin depends strongly on the rotation of the envelope, parameterized by $S$. 

For $S=0.4$, evolution is similar to non-magnetized models, and sub-critical models with $\beta=1$ and $\beta=100$ are nearly identical, and moreover they behave in a similar way for every value of the initial black hole spin. 
The accretion rate smoothly grows  to the value of $\sim 50 M_{\odot}s^{-1}$ (Figures \ref{fig:evol_magA03}, \ref{fig:evol_magA05}, and \ref{fig:evol_magA085}, cf. purple lines solid and dashed) and then it starts to decrease. Spin increases at any stage of the simulations only for $A_0=0.3$ and in this model the final black hole mass is the highest. It is also higher than in the case of non-magnetized envelope.  

In case of critical and super-critical rotation, the evolution pattern for various initial  parameters is more complicated. 
Firstly, the $\beta$ value affects the appearance the oscillations in the accretion rate. For a less magnetized envelope these oscillations do not appear for critical rotation, and both accretion rate and black hole mass evolution look similar to sub-critical case. The evolution is smooth, accretion rate grows to the values below $\sim 50 M_{\odot}s^{-1}$, which are slightly smaller than values reached in case of sub-critical rotation, and the shape of the evolution is very similar. The achieved final black hole masses are somehow smaller than in the sub-critical case.  However, the difference is notable  in case of the spin. Critical initial rotation of the envelope for every value of $A_0$  results in maximal spins achieved during simulation, which are significantly larger than the initial ones.

For super-critical rotation of the envelope, the oscillations of the accretion rate appear around $0.4\textrm{s}$ for every $A_0$ parameter. The accretion rate growth is steepest for $A=0.3$. For every $A_0$ value, the oscillation is not related with the $\dot{M}$  maximum value, it occurs later during the simulation and the peak is at about $\sim 45 M_{\odot}s^{-1}$. Final black hole masses are lower than in critical and sub-critical cases, but higher for $\beta=100$ than in the corresponding model with $\beta=1$. Also, the maximal value of the spin is higher than for the stronger magnetization, but still  in none of the cases maximal spin is reached and spins are generally smaller than in the case of non-magnetized envelope.  

For $\beta=1$, in both critically and super-critically rotating envelopes during the growth of $\dot{M}$ we can observe oscillations of its value. After the peak of $\dot{M}$, the accretion rate smoothly decreases. In case of the model parameterized with $A_0=0.85$ and $S=1.4$, the accretion rate evolution is different than for all the other models. After initial oscillations of its value we observe a large spike at the time $\sim 0.5s$. Then it drops to the previous level, saturates on it for some time, grows smoothly, and then starts to decrease. The final black hole masses form $\beta=1$ are always higher than in case of non-magnetized accretion (cf. Table \ref{tab:models}), but somewhat smaller than in the case of $\beta=100$. In particular, for the initial parameters $A_0=0.85$ and $S=1.4$, we get $M_{BH}= 15.2 M_{\odot}$,  compared to 11.9 $M_{\odot}$ for non magnetized model and 15.70$M_{\odot}$ for $\beta=100$. Final black hole spins,  on the other hand, are on average lower, e.g. $A_0=0.85$ and $S=1.4$ we have $A_{final}=0.51$ versus 0.65, for HS-M-14 (magnetized) and HS-T-14 (non-magnetized) models, respectively. Plots illustrating the evolution of $\dot M_{in}$, $A$ and $M_{BH}$ are shown in Figures from \ref{fig:evol_magA03} to \ref{fig:evol_magA085}. Final and maximal values of those parameters are listed in Table \ref{tab:models}.

 \subsubsection{Weakly magnetized envelope: $\beta=100$}
 \label{sec:beta100}

 For model $A_0=0.3$ and $S=0.4$, the density, radial Mach number and angular momentum during the collapse evolve similarly as in the case without magnetic field. They are spherically symmetric thought most of the time and  at the end of the simulation they tend to flatten toward the equator. This behavior is the same for every value of the initial spin. 
 Interestingly, this  course of evolution is maintained in simulations with $S=1.0$. It differs then from the simulations of non-magnetized envelope and from the $\beta=1$ case. 
 Situation changes in case of the models with $S=1.4$. Evolution of the profiles is similar for every $A_0$ value. At first $\sim 0.4 \textrm{s}$ there are some  deviations from spherical symmetry. Contours of $A_{\phi}$ are disturbed at the central part of the grid. In the radial Mach number profile, the sonic surface has an eight-like shape. Those features start to change around $t= 0.4 \textrm{s}$, which corresponds to the oscillation in accretion rate. Sonic surface shape starts to vary, it gets separated and splits into two parts: one of a small circular shape at the innermost part of the grid  and one at the scales of $\sim 200 r_g$. At later time, the inner contour gets accreted and the outer gets spherically symmetric and moves toward the outer part of the grid.  Contours of $A_{\phi}$ go back to the original configuration and all the profiles are smoothed and spherically symmetric.
This is an effect of evacuating most of the mass from the envelope. 
 
\begin{figure*}
  \begin{tabular}{cccc}
 \includegraphics[width=0.33\textwidth]{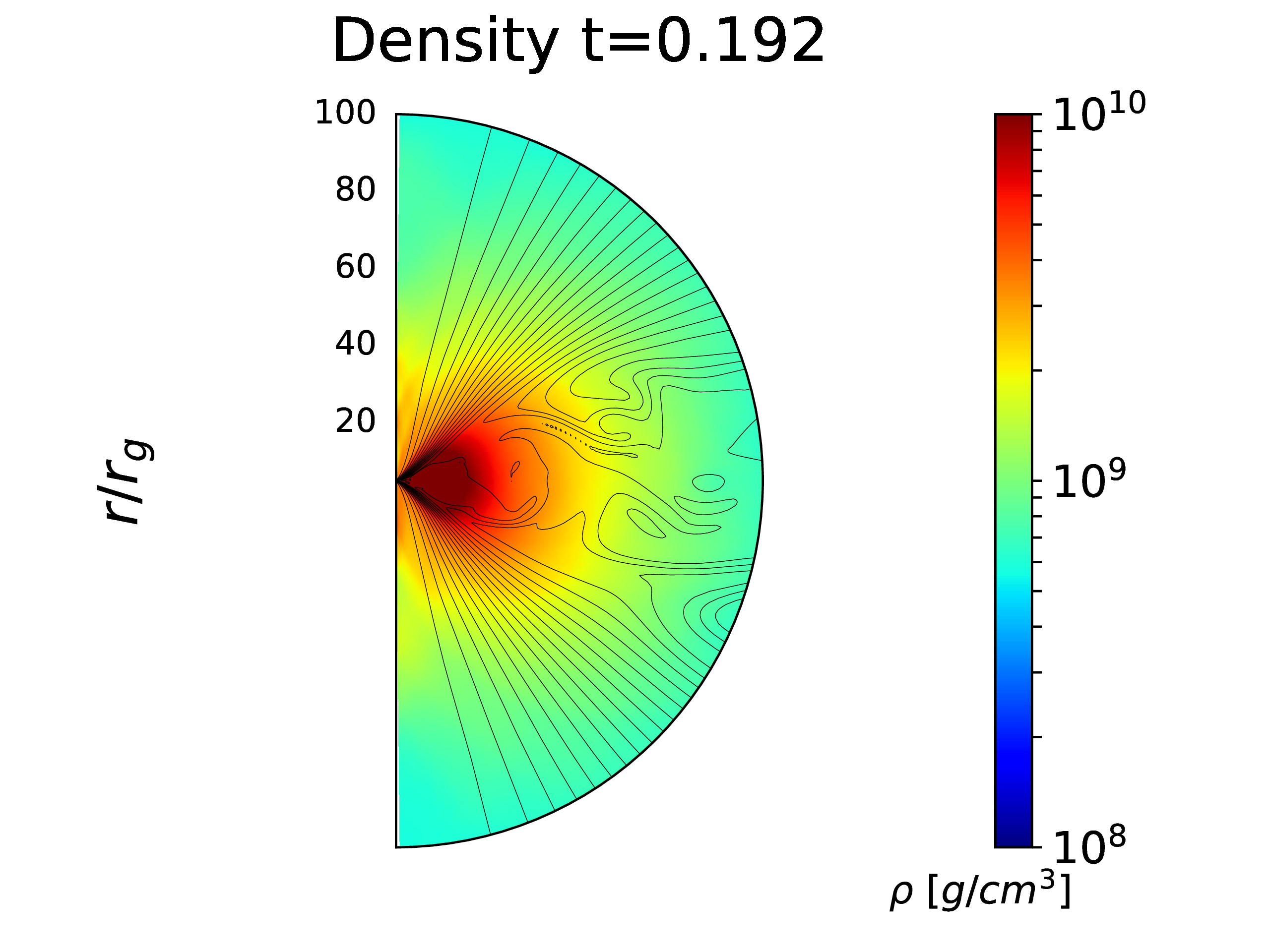}
 \includegraphics[width=0.33\textwidth]{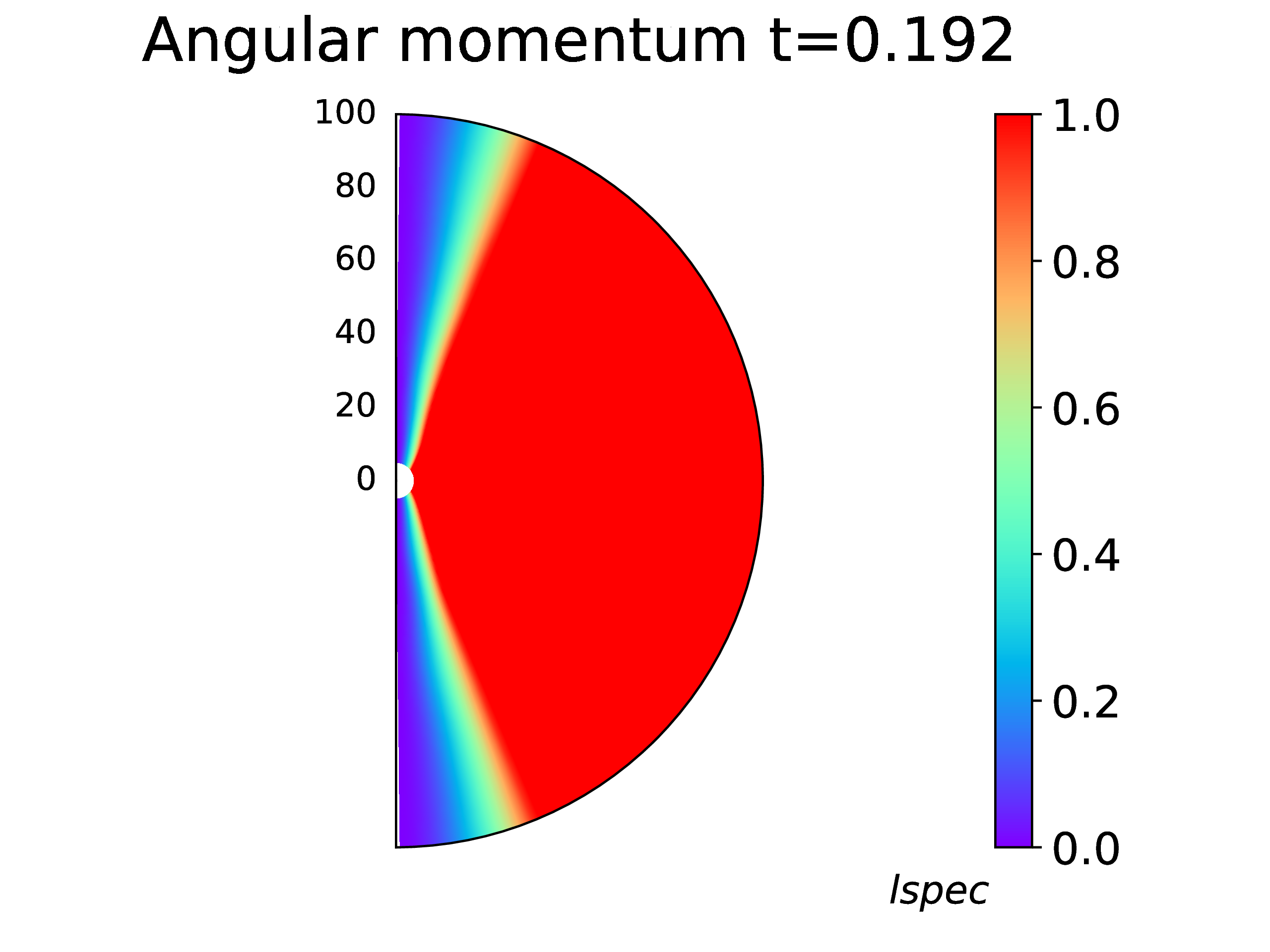} &
 \includegraphics[width=0.33\textwidth]{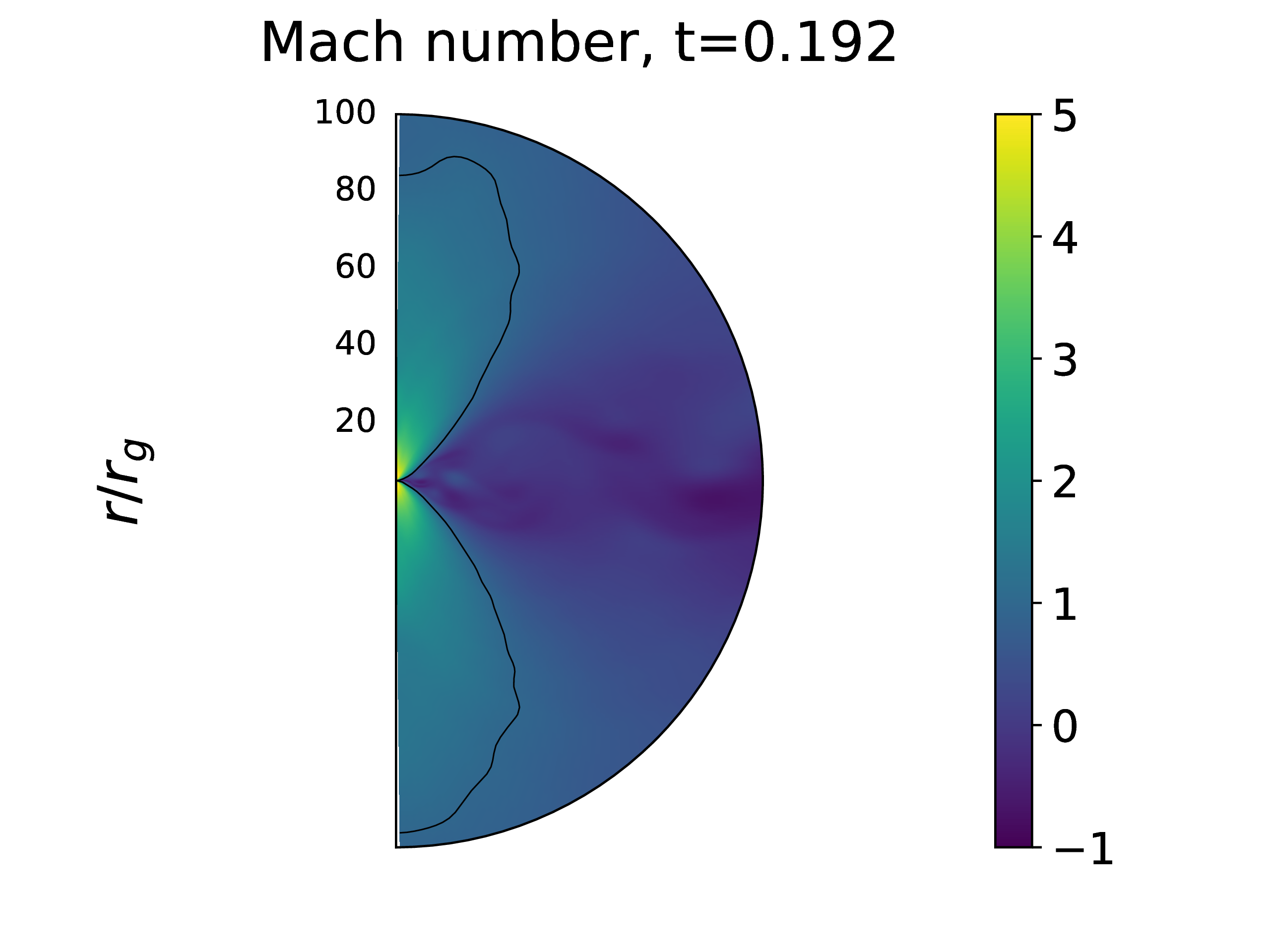} &
\\ 
 \includegraphics[width=0.33\textwidth]{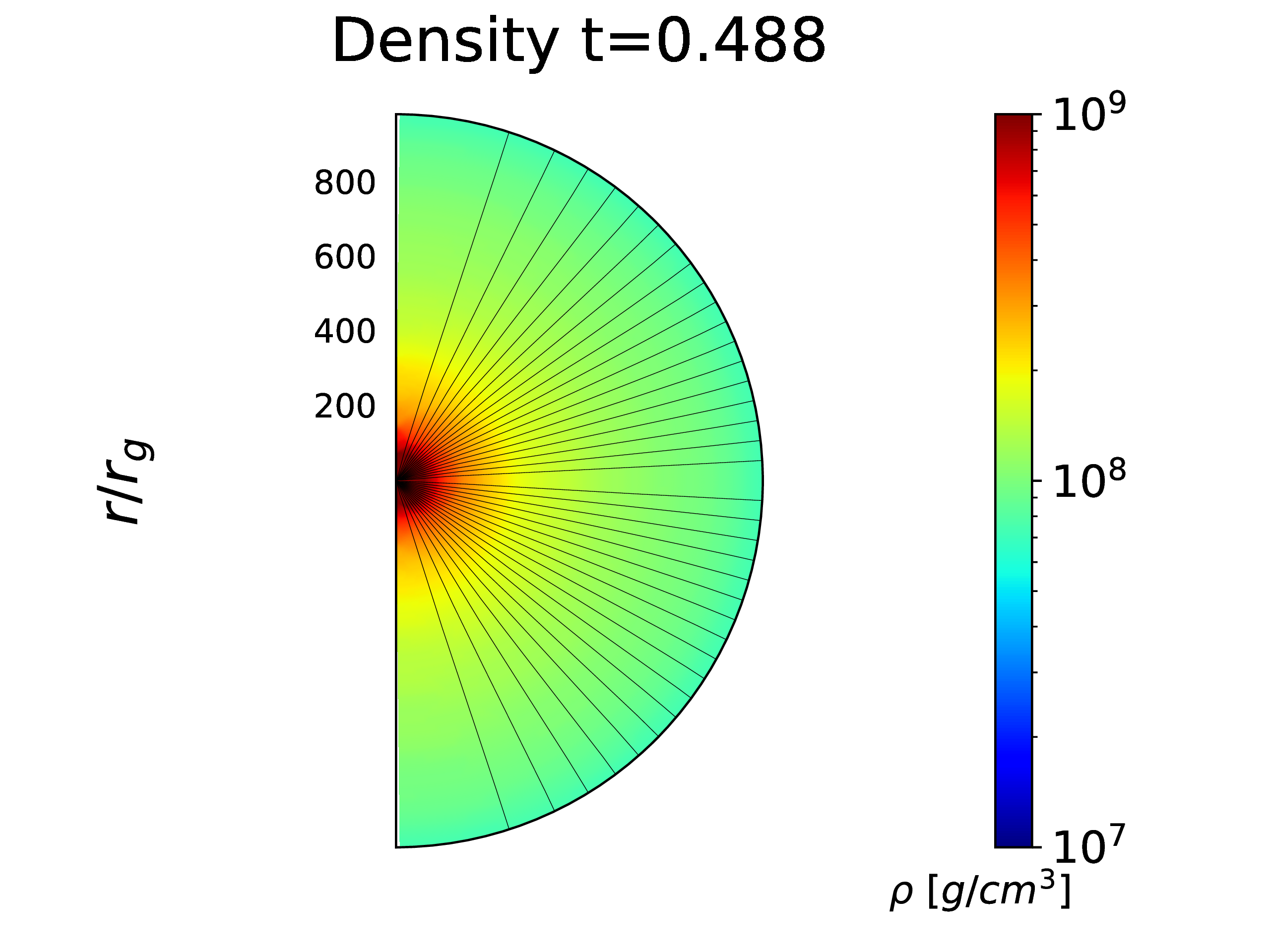}
 \includegraphics[width=0.33\textwidth]{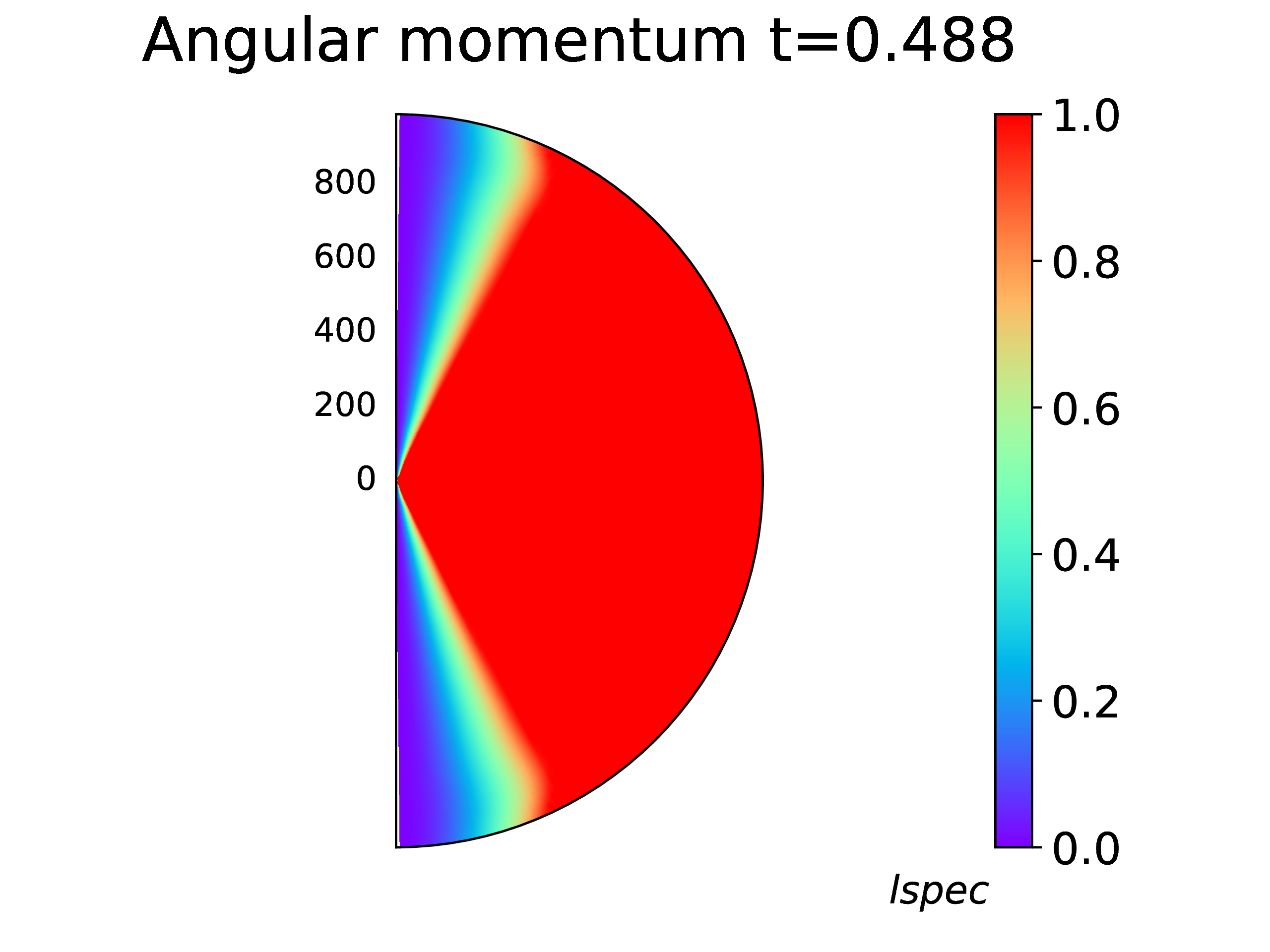} &
 \includegraphics[width=0.33\textwidth]{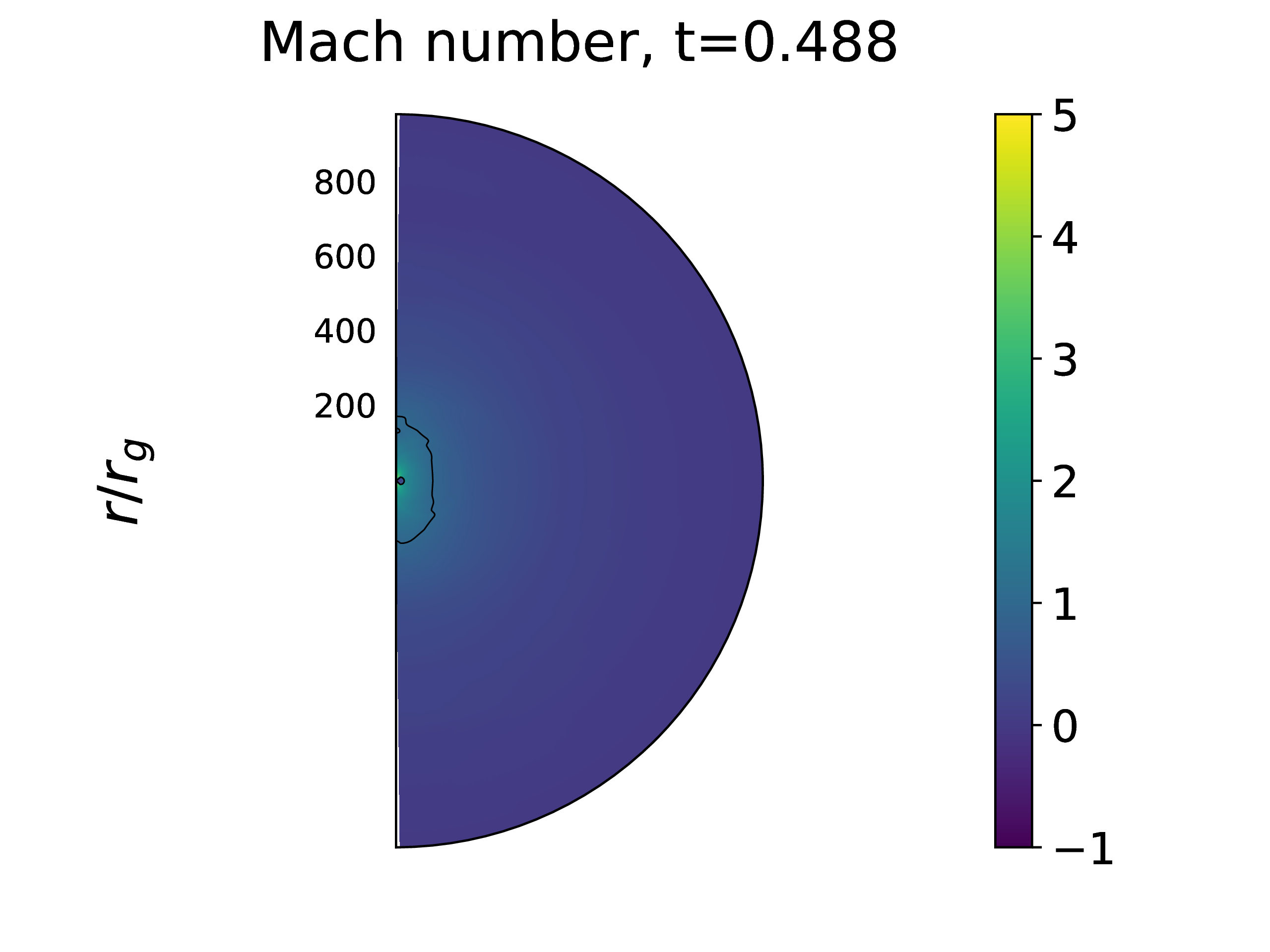} &
\\
 \includegraphics[width=0.33\textwidth]{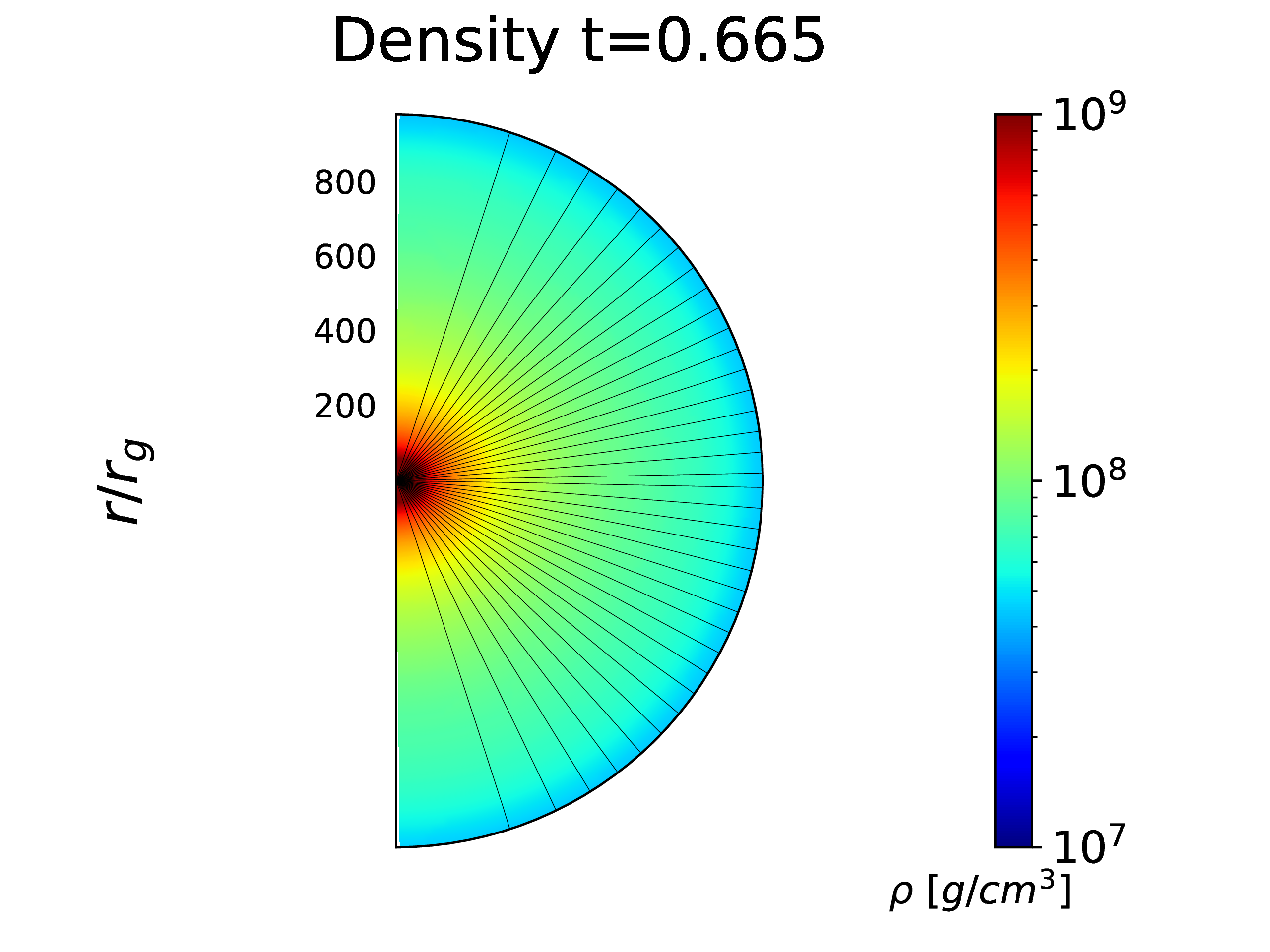}
 \includegraphics[width=0.33\textwidth]{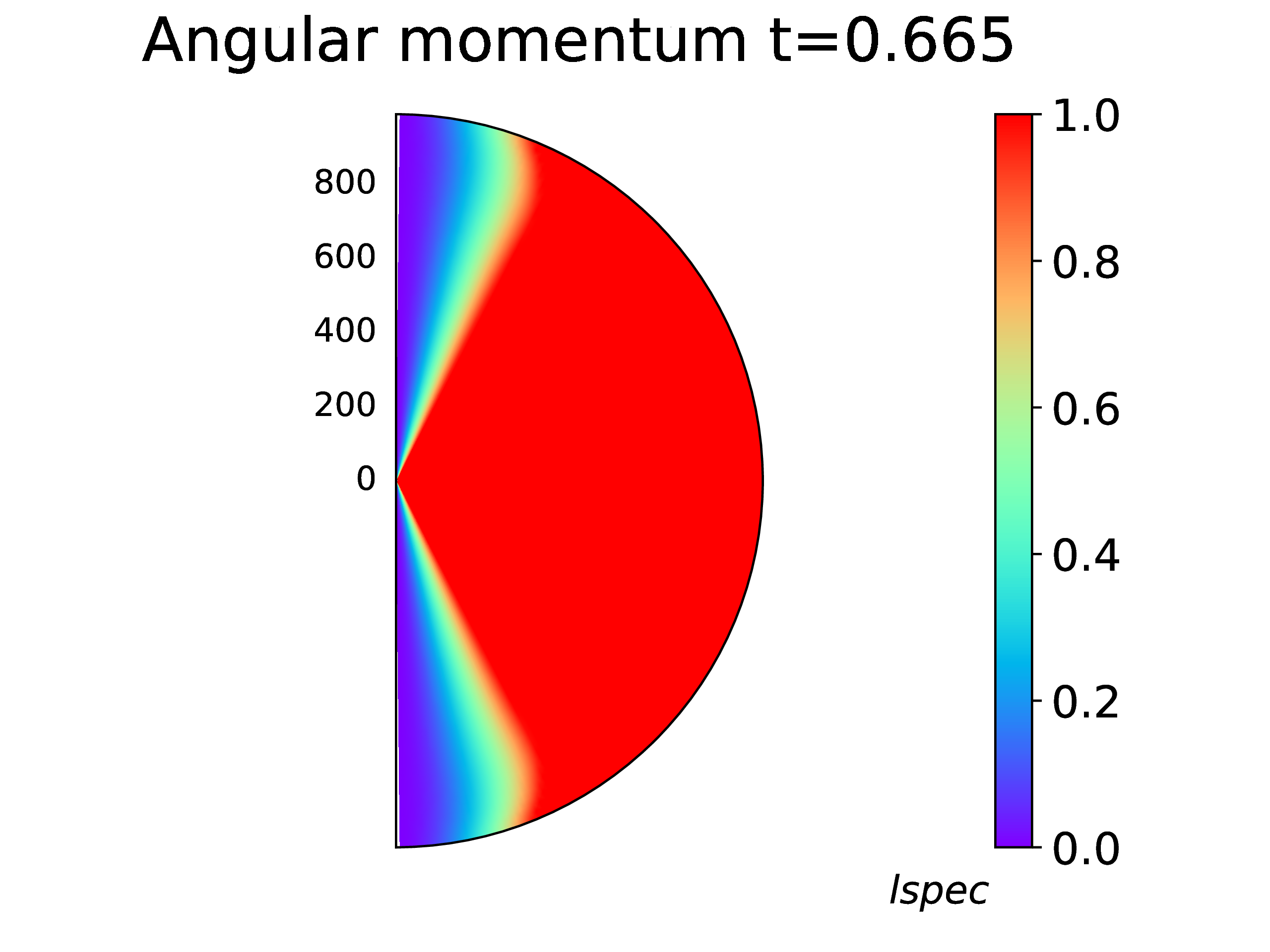} &
 \includegraphics[width=0.33\textwidth]{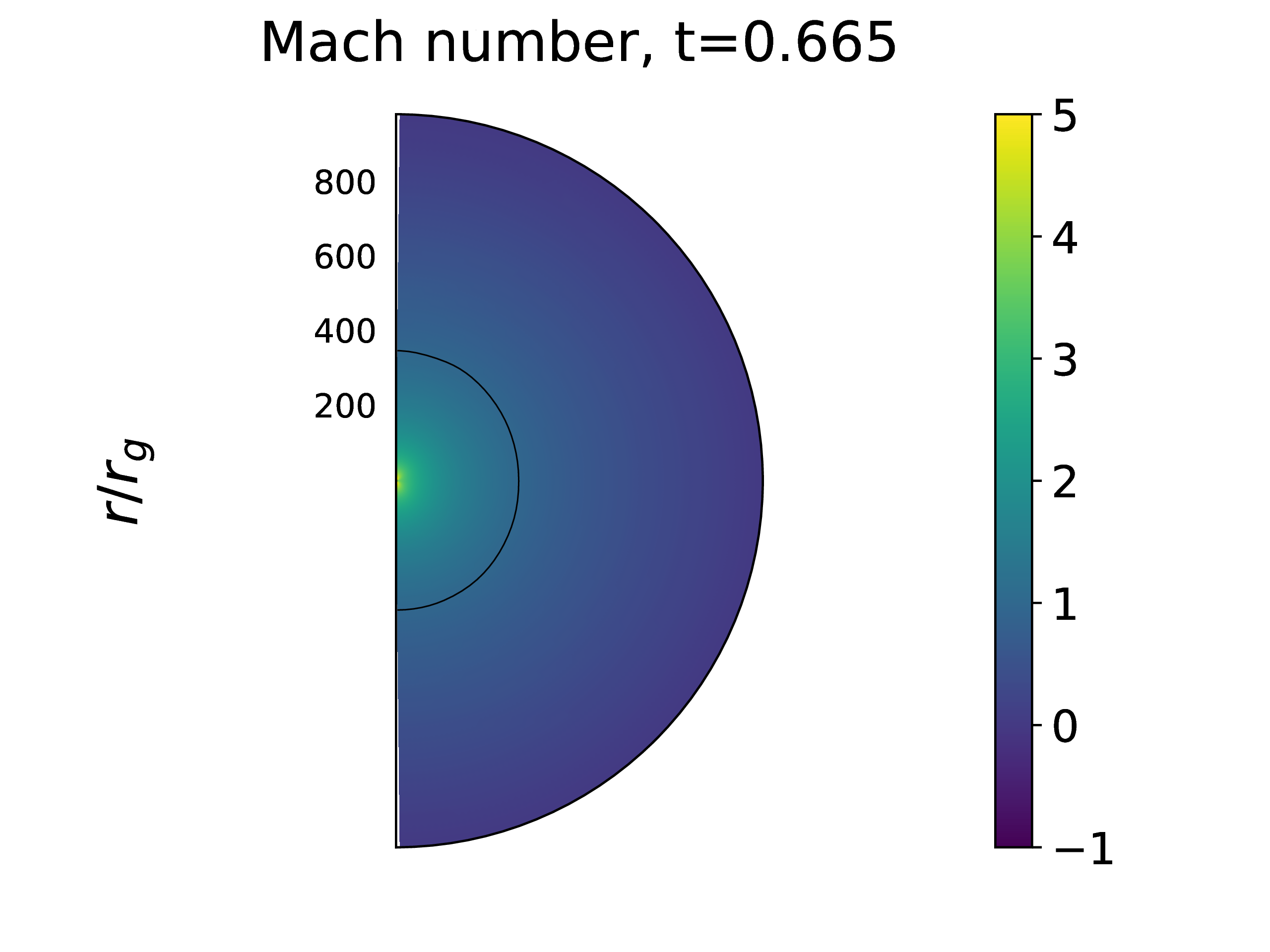} &  
\end{tabular}
\caption{The results, from left to right, for the Density with overplotted contours of magnetic field $A_{\phi}$, Specific Angular Momentum, and Mach Number  distributions, for the model with  magnetic fields and initial black hole spin of $A_{0}=0.3$. The rotation parameter was $S=1.4$ and $\beta=100$ (NS-LM-14).
  The color maps are taken at times $t=0.192\textrm{s}$, $t=0.488\textrm{s}$ and $t=0.665 \textrm{s}$ which corresponds to time before, right after  and after the spike in accretion rate. (Note that first row shows a different spatial scale of 100 $r_{g}$, for which also density scale is adjusted).
  In addition, contour of $M=1$ is shown with a black line.
  }
     \label{fig:beta100profiles14}
\end{figure*}

 We present selected profiles in the Fig.\ref{fig:beta100profiles14}.
 
The presence of magnetic field and its strength influences the results in many different ways. In general, it affects the global parameters evolution and their dependence on the initial conditions. In particular, we get the highest peak value of accretion rate for non-magnetized envelope with $A_0=0.3$ and $S=1.4$ ($\sim 80 M_{\odot}\textrm{s}^{-1})$. The peak value appears during oscillations and does not persist, so it does not affect significantly the value of the final black hole mass. Presence of magnetic field modifies the shape of $\dot{M}$ evolution, the least affected case is sub-critical rotation of the envelope and for super-critical rotation we see substantial changes.  Moreover, for magnetized envelopes we get significantly smaller maximal spins of the black hole. There is no magnetized model in which $A=1$ is achieved. Spins in simulations with weaker magnetic field are higher than for $\beta=1$. Interestingly, the highest black hole masses are reached for weaker magnetic field, and they are slightly higher than those in strongly magnetized envelopes, while the models without magnetic field result in the significantly smaller $M^f_{BH}$ (see details in Table \ref{tab:models}). Some particular sets of initial parameters do not follow this general tendencies. They were described in previous sections.

 We attribute this result to the action of black hole rotation,
 which winds up the magnetic field lines, and launches Poynting dominated jets
 more efficiently, while the uncollimated wind outflows are decaying.
 The black hole spin does not affect the results directly, but its action is moderated by the magnetic fields.

\subsubsection{Strongly magnetized envelope: $\beta=1$}
\label{sec:beta1}

In this subsection we discuss in more detail distributions of density and magnetic field 
angular momentum and radial Mach number.
We focus on the representative models in selected, characteristic snapshots representative for simulations, which correspond to changes in the $\dot{M}$ plot.

\begin{figure*}
\begin{tabular}{ccc}
\includegraphics[width=0.33\textwidth]{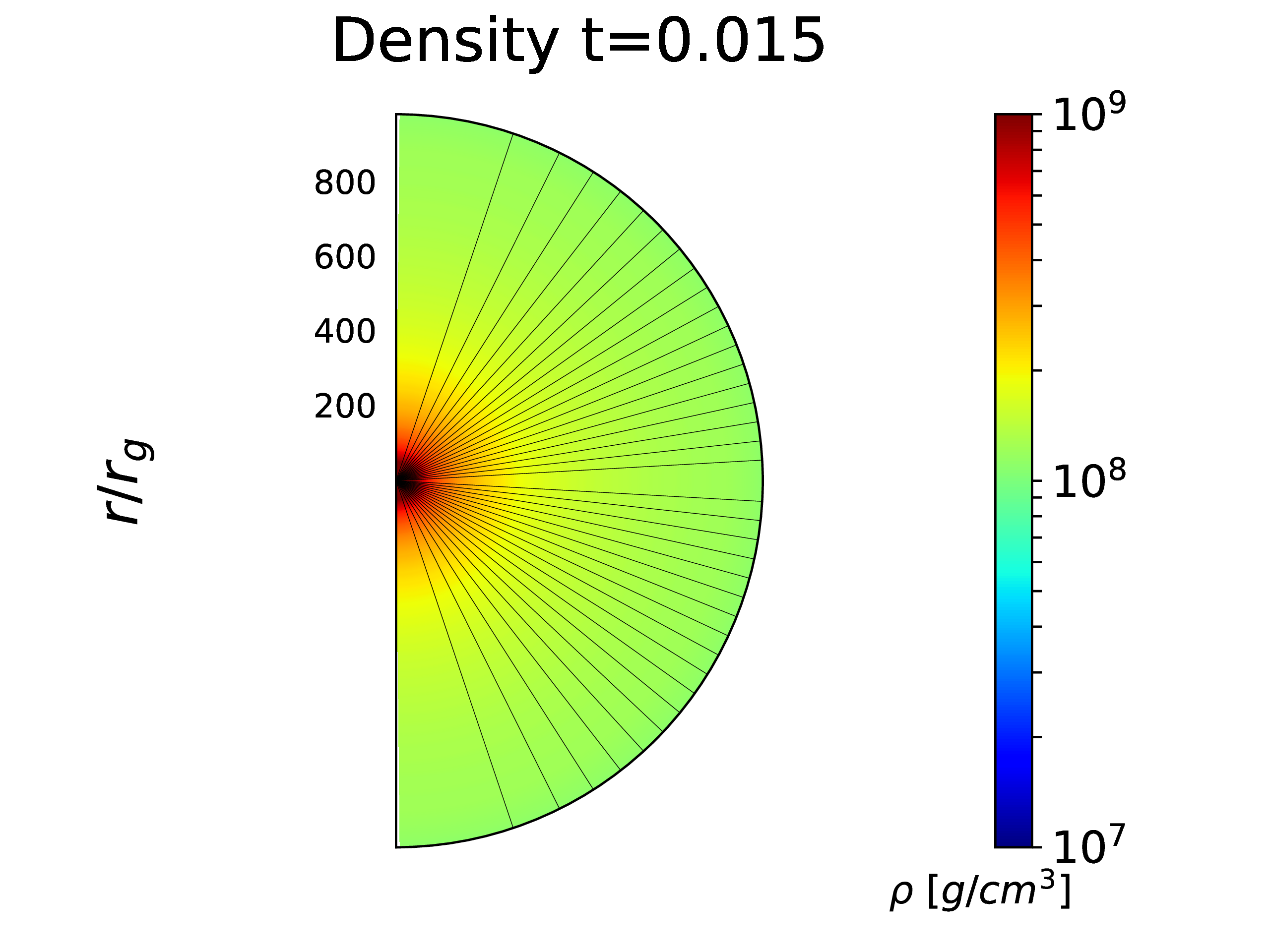} &
 \includegraphics[width=0.33\textwidth]{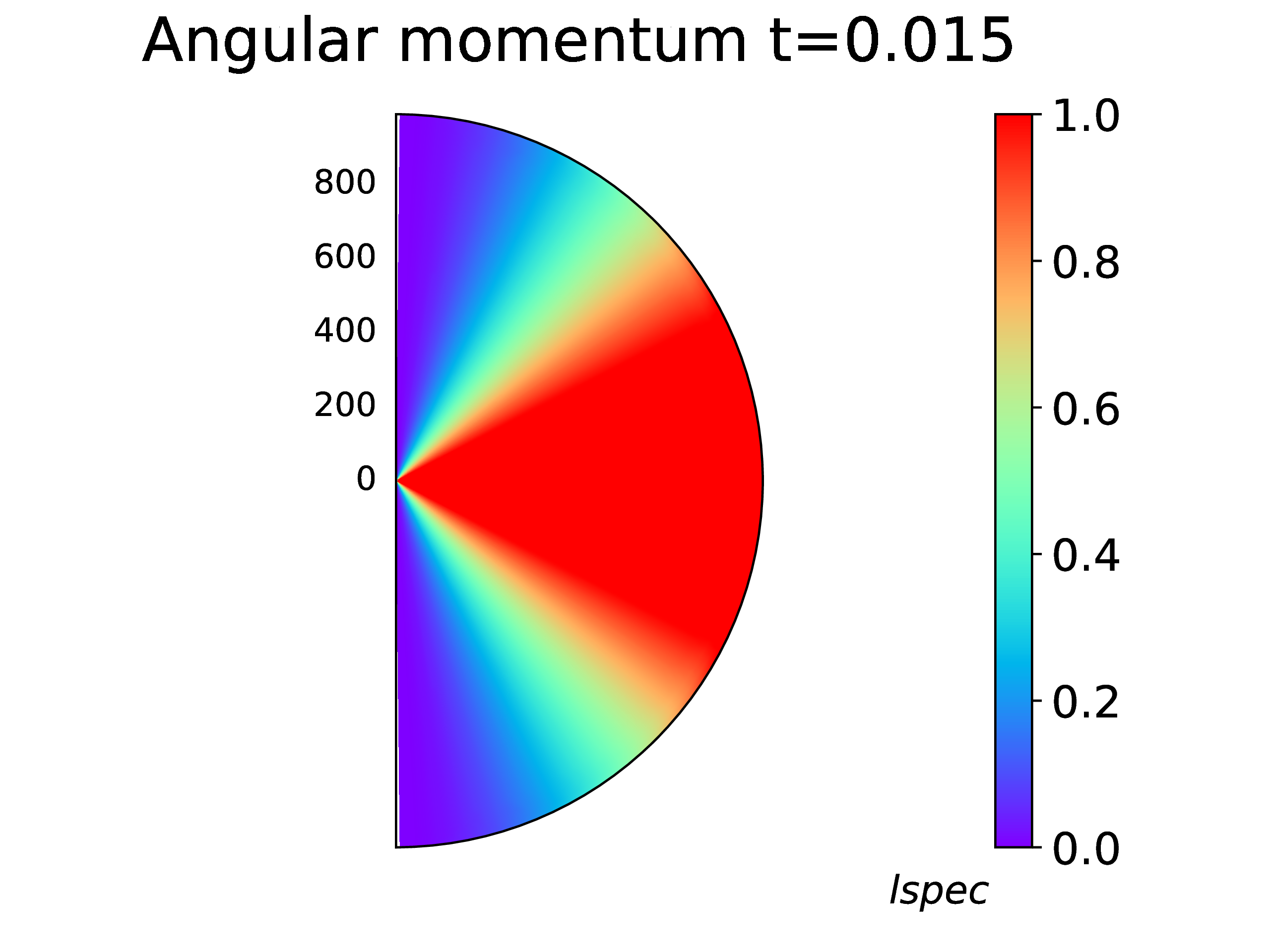} &
 \includegraphics[width=0.33\textwidth]{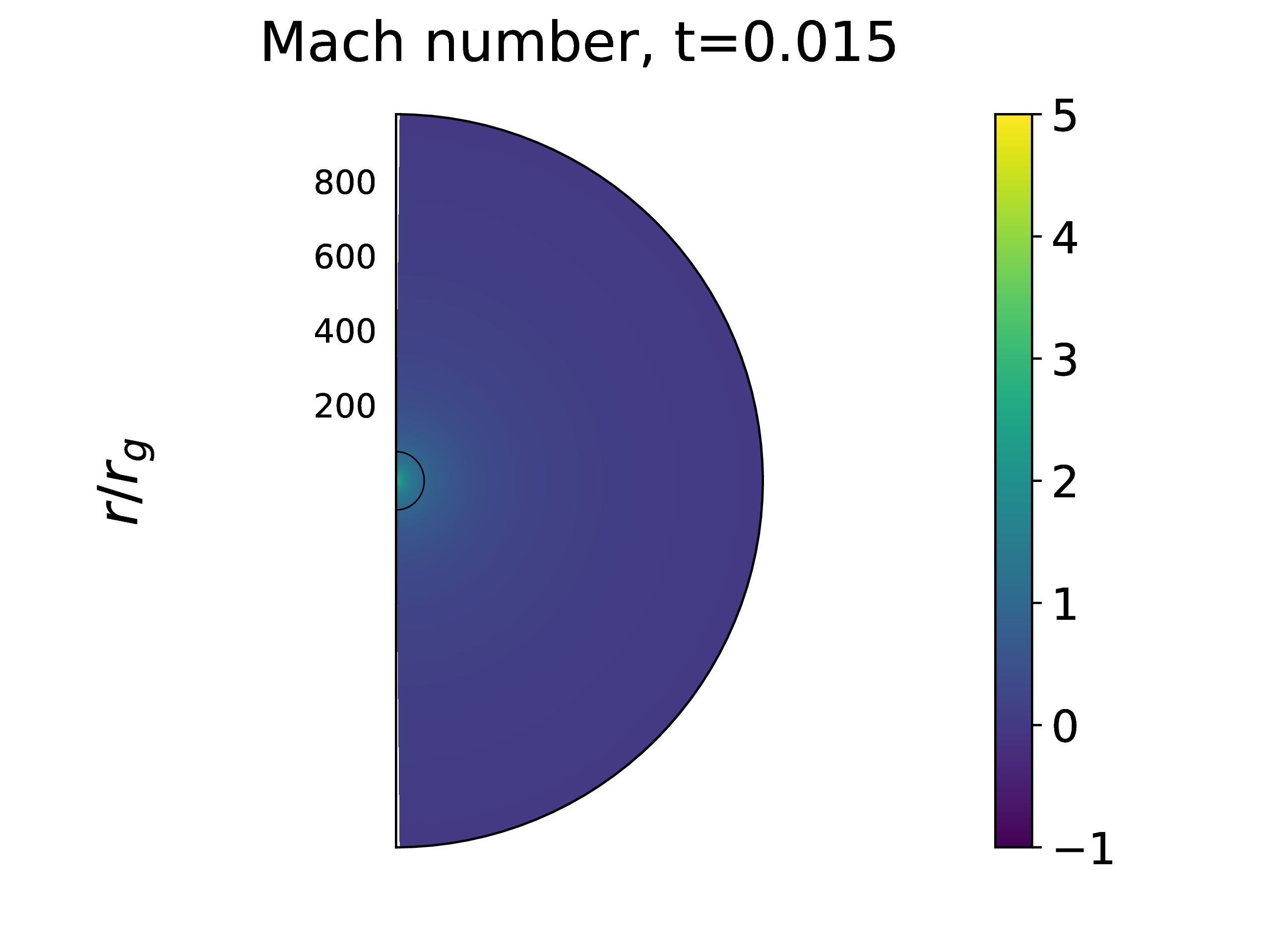}
\\
\includegraphics[width=0.33\textwidth]{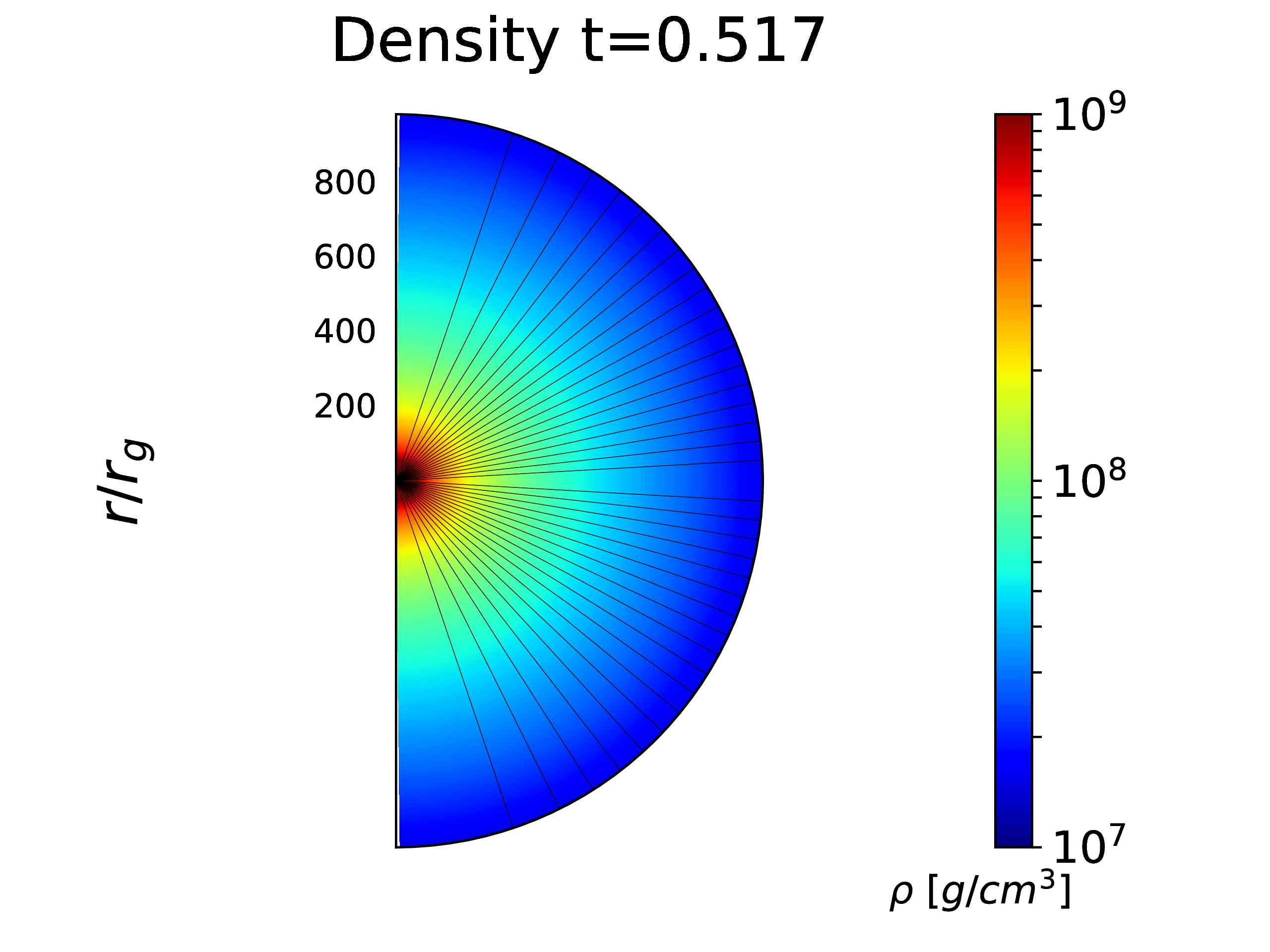} &
 \includegraphics[width=0.33\textwidth]{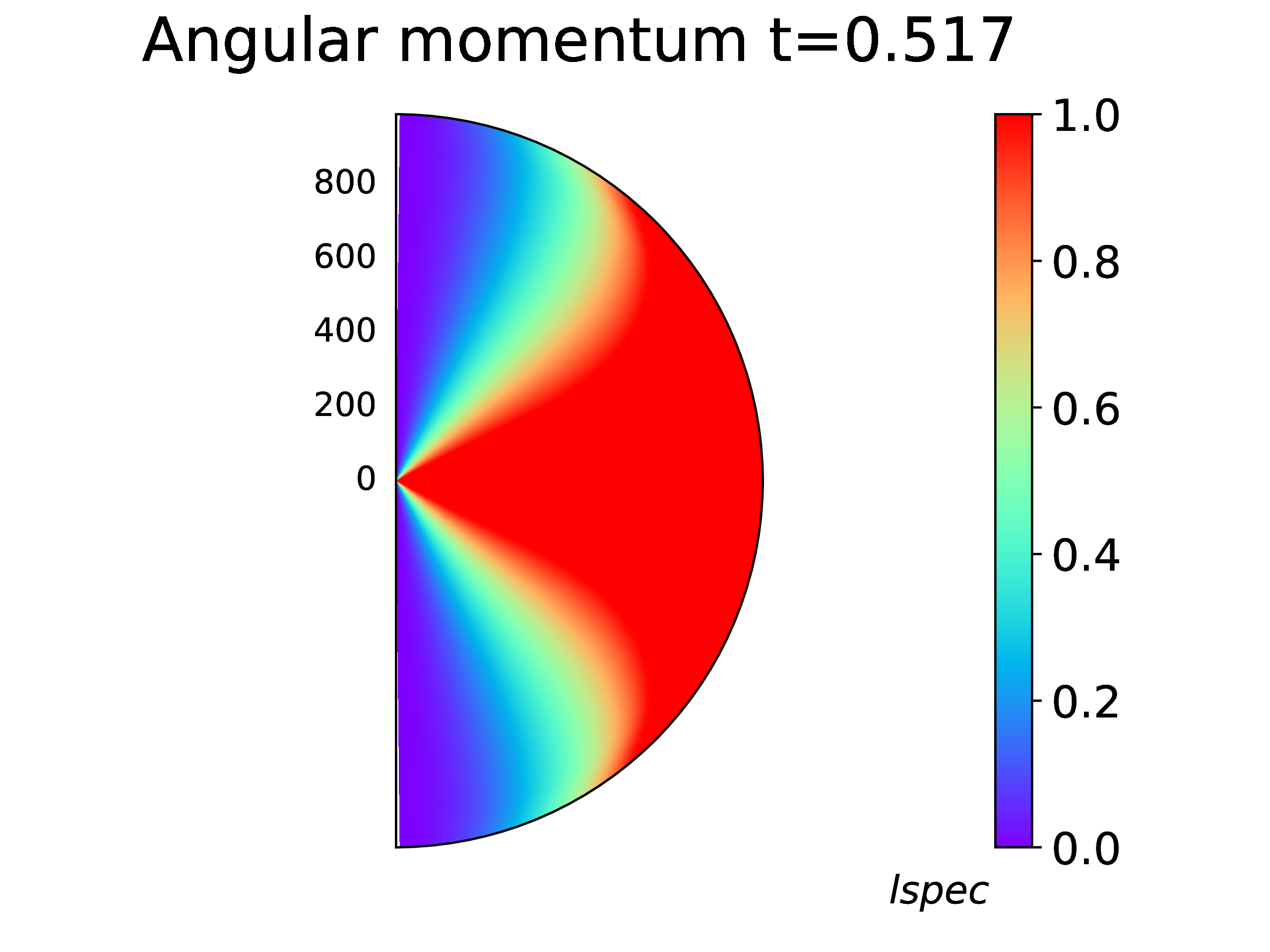} &
 \includegraphics[width=0.33\textwidth]{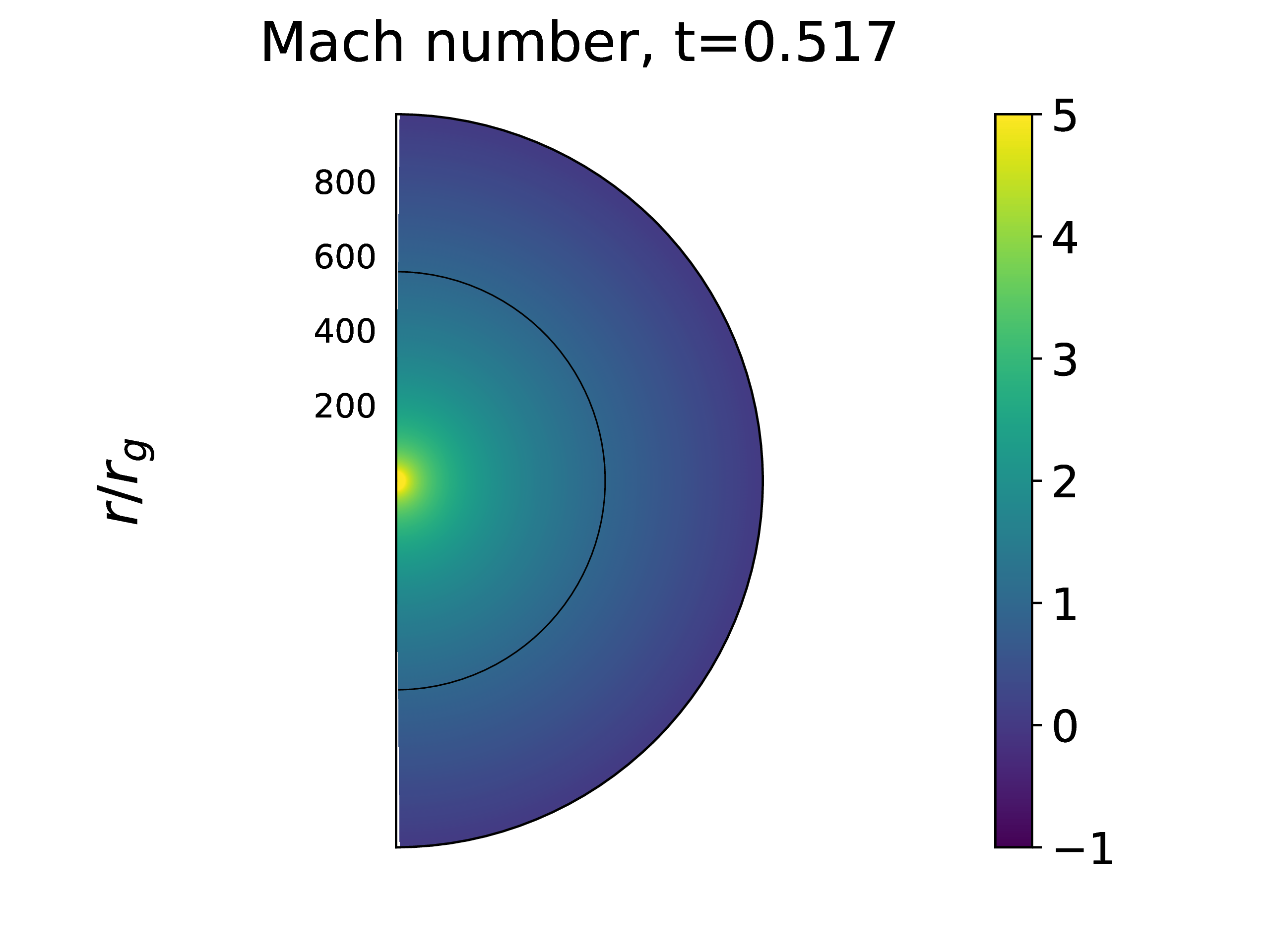}
\\
 \includegraphics[width=0.33\textwidth]{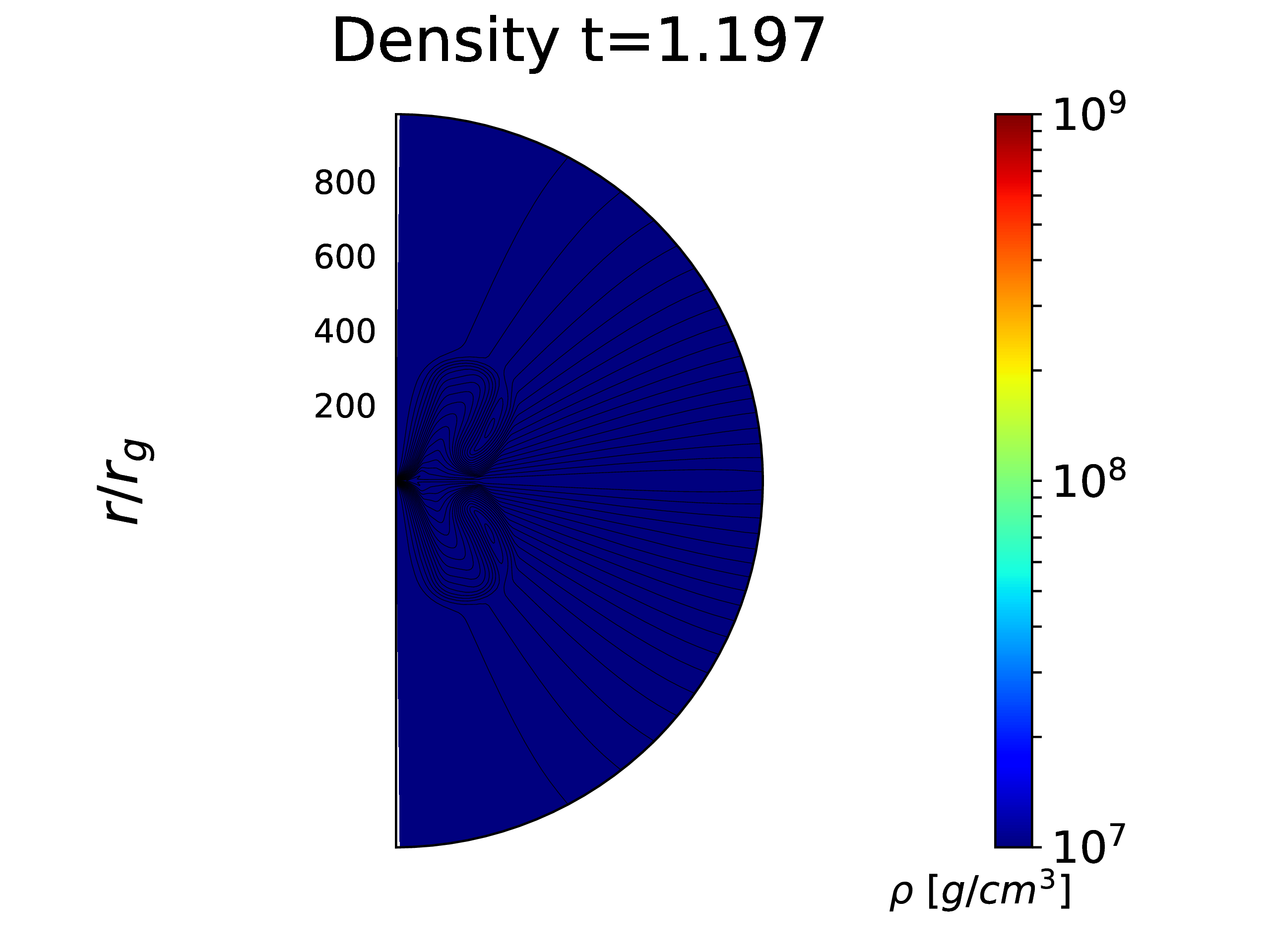} &
 \includegraphics[width=0.33\textwidth]{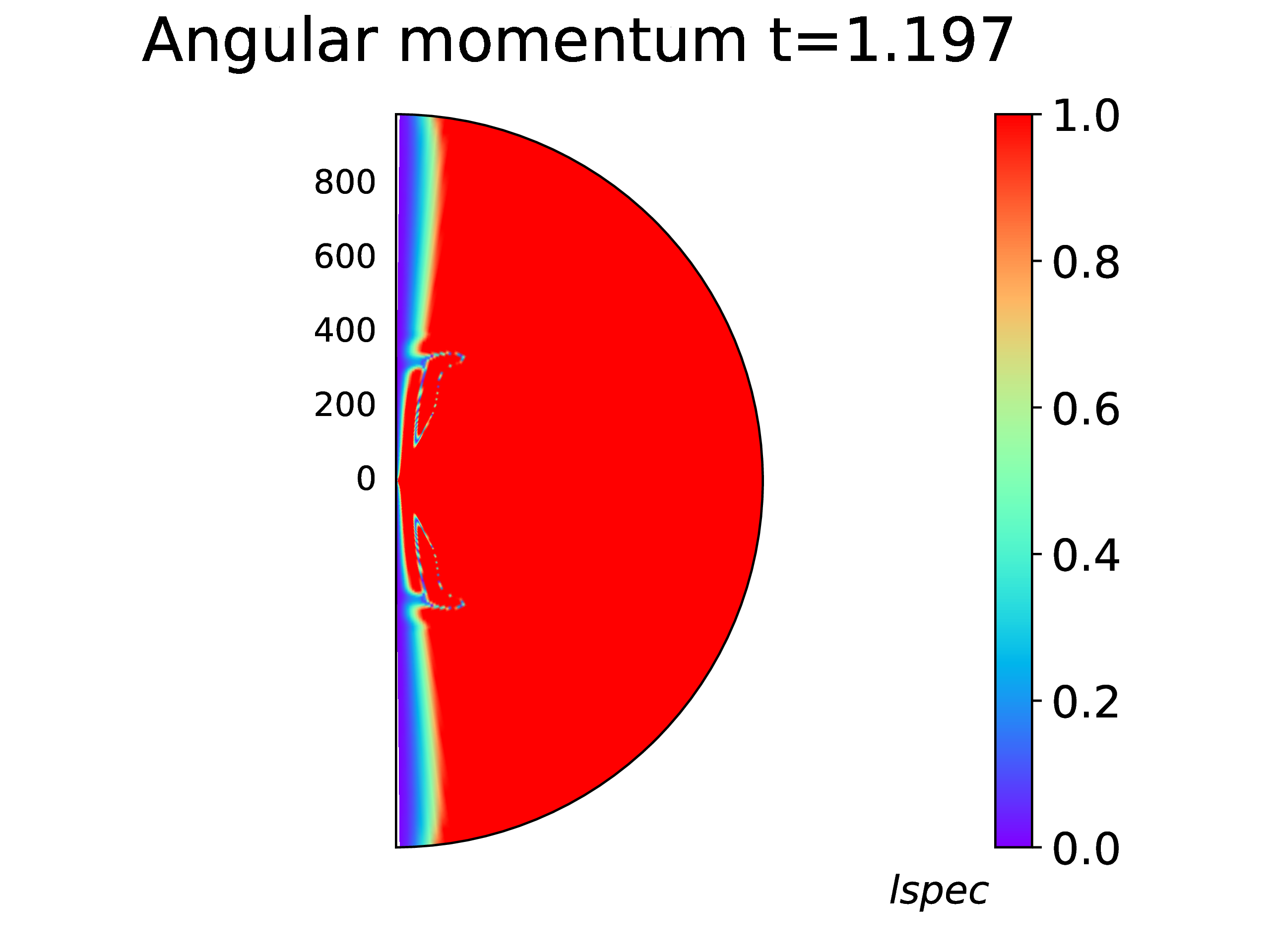} &
 \includegraphics[width=0.33\textwidth]{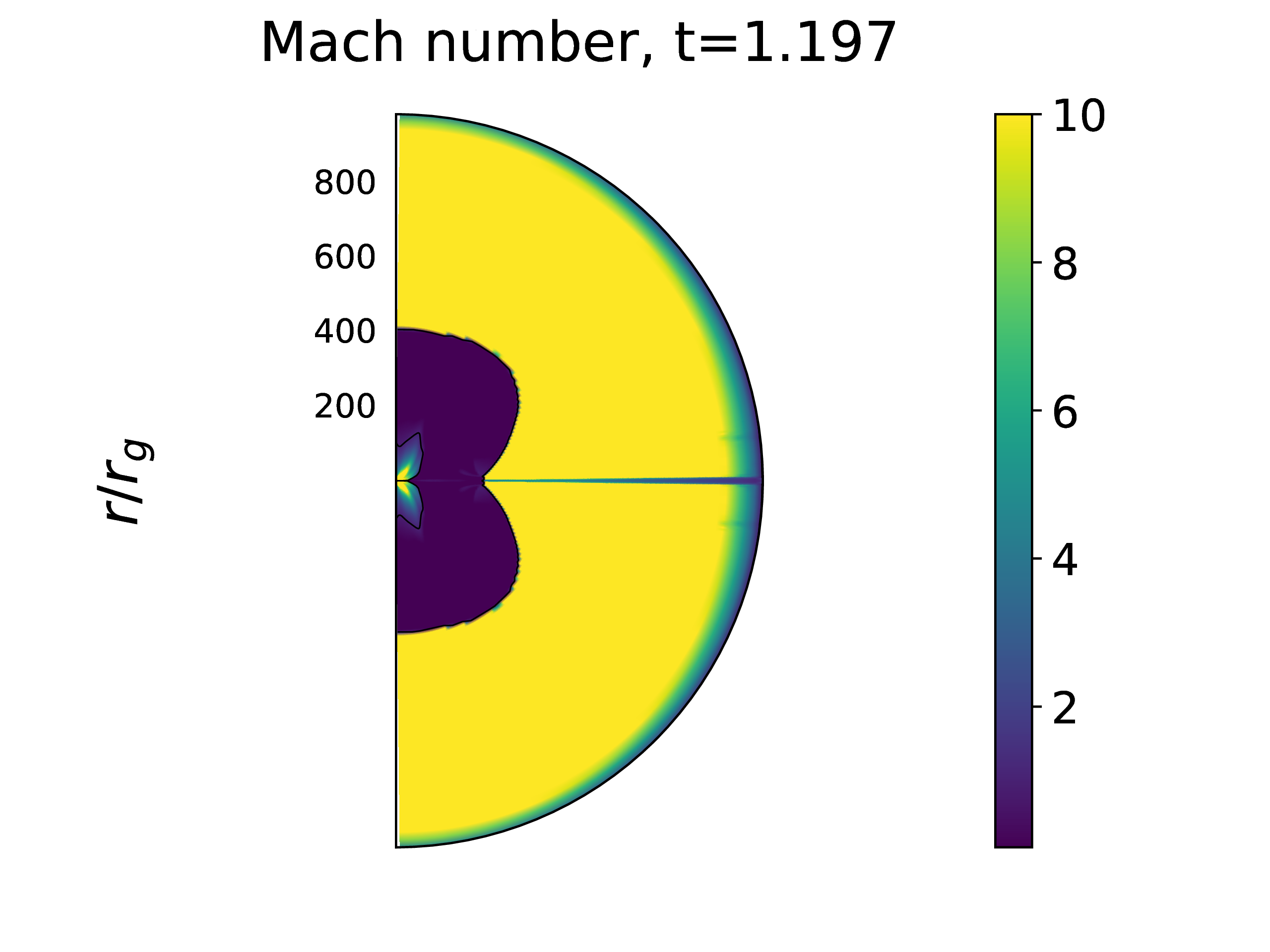}
\\
 \includegraphics[width=0.33\textwidth]{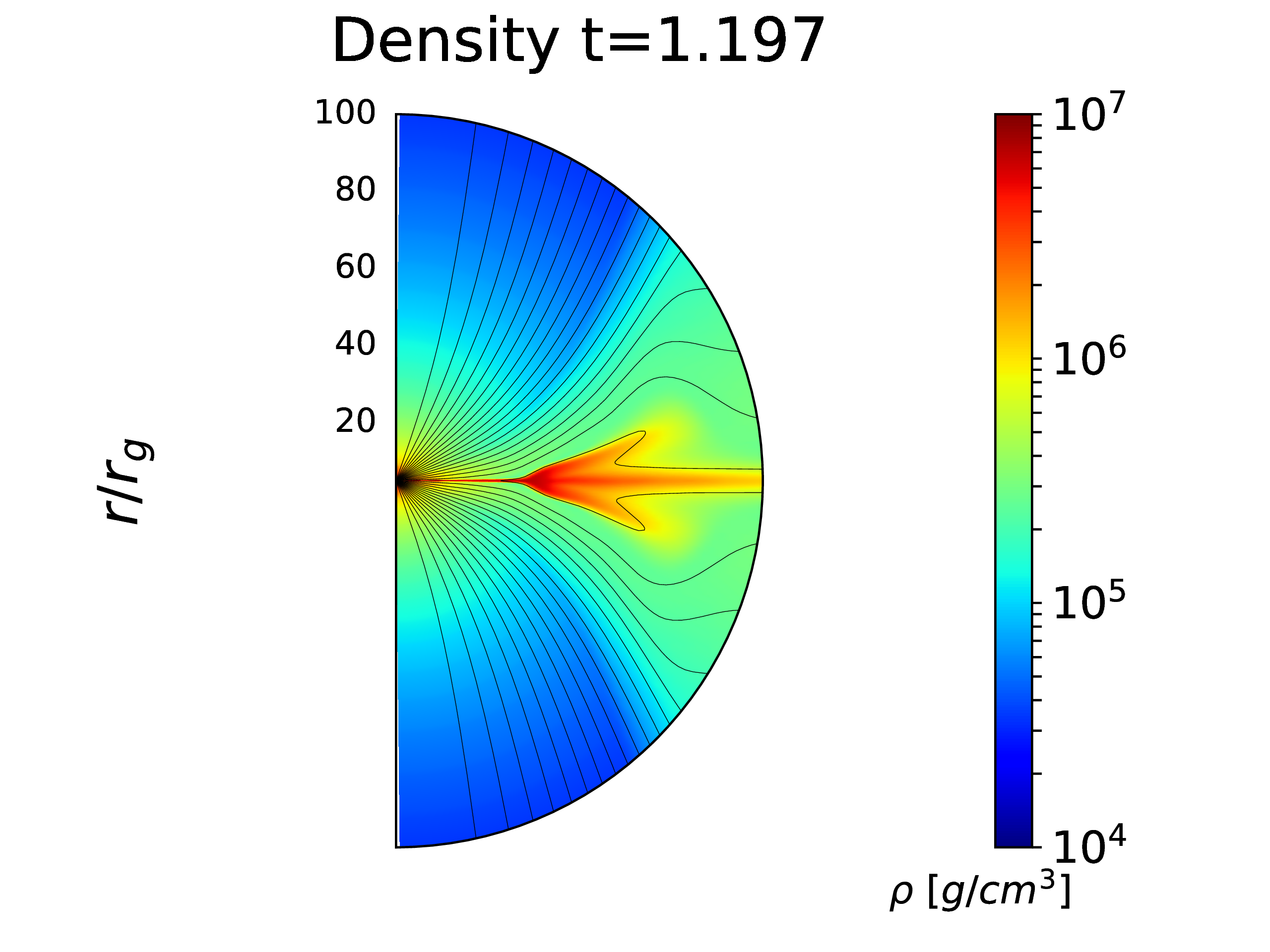}  &
  \includegraphics[width=0.33\textwidth]{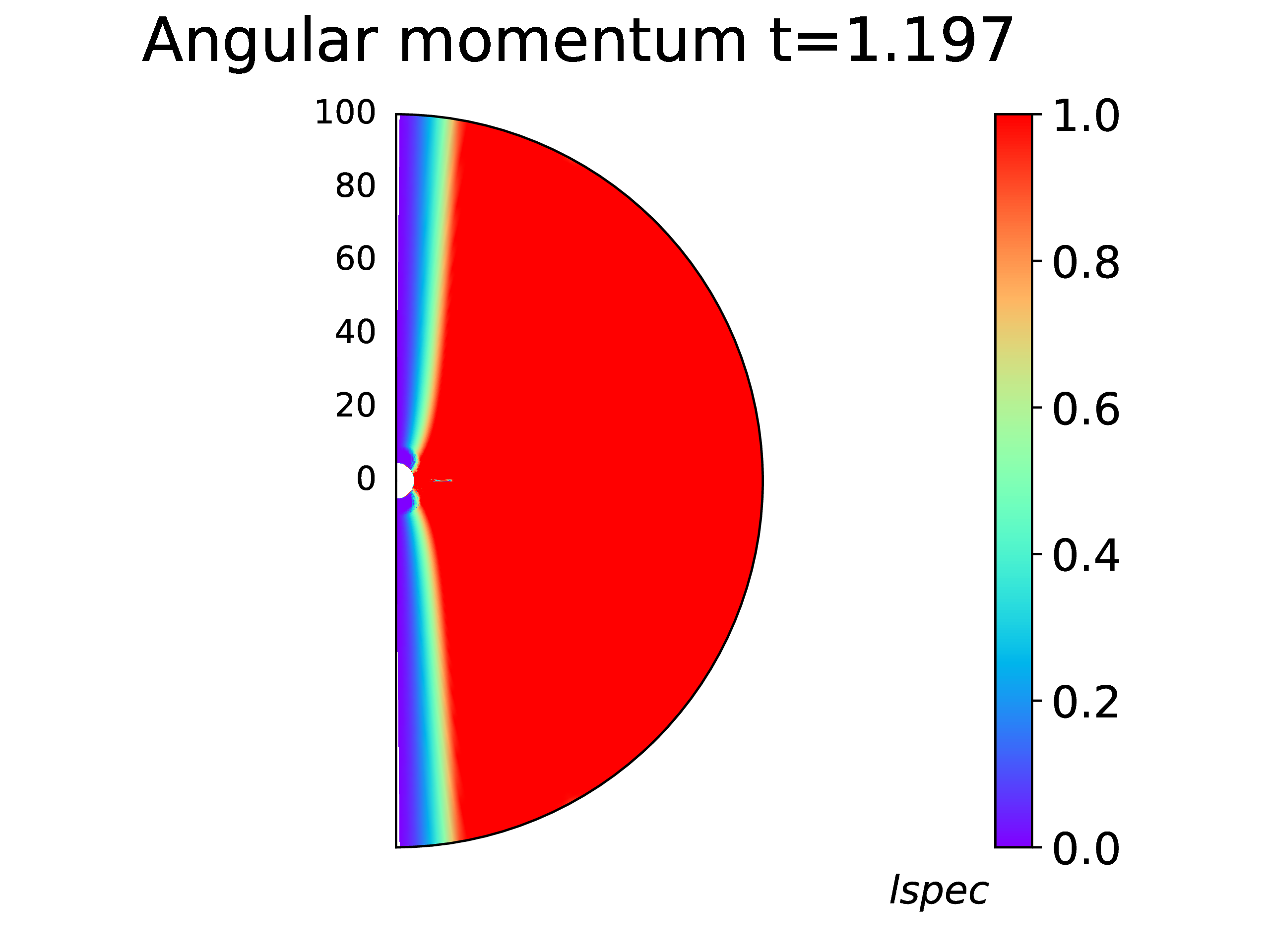}   &
  \includegraphics[width=0.33\textwidth]{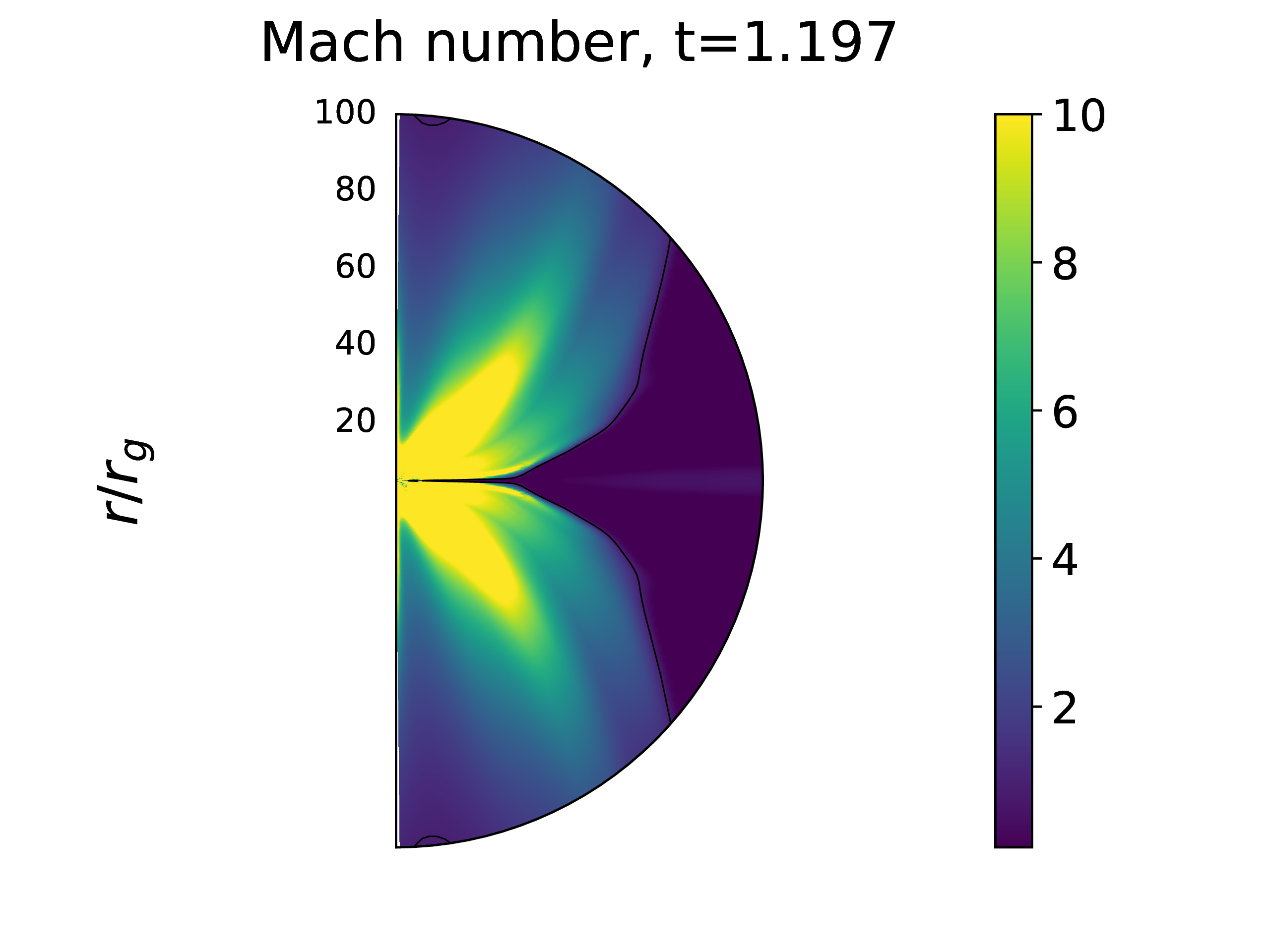}
  \end{tabular}
\caption{The results, from left to right, for the \textit{Density, Specific Angular Momentum, and Mach Number} distributions, for the model with  magnetic field characterized by $\beta=1$, and initial black hole spin of $A_{0}=0.3$. The rotation parameter was $S=0.4$ (model NS-HM-04).
First row presents profiles at the beginning of the simulation.In the second row we show profile from intermediate stage of the simulations: $t=0.517$. In the third and the fourth row we present the color maps taken at time $t=1.197$ which corresponds to final stages of  the simulation.
In addition, contour of $M=1$ is shown with a black line. 
      }
      \label{fig:A03S04Beta1_profile}
\end{figure*}

In Figure \ref{fig:A03S04Beta1_profile} we present selected profiles of density, Mach number and specific angular momentum for simulations with $S=0.4$ and $A_{0}=0.3$. The flow is magnetized and magnetic field lines are over-plotted on density maps. Here we show the models with lowest rotation parameter. For most of the simulation time, the density distribution is spherically symmetric, similarly to the $S=0.4$ case where the simulations were without magnetic field. The shape of the contours of $A_{\phi}$ stays radial and is unchanged during that time. The matter slowly accretes radially towards the black hole, and the density in the whole domain uniformly decreases, and the flow becomes supersonic, as can be seen by comparison between first and second snapshots.
The flow structure changes around the time $1.1 \textrm{s}$. Now the density profile flattens towards the equator and the shape of the $A_{\phi}$ contours gets disturbed. At the end of the simulation there is a structure created which is symmetrical about equator and manifests a pile-up of matter in the density profile, with turbulent magnetic field depicted by the contours of $A_{\phi}$. The radial Mach number profile shows that the flow here is highly subsonic, and the angular momentum increase means it is rotating down to the black hole horizon. 
The last row in this Figure shows the state of the simulation at time 1.19 s, zoomed in to the innermost part of the flow. Here we notice that the very thin region located at the equator is in fact rotating, while the material with small angular momentum and supersonic speeds can reach the black hole and disappear under the black hole horizon. At the same time, open magnetic field lines start forming, while they bend towards the polar axis. This evacuated region can be envisaged as the place where the jet is could be born in the collapsar's envelope. The amount of angular momentum in this model is however not sufficient to sustain the jet for a very long time.

\begin{figure*}
\begin{tabular}{ccc}
 \includegraphics[width=0.33\textwidth]{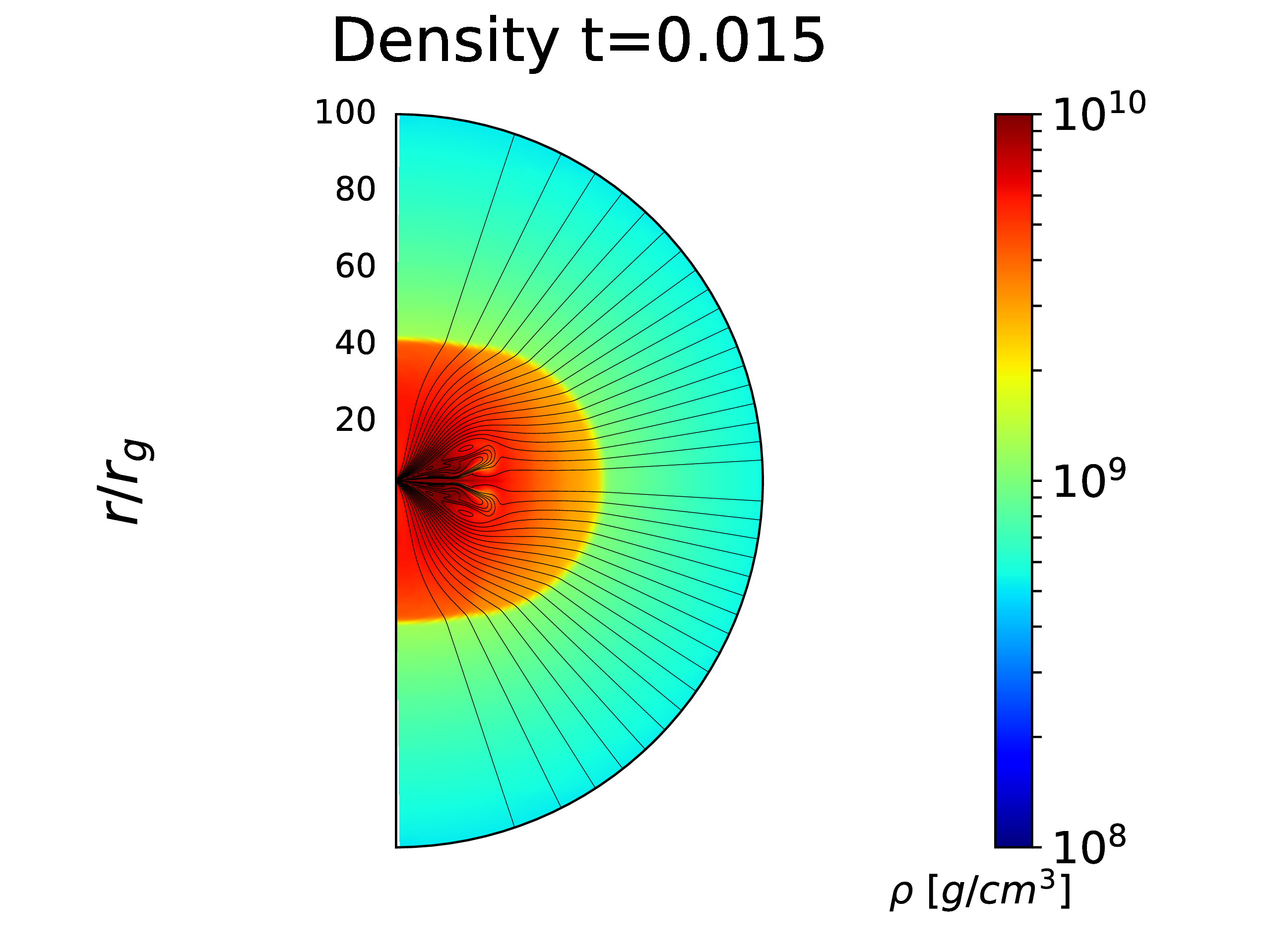}
 \includegraphics[width=0.33\textwidth]{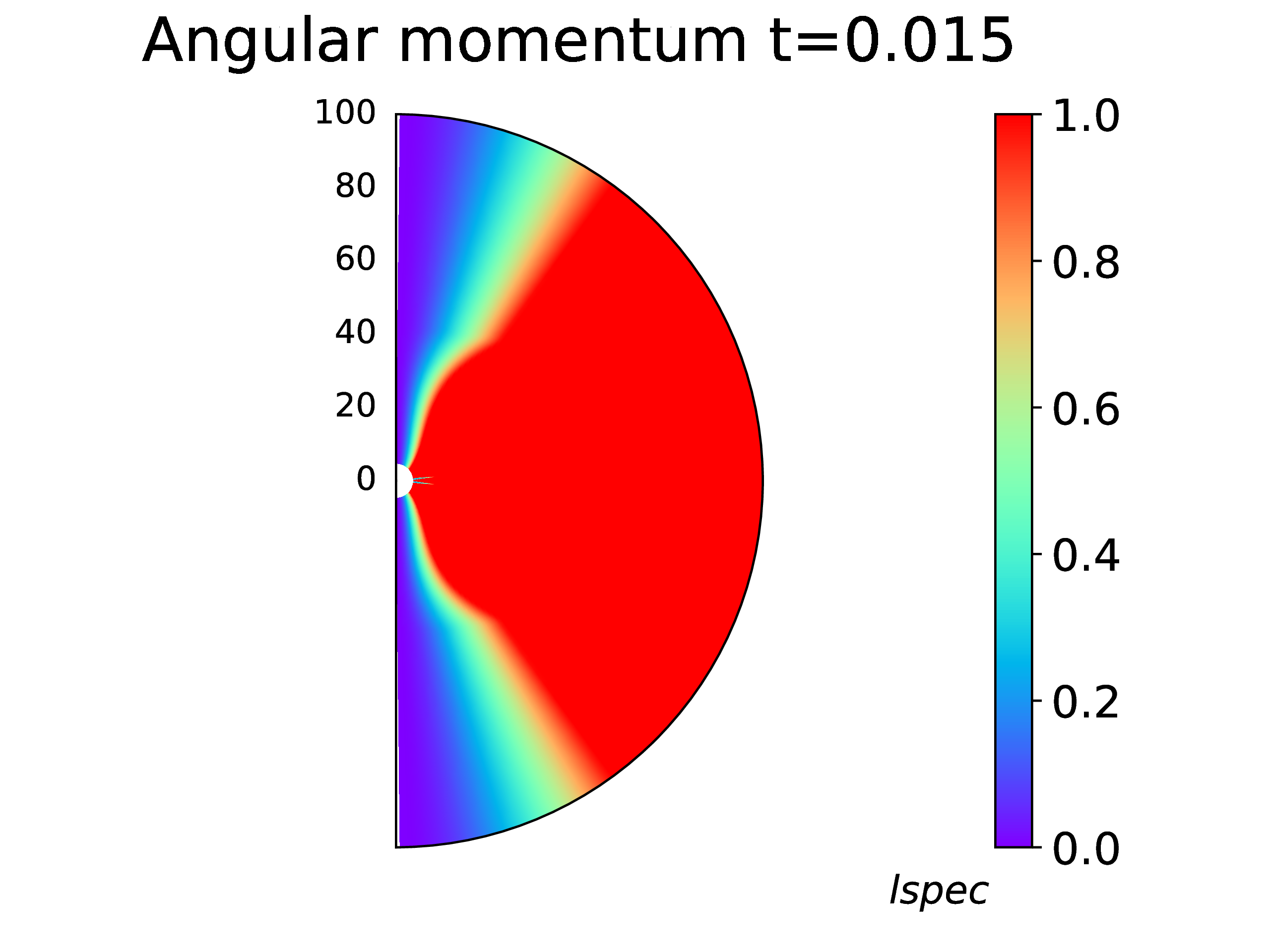} &
 \includegraphics[width=0.33\textwidth]{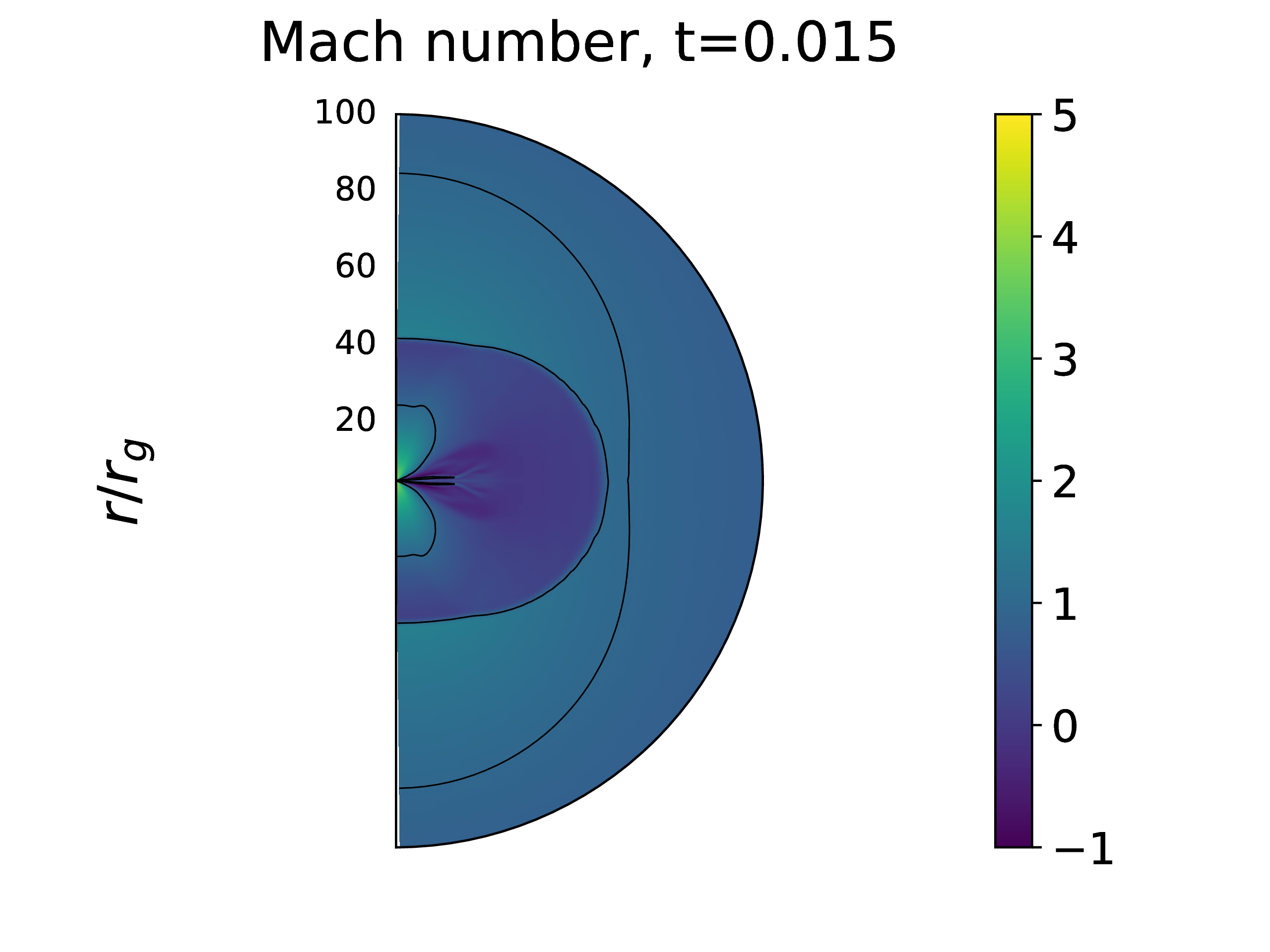} &
\\
 \includegraphics[width=0.33\textwidth]{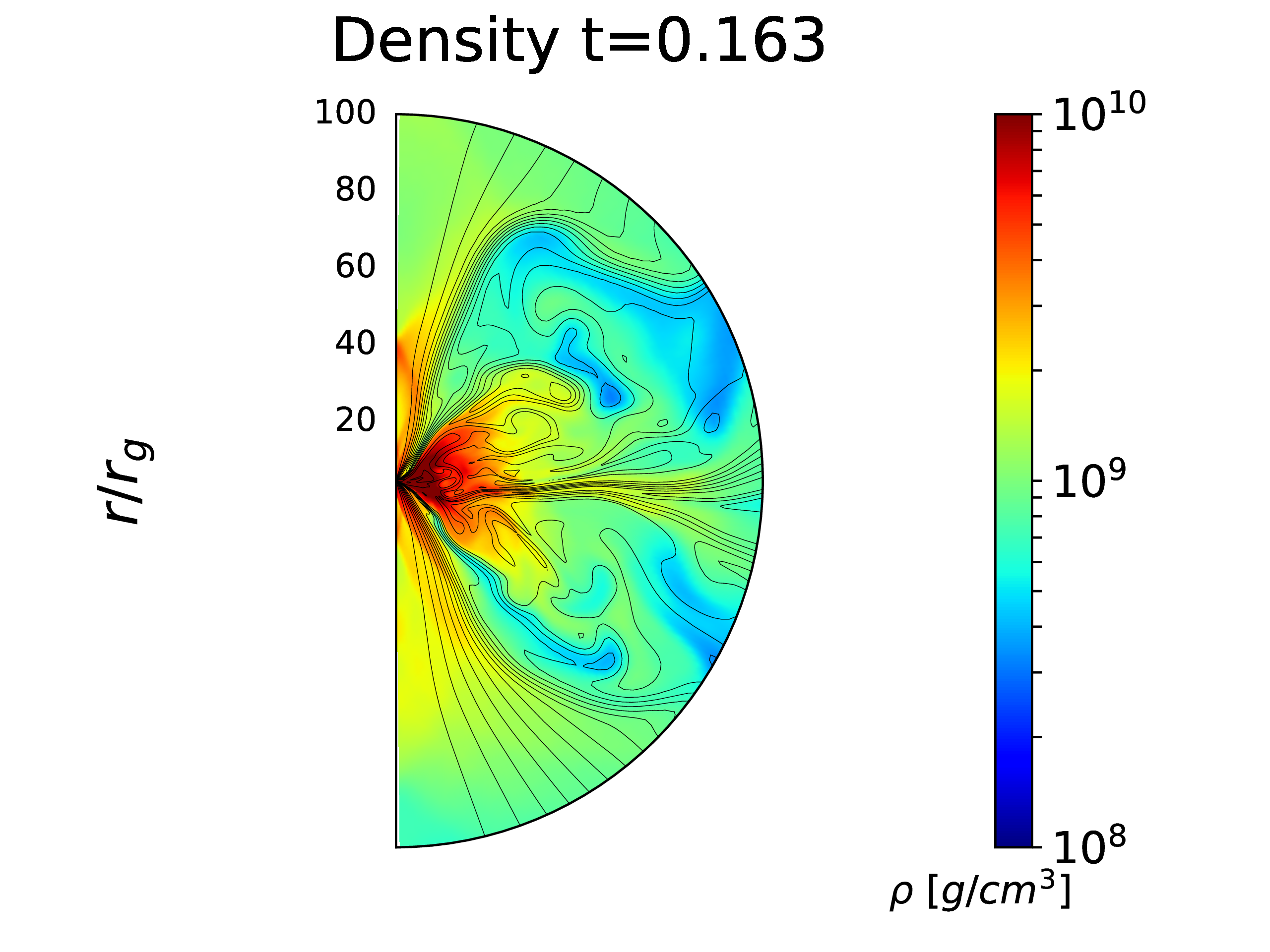}
 \includegraphics[width=0.33\textwidth]{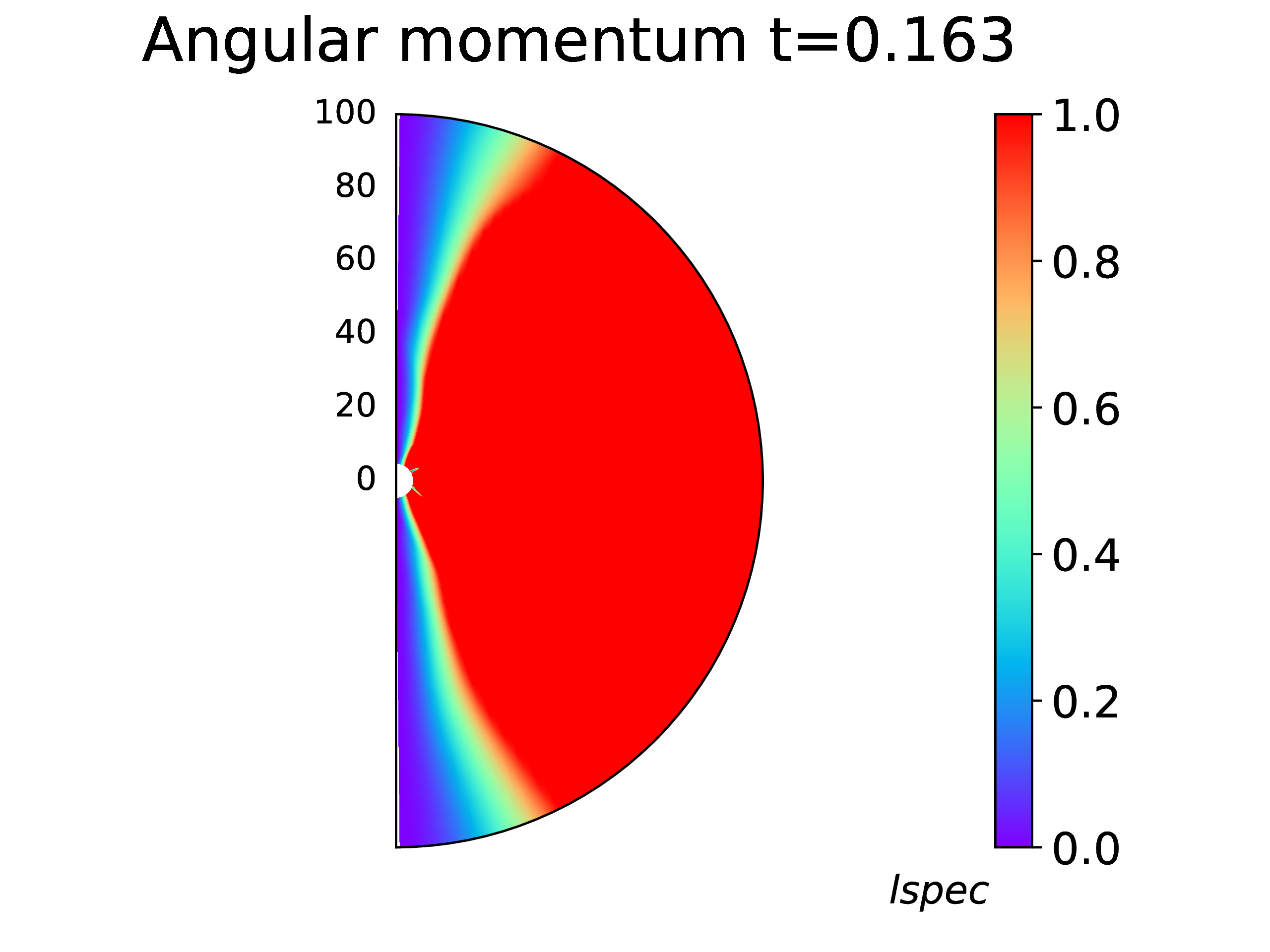} &
 \includegraphics[width=0.33\textwidth]{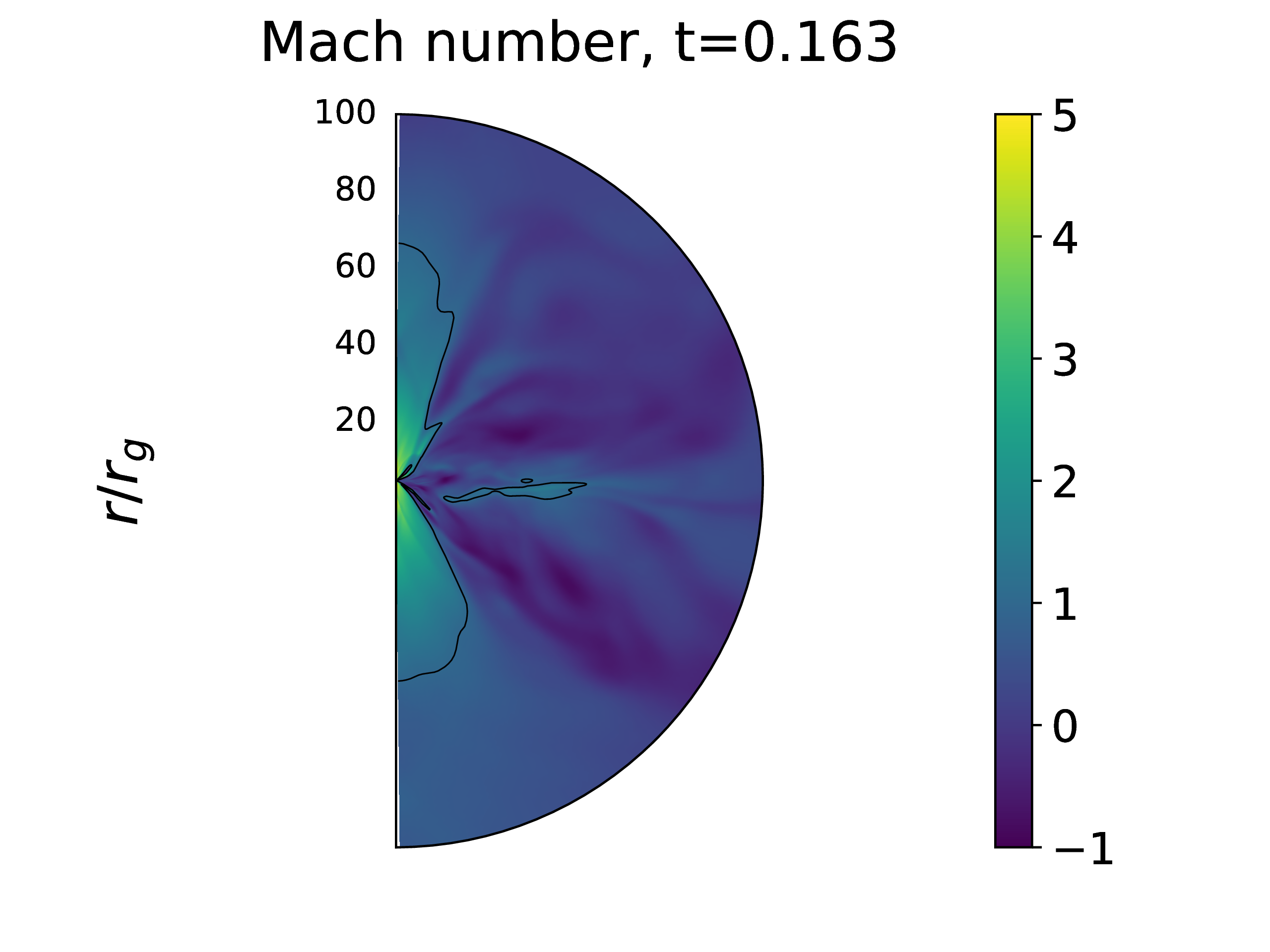} &
 \\
 \includegraphics[width=0.33\textwidth]{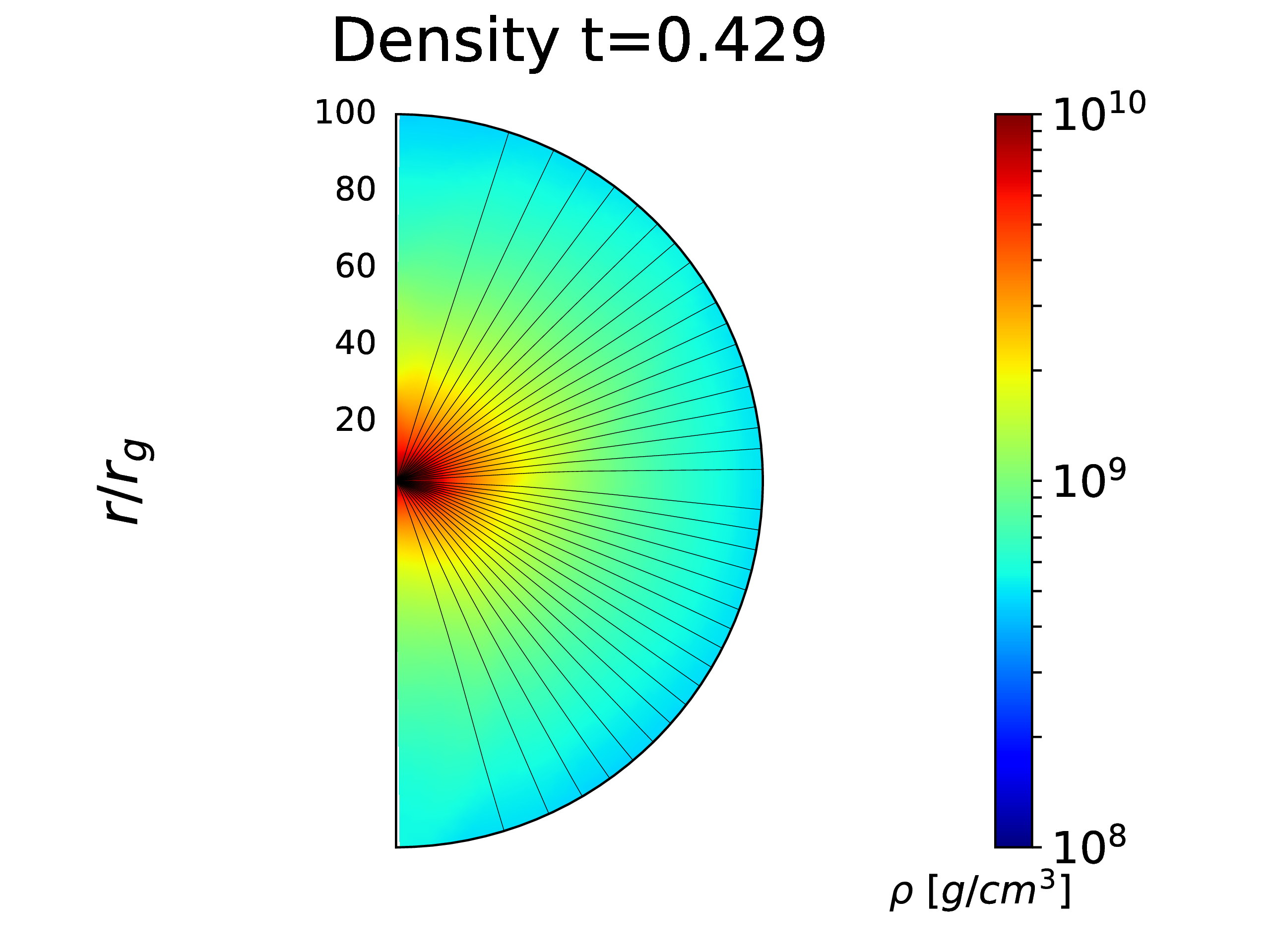}
 \includegraphics[width=0.33\textwidth]{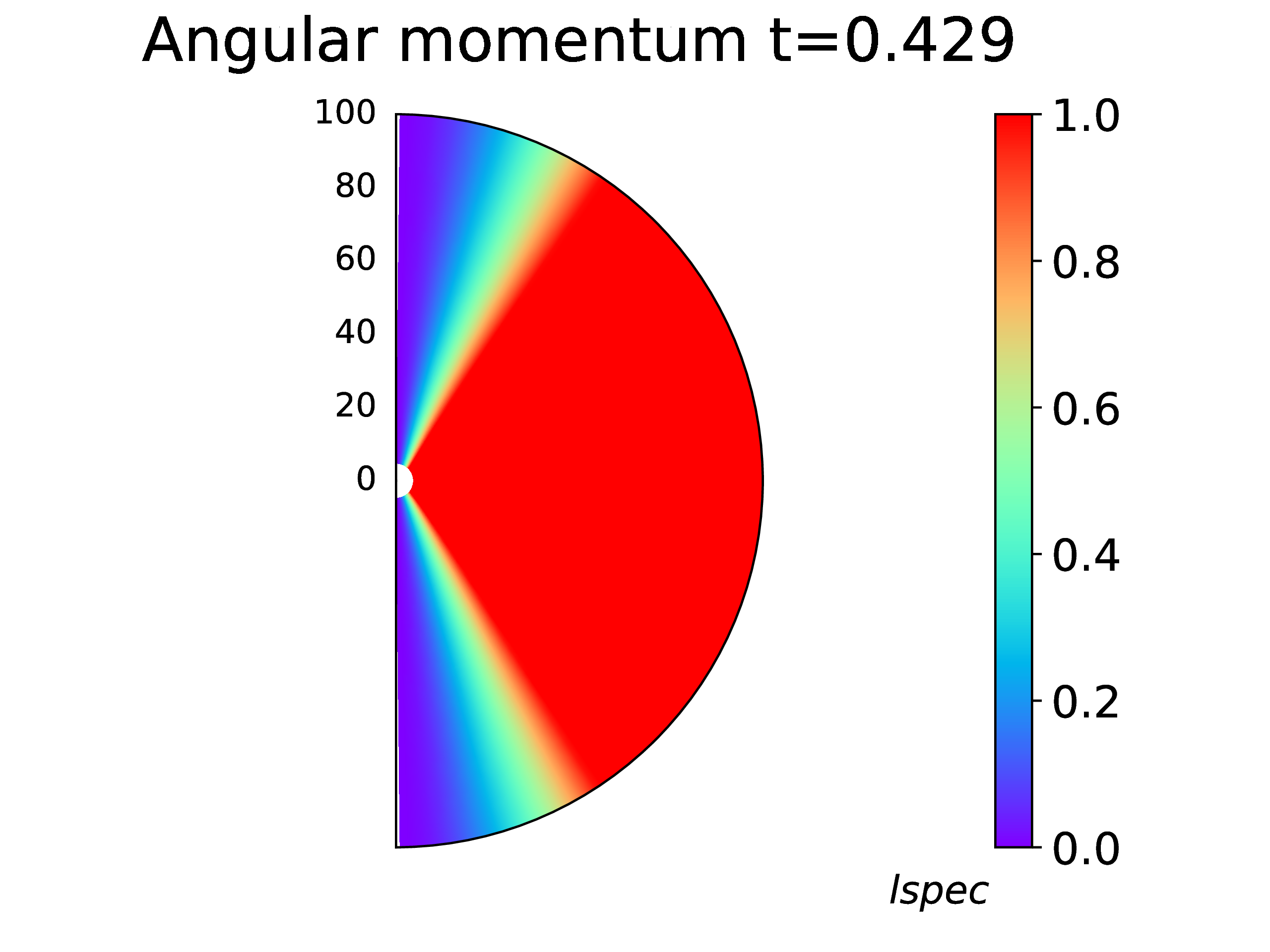} &
 \includegraphics[width=0.33\textwidth]{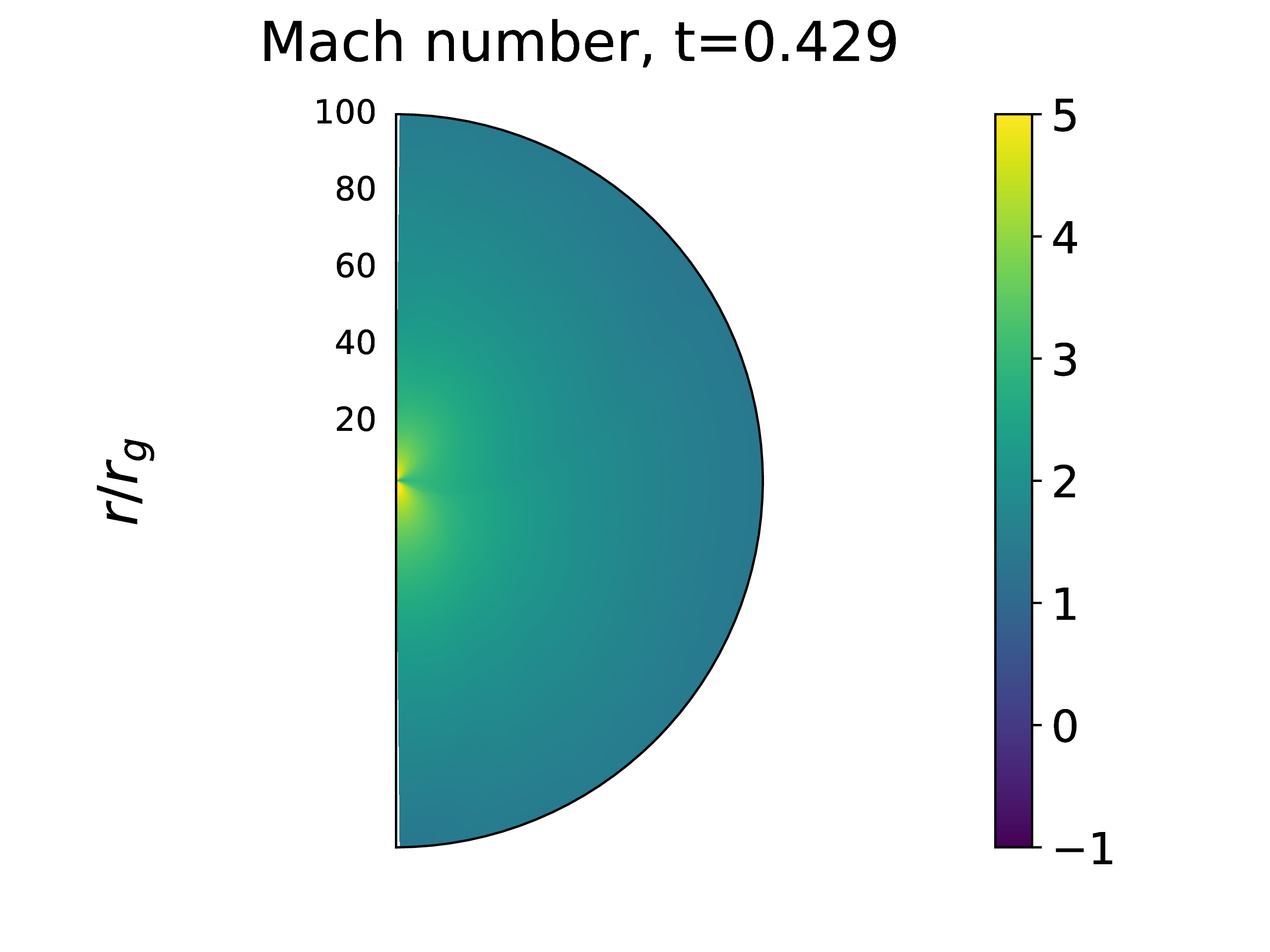} &
\end{tabular}
\caption{The results, from left to right, for the density with contours of $A_{\phi}$, specific angular momentum, and Mach Number
distributions, for the model with  magnetic field normalized by $\beta=1$, and initial black hole spin of $A_{0}=0.5$. The rotation parameter was $S=1.0$ (model LS-HM-10).
First row shows profiles at the beginning of the simulation, middle row profiles are taken at time $t=0.163$ \textrm{s}, which corresponds to time of oscillations of $\dot{M}$, and lower panel shows profiles after the end of the oscillations.
In addition, contour of $M=1$ is shown with a black line. 
  }
     \label{fig:A05S10Beta1_profile}
\end{figure*}

\begin{figure*}
\begin{tabular}{cccc}
 \includegraphics[width=0.33\textwidth]{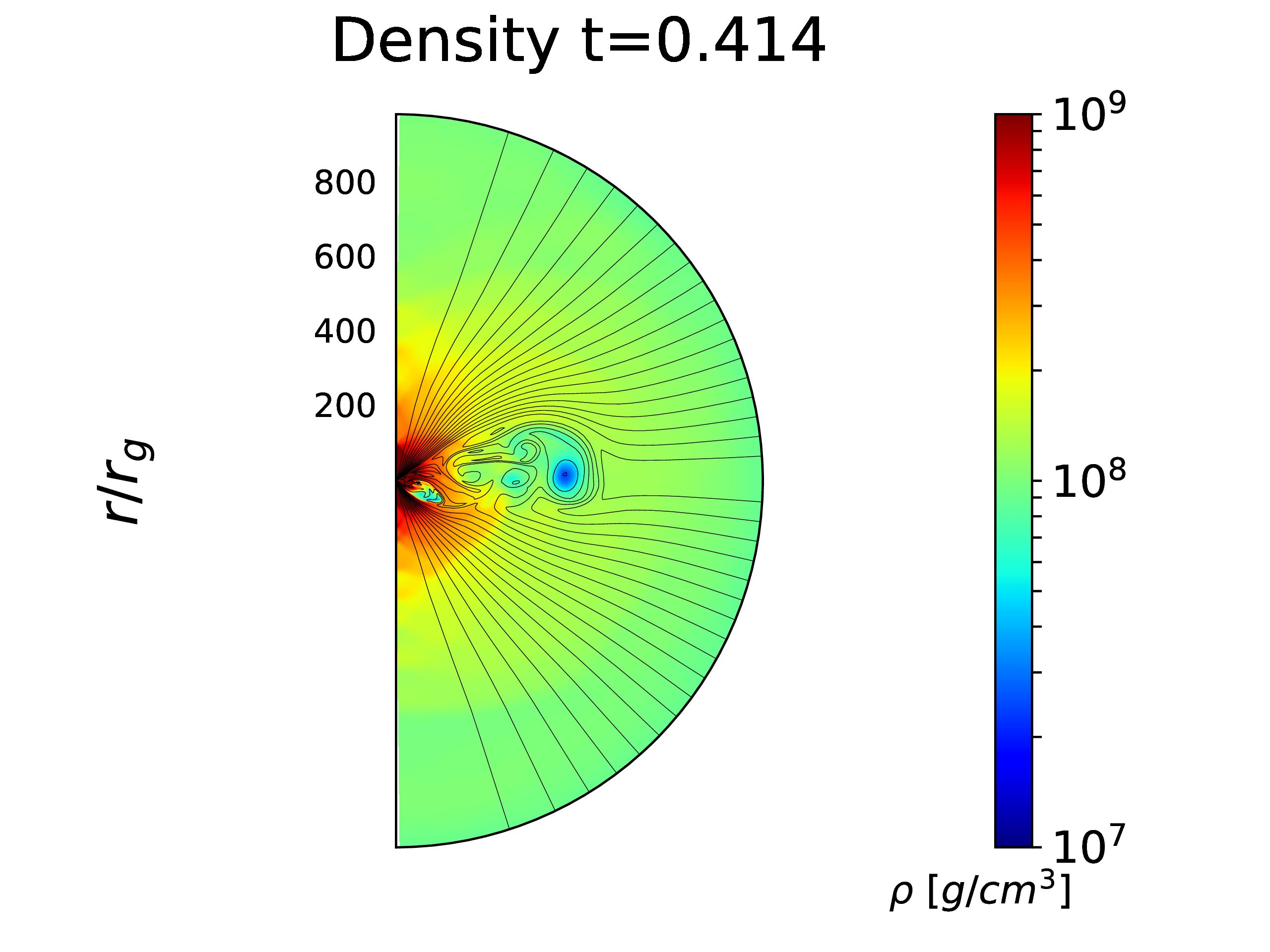}
 \includegraphics[width=0.33\textwidth]{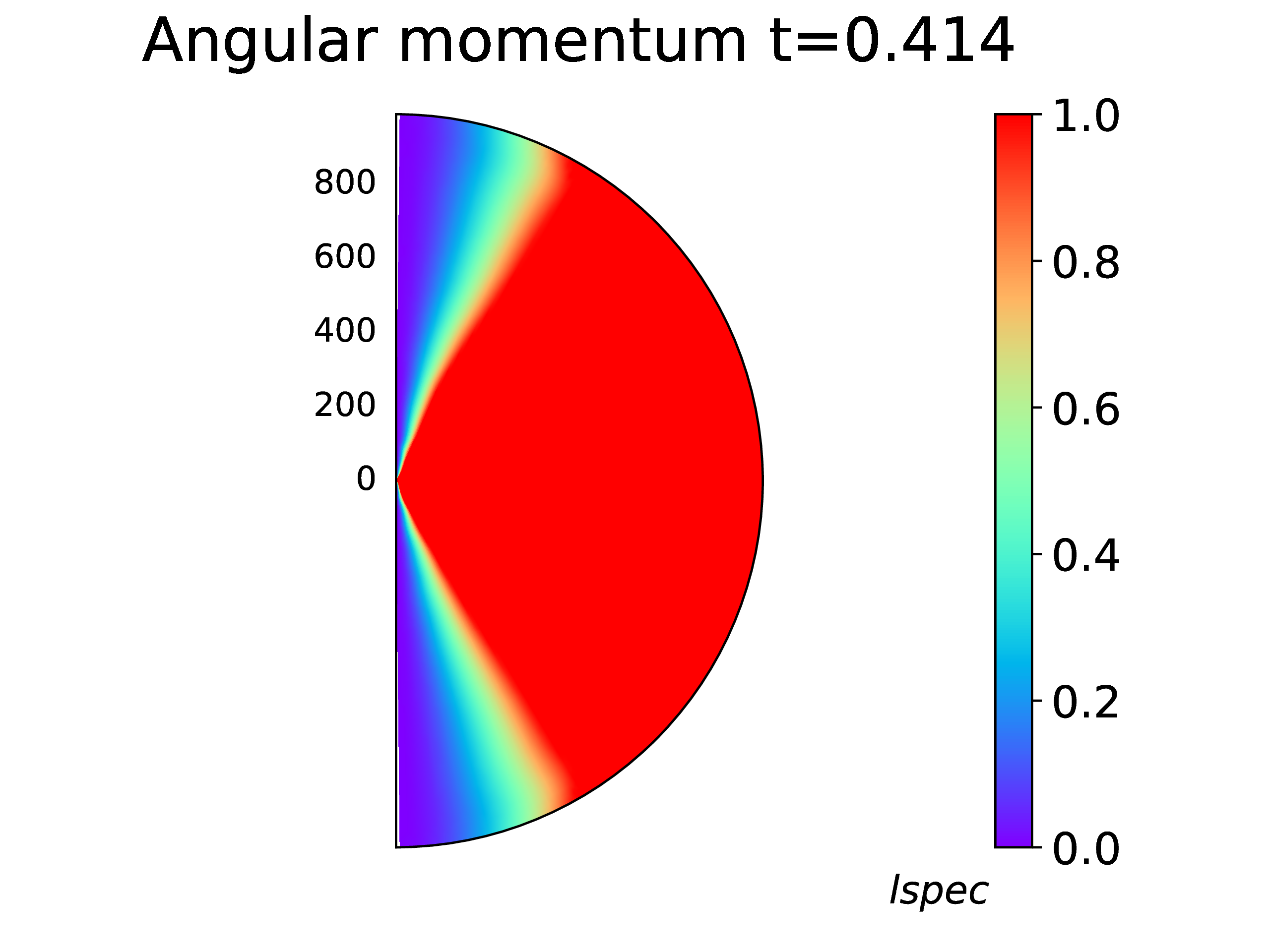} &
 \includegraphics[width=0.33\textwidth]{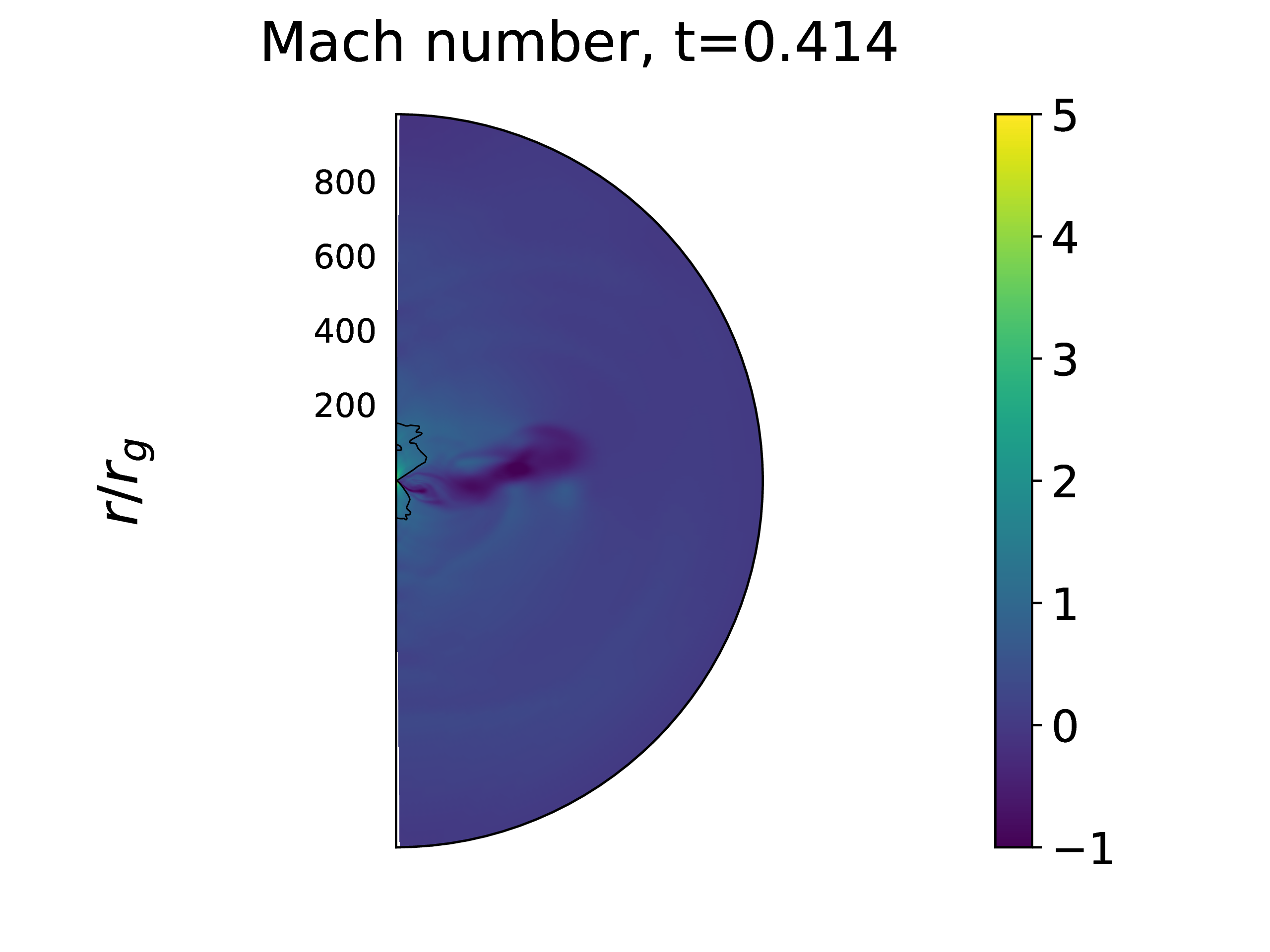} &
\\ 
 \includegraphics[width=0.33\textwidth]{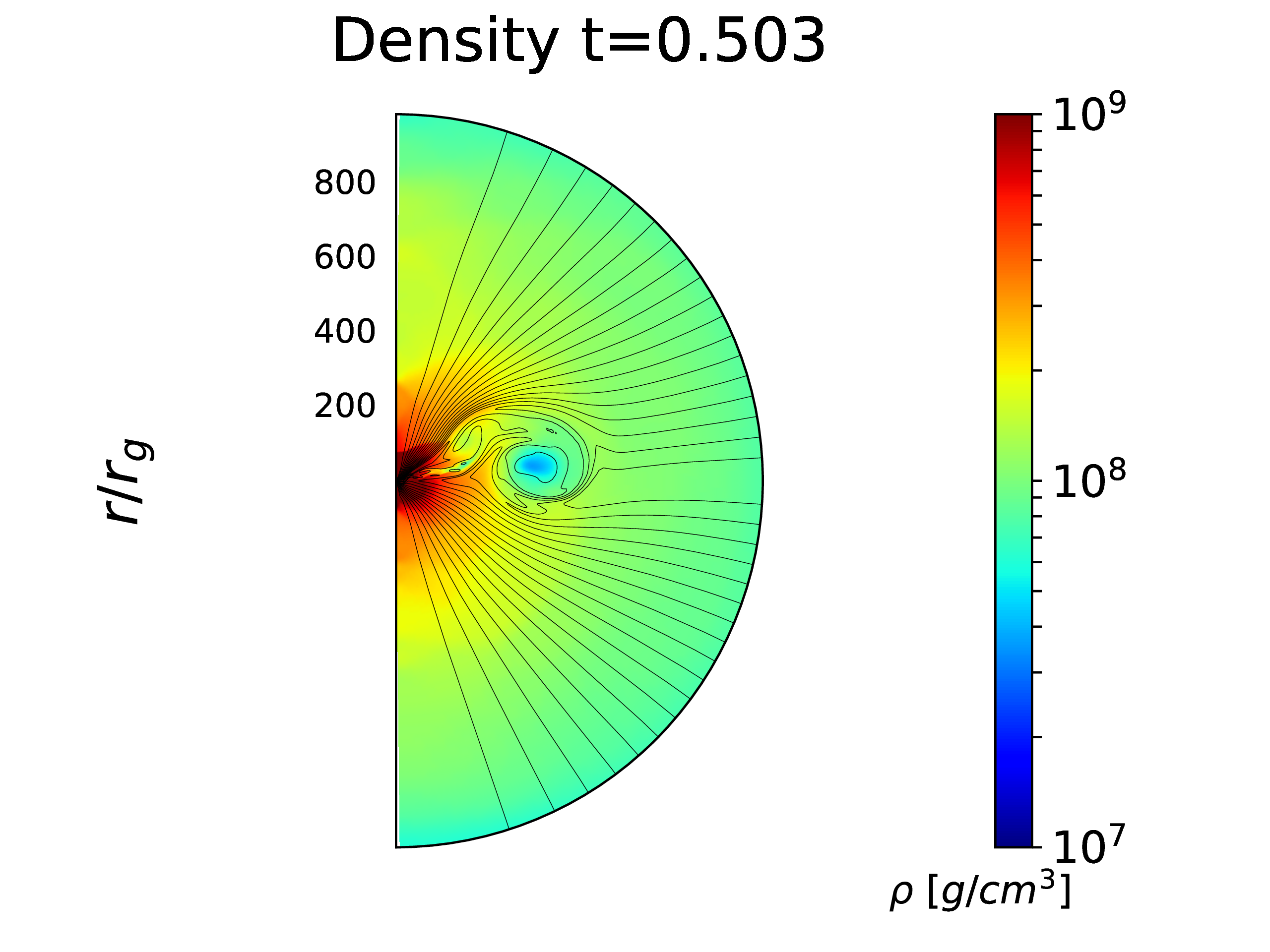}
 \includegraphics[width=0.33\textwidth]{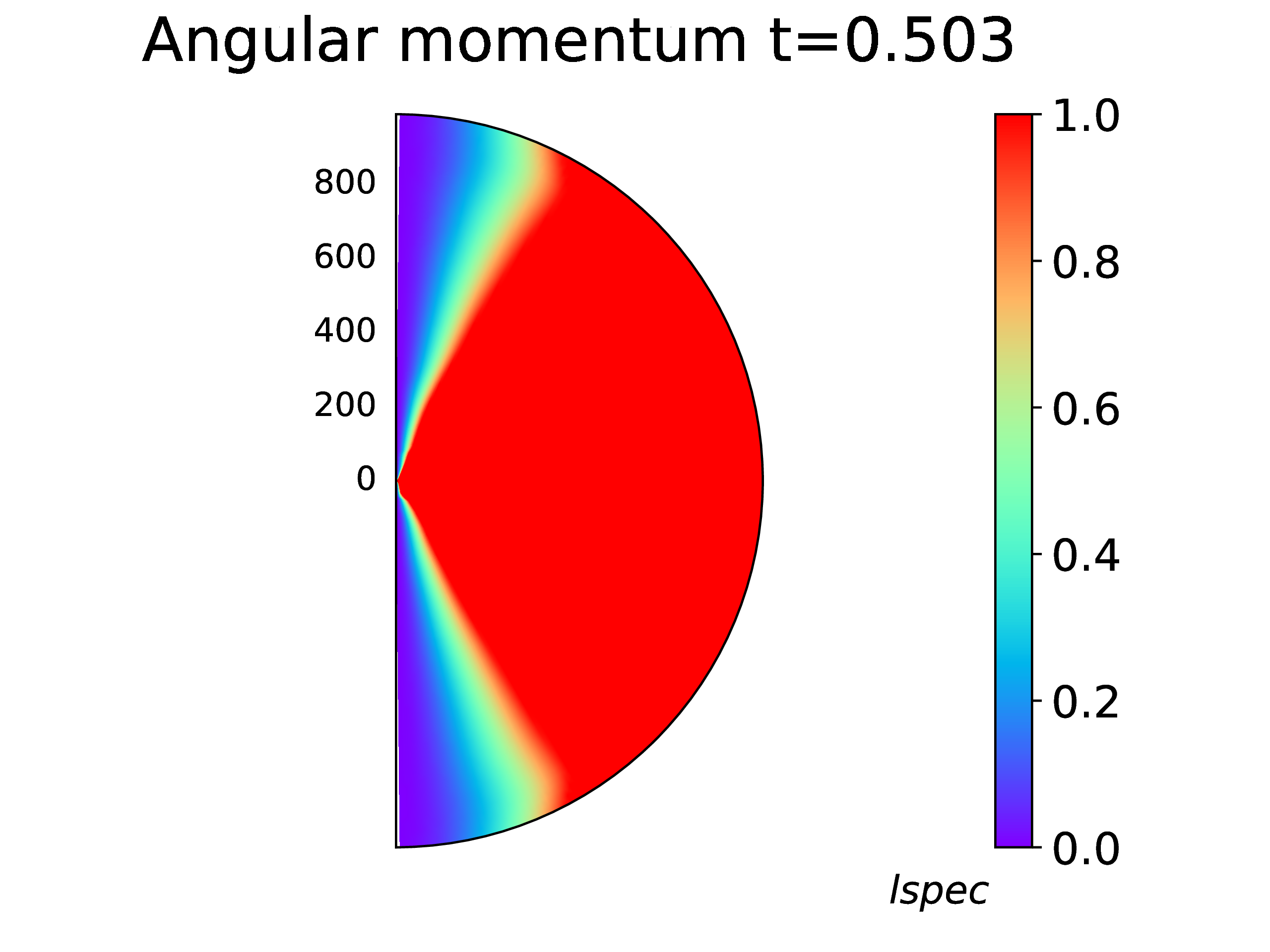} &
 \includegraphics[width=0.33\textwidth]{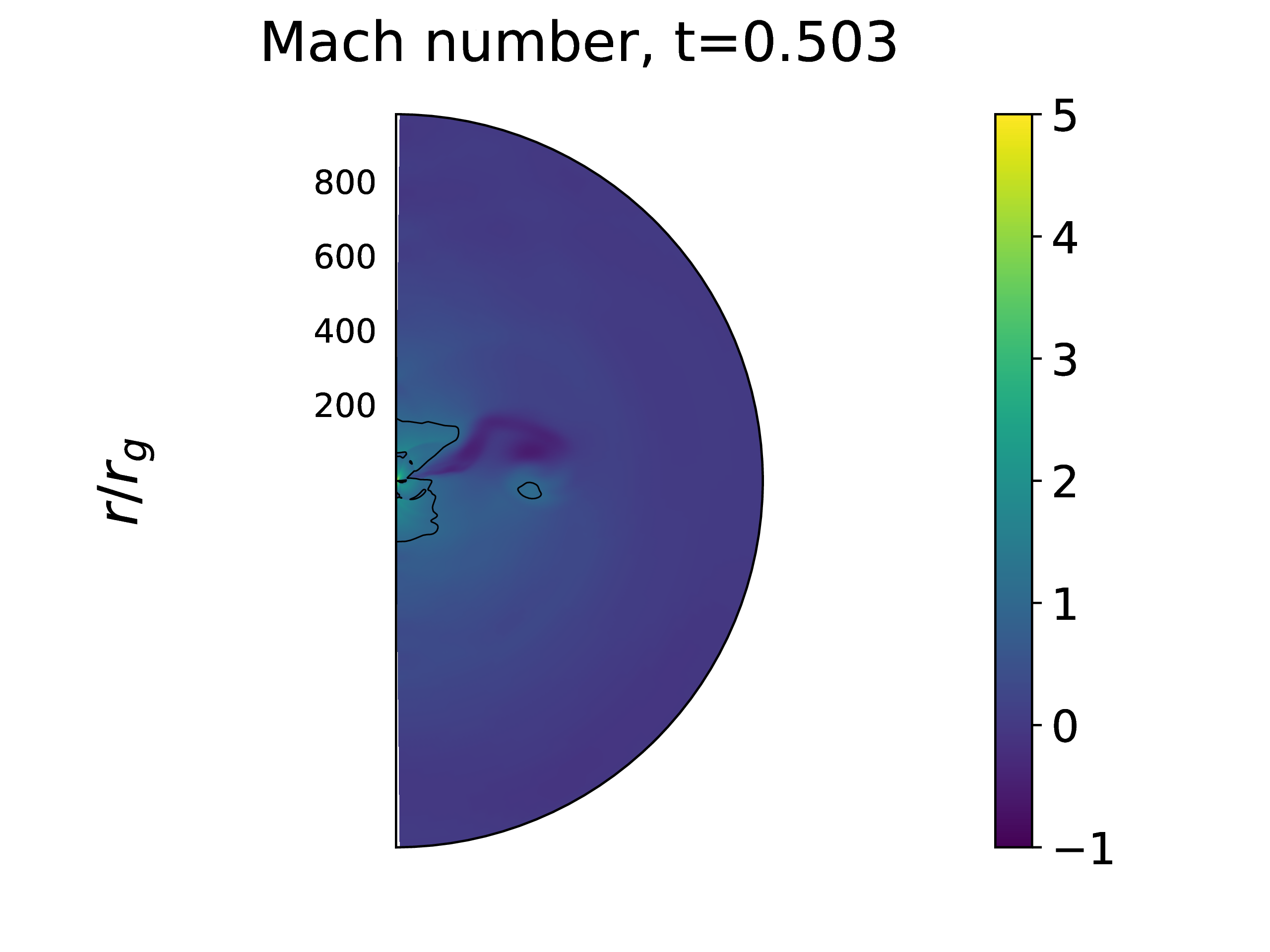} &
\\
\includegraphics[width=0.33\textwidth]{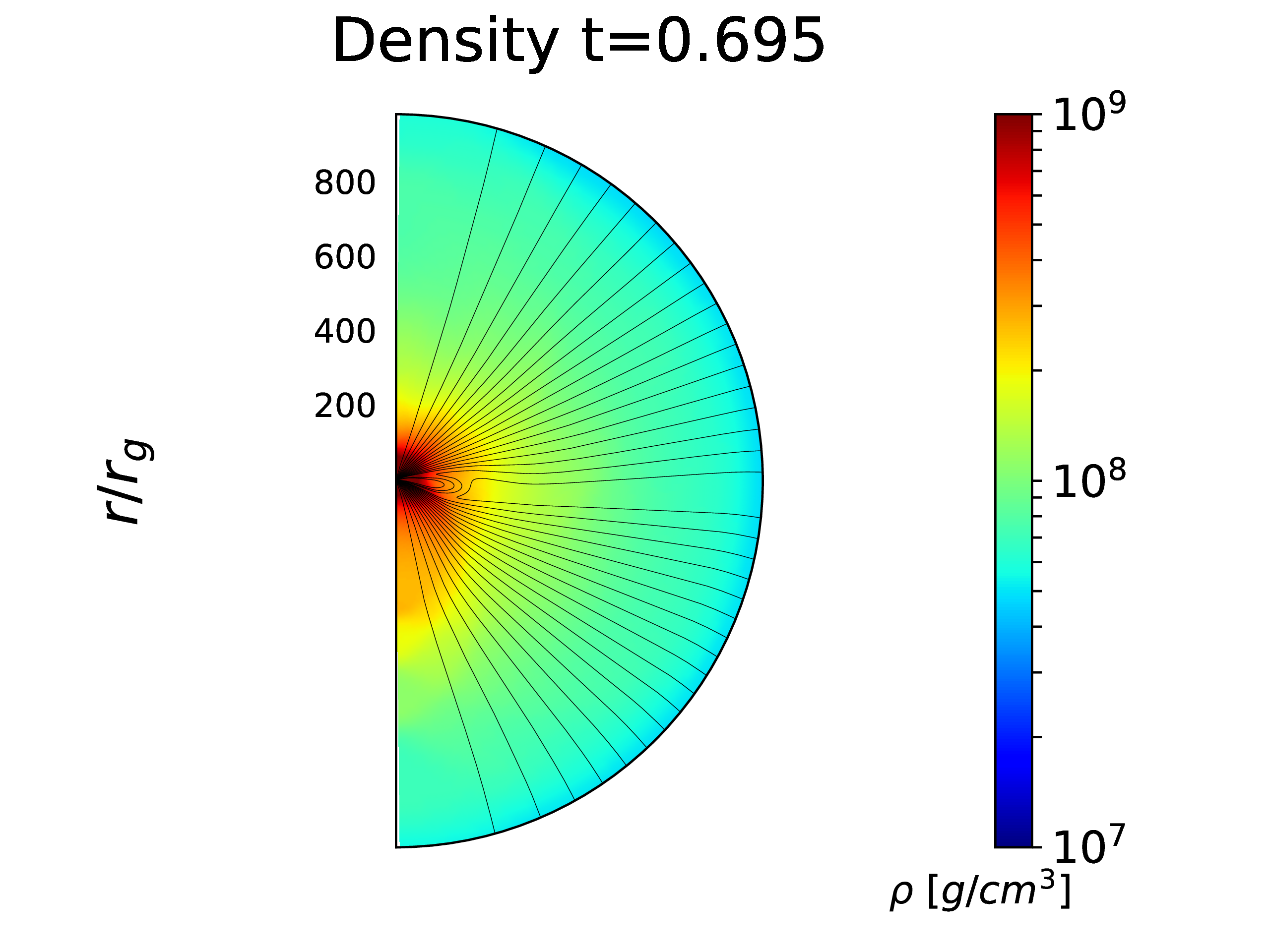}
 \includegraphics[width=0.33\textwidth]{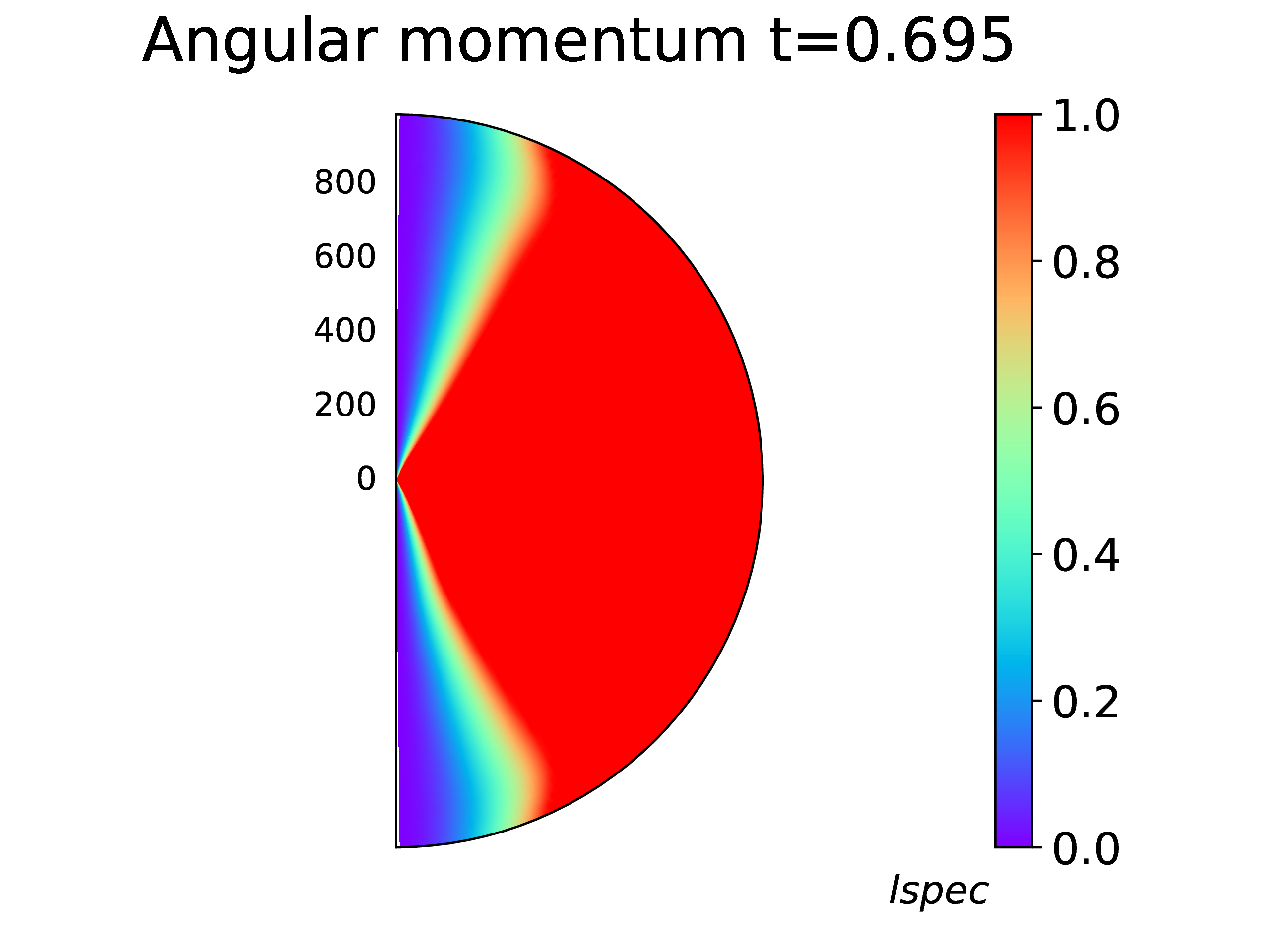} &
\includegraphics[width=0.33\textwidth]{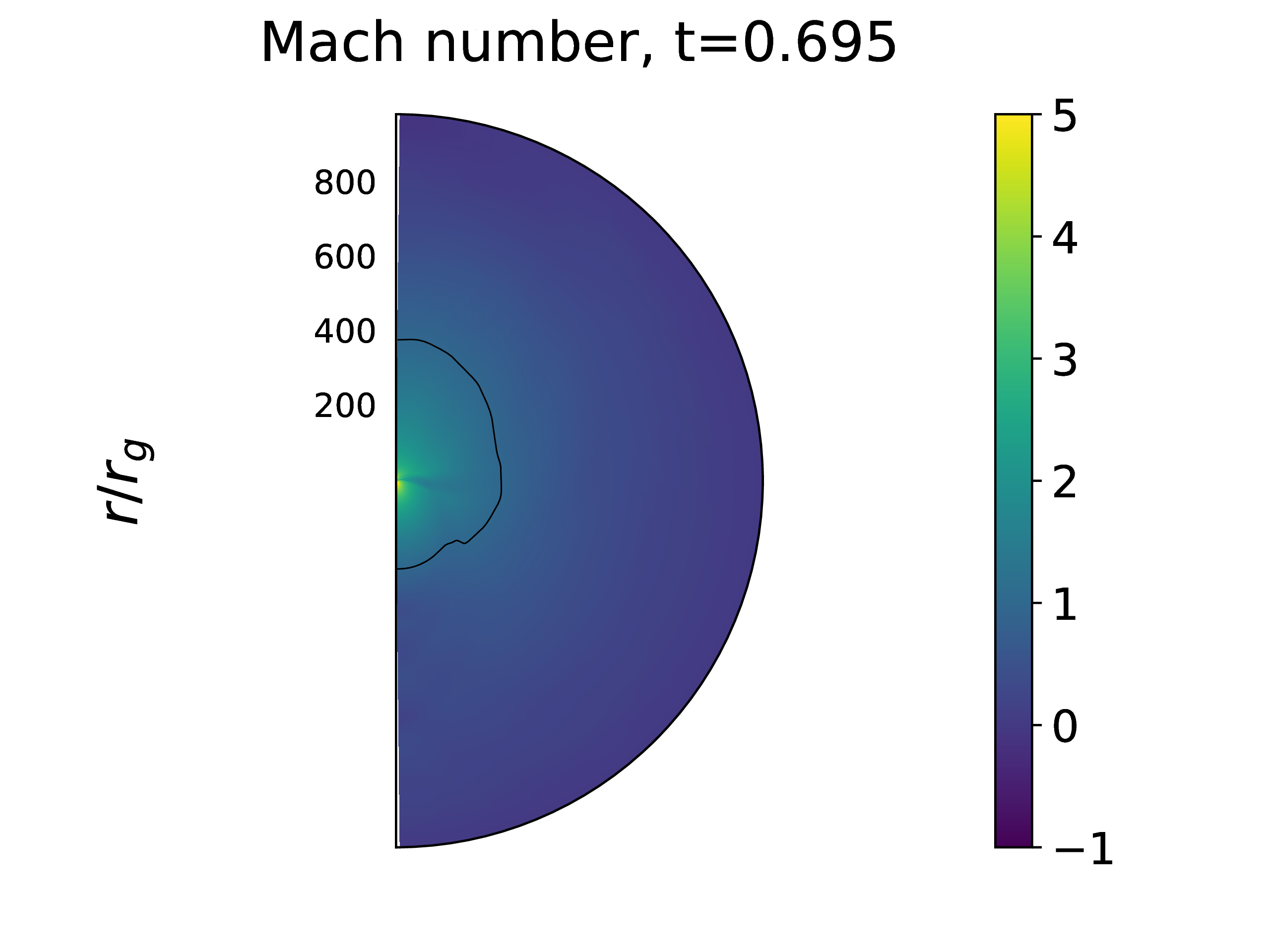} &
\end{tabular}
\caption{The results, from left to right, for the Density with overplotted contours of magnetic field $A_{\phi}$, Specific Angular Momentum, and Mach Number  distributions, for the model with  magnetic field characterized by $\beta=1$ and initial black hole spin of $A_{0}=0.85$. The rotation parameter was $S=1.4$ (model HS-HM-14).
  The color maps are taken at times $t=0.414$, $t=0.503$ and $t=0.695$ which corresponds to time before, in the time and after the spike. 
  In addition, contour of $M=1$ is shown with a black line.
  }
     \label{fig:A085S14Beta1_profile}
\end{figure*}

In Figures \ref{fig:A05S10Beta1_profile}, and \ref{fig:A085S14Beta1_profile} we show the magnetized models with $S=1.0$ and $S=1.4$. First, we present profiles for representative case: $A_0=0.5$ and $S=1.0$. The magnetic field structure, depicted by contours of constant $A_{\phi}$, is distracted at the time when oscillations of accretion rate value occur, while when they stop, the structure of magnetic field returns to its original shape (see Fig. \ref{fig:A05S10Beta1_profile}). 
As the Mach number profile shows, at the beginning of the simulations, the multiple sonic surfaces appear.  Density and angular momentum profiles also follow those structures.
Then, during the low amplitude oscillations, the multiple sonic surfaces  merge and form a single structure as is shown in the second row of Fig. \ref{fig:A05S10Beta1_profile}. The sonic point is very close to the black hole at the equator, while the shape of the sonic surface elongates with the latitude and gets disturbed, becoming an 'eight-shape'. The magnetic field lines close to the polar region are bent and tend to an open field lines configuration.  As mentioned above, this may suggest that flow in the collapsing star will try to launch a magnetically supported jet along the evacuated low density funnel. 
The angular momentum is the flow is however rather low everywhere, and only in localized places the turbulence allows the matter to rotate faster than critical speed.
Finally after oscillations it becomes again spherically symmetrical and moves towards the outer boundary. Density and angular momentum profiles are smoothed and as mentioned above the shape of $A_{\phi}$ contours turn back to the initial state. Those profiles are shown in the lower row of Fig. \ref{fig:A05S10Beta1_profile}.

Situation starts to change for critical initial rotation -- $S=1.0$.
In Figure \ref{fig:A03S04Beta1_profile} we present selected profiles of density, Mach number and specific angular momentum for simulations with $S=0.4$ and $A_{0}=0.3$. The flow is magnetized and magnetic field lines are overplotted on density maps. Here we show the models with lowest rotation parameter. For most of the simulation time, the density distribution is spherically symmetric, similarly to the $S=0.4$ case where the simulations were without magnetic field. The shape of the contours of $A_{\phi}$ stays radial and is unchanged during that time. The matter slowly accretes radially towards the black hole, and the density in the whole domain uniformly decreases, and the flow becomes supersonic, as can be seen by comparison between first and second snapshots.
The flow structure changes around the time $1.1 \textrm{s}$. Now the density profile flattens towards the equator and the shape of the $A_{\phi}$ contours gets disturbed. At the end of the simulation there is a structure created which is symmetrical about equator and manifests a pile-up of matter in the density profile, with turbulent magnetic field depicted by the contours of $A_{\phi}$. The radial Mach number profile shows that the flow here is highly subsonic, and the angular momentum increase means it is rotating down to the black hole horizon. 
The last row in this Figure shows the state of the simulation at time 1.19 s, zoomed in to the innermost part of the flow. Here we notice that the very thin region located at the equator is in fact rotating, while the material with small angular momentum and supersonic speeds can reach the black hole and disappear under the black hole horizon. At the same time, open magnetic field lines start forming, while they bend towards the polar axis. This evacuated region can be envisaged as the place where the jet is could be born in the collapsar's envelope. The amount of angular momentum in this model is however not sufficient to sustain the jet for a very long time.

As mentioned before, for initial parameters $A_0=0.85$ and $S=1.4$, the evolution of accretion rate is different from all the other models with $S=1.4$ and $S=1.0$.   In Figure \ref{fig:A085S14Beta1_profile} we show distribution of density with $A_{\phi}$ contours, angular momentum profiles and radial Mach number. The snapshots were taken before, around the time of, and after the highest spike in the accretion rate. It is visible that in this case density profile disturbances appear on much larger scales than in case of the models with sub-critical and critical initial rotation. Before the spike there are present disturbances in density profile with higher values  in central regions and $A_{\phi}$ contour structure close to equator plane with smaller density inside  Sonic surface before the spike is perturbed. Near the equatorial plane there is an area with lower $M_{r}$ values. What differs those profiles from previously presented is a scale of those  disturbances. For  $A_0=0.85, S=1.4$ model they are reaching $r\sim 500 r_g$, for models with critical initial rotation of the envelope, maximal radius which they are reaching was $\sim 200 r_g$. Spike in the $\dot{M}$ corresponds to accretion of the matter from the poles. After this spike, accretion rate stays constant for  around $0.1 \textrm{s}$. During that time reminding disturbances  in the density profile are  accreted, $A_{\phi}$ contours are getting back to the initial shape and  sonic surface is transforming to spherically symmetrical. At further stages of the simulation all profiles are getting smoothed and sonic surface moves towards outer boundary of the grid. In the case of the other simulations with $S=1.4$, a similar spike in $\dot{M}$ does not occur.  Evolution of the profiles is slightly different, in particular the $A_{\phi}$ structure is less stable and loses it's shape, but it also extends at its maximum to $r\sim 500 r_g$.

\subsection{Toroidal component of the magnetic field}
\label{sec:toroidal}

  We start simulations with $A_\phi$ being the only non-zero vector potential component, which means that the absence of $B_\phi$ component. However, the toroidal magnetic field component appears shortly after simulation starts.  We examined the $B_\phi$ distribution for characteristic models, presented in the previous Sections \ref{sec:beta1} and \ref{sec:beta100}. 
\begin{itemize}
    \item \textbf{Model NS-HM-04 ($\beta=1$, $A_0=0.3$, and $S=0.4$).} For most of the evolution the flow is spherically symmetric with the strongest toroidal magnetic field developing at the central part of the flow close to the black hole.
Around time $t=0.3\textrm{s}$, a sphere of weaker magnetic field starts to move inwards. 
Around time $t=0.7 \textrm{s}$ it stops around  $r=700 r_g$. This feature maintains its position until $t=1 \textrm{s}$. Then it continues to move inwards and at time $t=1.197\textrm{s}$ the flow loses its spherical symmetry, for the toroidal field as well as other  quantities.
    \item \textbf{ Model LS-HM-10 ($\beta=1$ $A_0=0.5$ and $S=1.0$).} Shortly after beginning of the simulation, the toroidal component of the magnetic field appears. It follows similar small scale structures as in density profile presented in Fig. \ref{fig:A05S10Beta1_profile}. The structure relaxes around $t=0.4 \textrm{s}$.  From this moment the flow evolution is spherically symmetric. Around $t=0.5 \textrm{s}$ a sphere with lower value of $B_{\phi}$ starts to accrete inwards. It reaches $r=400 r_g$ at $t=1.449 \textrm{s}$ and then the flow loses its symmetry. 
    \item  \textbf{Model HS-HM-14 ($\beta=1$ $A_0=0.85$ and $S=1.4$).} Toroidal component of the magnetic filed which appears shortly after beginning of the simulation has similar shape to the profile of density seen in Fig. \ref{fig:A085S14Beta1_profile}. The low density regions near the equator correspond to high toroidal magnetic field. This kind of a structure is maintained until $t\sim 0.7\textrm{s}$. Then he flow recovers its spherical symmetry which is retained until $t\sim 1.4 \textrm{s} $, and a sphere with lower magnetic field strength is moving inwards. Then the symmetry is broken. 
    \item \textbf{Model NS-LM-14 ($\beta=100$ $A_0=0.3$ and $S=1.4$).} In this simulation with weaker magnetic field we observe disturbances in the innermost part of the flow (below $r\sim 100 r_g$) at the beginning of the simulation, when the high density regions are more highly magnetized. In this case however, the matter distribution does not follow the $B_{\phi}$ distribution. Around $t=0.488 \textrm{s}$ the flow retains spherical symmetry. At $t \sim 0.7 \textrm{s}$ a sphere with weaker magnetic field starts moving inwards, and then at $t\sim 0.95 \textrm{s}$ this sphere stops at $r=750 r_g$  and remains there until the end of the simulation.
\end{itemize}

As noted above, the qualitative evolution of toroidal magnetic field component depends strongly on initial parameters of the model.
 Our results do not depend significantly on the adopted grid resolution.
  We have run additional simulation with twice bigger resolution in both $r$ and $\theta$ directions, namely 512x512 points, and compare results quantitatively.
  We repeated simulations with $\beta=100$, $A_0=0.3$ and $S=1.0$  with this higher grid resolution. The comparison of the evolution of accretion rate, spin and black hole mass shows, that during the first $\sim 0.6$ seconds of the collapsar evolution, when the accretion rate increases, the black hole mass and spin increase is exactly the same. Both runs resulted in the same maximum spin value.
 The mean relative difference of black hole mass, accretion rate and spin throughout the simulation is about $0.002\%$. Final black hole mass difference is $0.8\%$ and is higher for higher resolution.
There were no big differences in the profiles of density, radial Mach number and angular momentum, either, as checked at the characteristic moments of the collapsar evolution.

\section{Discussion}
\label{sec:diss}
In this Section, we discuss the main assumptions, results, implications, and limitations of our models.
Our simulations are divided into two sets with the \textit{Magn} class of models referring to a moderately magnetized accreting cloud, and the \textit{Therm} class where the cloud magnetization is neglected.
Table~\ref{tab:in} presents summary of the models and their initial parameters, namely the black hole spin and rotation of the envelope, and the final black hole mass and spin at the end of simulation. Also, the maximum accretion rate reached during the simulation is mentioned in Table~\ref{tab:in}.

 We note that the normalization of rotation in the envelope is not done fully independently from the black hole spin value at the initial state. In reality these two quantities should depend on each other. The angular momentum adopted in the simulation is assumed to be a fraction ($S$) of the momentum at the last stable circular orbit. The radius of this orbit depends on the spin value, therefore the models with the same $S$ parameter and different $A_0$ value have different magnitudes of angular momentum in the envelope. 
However, the picture is more complex and our models show a non-linear behavior of black hole spin and accretion rate evolution, depending on the initial setup.

The Alfven velocity and the sound speed in the flow are in general related through through $\sqrt{\beta}=\frac{c_{s}}{v_A}$. Initial conditions in our simulations prescribe the magnetic field potential as a function of $\theta$ only, and the field strength is  normalized with $\beta$ parameter (Eq. \ref{eq:mag}). The sound speed is  given by Eq. \ref{eq:cs} with thermodynamical parameters resulting from the Bondi solution and depends on $r$. We checked that at the beginning of the simulation the ratio of $\frac{c_{s}}{v_A}$ calculated for both magnetic field normalizations gives $\frac{\beta_{100}}{\beta_{1}}=10$ in every point of the grid, to confirm the viability of embedding the transonic Bondi accretion flow in the poloidal magnetic field normalized to the gas pressure.

  There are no specific outflow / jets signatures in the magnetized models. A structure with dominating open magnetic field lines can be seen in the zoom-in view  up to r=100 $r_{g}$ of the model LS-HM-10. This transient feature is accompanied with the existence of magnetized, turbulent torus of high density, located in the equatorial plane.
  Nevertheless, the polar regions are also having large density, and do not resemble evacuated, low-density funnels present in other type of simulations (e.g. \citet{Sap2019ApJ}). These authors started their models with a magnetized accreting torus embedded in a quasi vacuum. In contrast, here we start our models with an accreting dense spherical cloud, so the jet break through such a  dense envelope would be very difficult. We can speculate that if the torus structure existed for a longer time, the evacuation of polar regions on free-fall timescale would eventually provide a clear structure for a long gamma ray burst engine with a jet. But it was not found in the currently presented simulations.

 Our model neglects self-gravity effects of the accreting cloud, and assumes that the dominant gravitational potential is enforced by the central black hole, whose point mass and spin is changing.
    The secondary effects and metric perturbations due to the envelope self gravity might be however important, especially ate the beginning of the collapse.
    To asses this effect quantitatively, we made an exemplary test calculation, limited to the non-magnetized accretion (we adopted initially non-spinning black hole, and critical rotation of the envelope).
We found that initially, for the first $\sim 3000 t_{g}$, the simulation results do not differ. 
Then the main change in the evolution of black hole mass and spin is starting about time of 5000 M, when both mass and spin increase much faster than in the simulation neglecting self-gravity. 
Finally, at time $\sim 30000 t_{g}$, the models merge again, and ultimately, the final mass and spin of black hole are the same as found before.
We note, that this test simulation still neglected higher-order terms in metric evolution, i.e. connection coefficients and derivatives. The precise investigation of self-gravity effects is postponed to a separate future work.

 In our simulations we use the initial condition of the relativistic Bondi solution, for the radial components, and on top of that we add a net angular momentum, centered on the equatorial plane.
    This initial condition describes a quasi-spherical accretion onto BH, and initially the assumption is made that the stellar envelope was in a steady state. 
However, because we do not add mass to the envelope during the time evolution, the accretion rate is dynamically changing. Moreover, by adding rotation to the flow, we relax the steady-state assumption, and as a result the mini-disk forms.
We use the adiabatic index of $\gamma=4/3$, that describes the microphysics in the stellar interior, including large contribution from the photon pressure, rather than the isothermal index of $\gamma=1$ which would be describing the proto-stellar disks.
Also, as argued above, the dynamically changing Kerr background metric sufficiently describes the evolution of gravitational potential due to the increase of BH mass and its spin changes. This is because the gravitational potential of the black hole dominates the spacetime, while the self-gravity of the mini-disk is rather a small effect. It could be treated as a small perturbation, but this effect is currently beyond the scope of the present work.
  
  The magnetic field assumed in our models is either normalized to $\beta=100$ which means thermally dominated accretion flows, or to $\beta=1$, which means actually quite strong magnetization and equilibrium ratio of the magnetic and gas pressures. 
The gas pressure is set by the solution of the Bondi transonic accretion flow, which is our initial condition for the simulation. If only the magnetic pressure does not add a high contribution to this flow structure, we can safely assume that the solution holds, and the gas falls supersonically to the black hole downstream of the sonic point, ultimately reaching the speed of light at the black hole horizon. We perturb this flow in time, by adding a slow rotation to the gas, hence our sonic point is moving, and the shape of the sonic surface is disturbed.  
However, when the magnetic pressure is substantially large, the initial solution for the transonic accretion may not be a good approximation. Already at the beginning, multiple critical points might exist in the flow. The possible existence of shocks in low angular momentum flows connected with
the presence of multiple critical points in the phase space has been studied from
different points of view during the last thirty years. However, the theoretical works which describe the fundamental properties of the low angular momentum accretion and which usually treat the steady solution of the equations, has so far been
carried only for non-magnetized and usually also non-viscous flows
\citep{Abramowicz1981, Abramowicz1990, Das2002, Das2012}.
Hydrodynamical models
of the low angular momentum accretion flows have been studied already in two
and three dimensions, e.g. by \citet{Proga2003a, Janiuk2008ApJ...681...58J, Janiuk2009}. 
In those simulations, a single, constant value of the specific
angular momentum was assumed, while the variability of the flows occurred due
to e.g. non-spherical or non-axisymmetric distribution of the matter. The level of
this variability was also dependent on the adiabatic index (see also \citet{Palit2019}.
However, these studies considered non-magnetized flow only, which limits the
applicability of these models to observational data, and to the stellar  evolution and pre-supernova studies 
\citep{2005Heger}.
The range of strength of the magnetic field varies from very
low values in interstellar medium to extremely high values $\sim 10^{15}$ G in the magnetars, which are extremely magnetized neutron stars. During
the accretion process, the magnetic field in the star is likely to be amplified in the innermost region of the black hole, so negligible amount of magnetic field in those regions, where the shock and the sonic point could be located, is thus not anticipated. The only method however, to describe the effect of strong magnetic fields in the context of transonic accretion with low angular momentum, is via numerical simulations, because no analytic solution of this problem exists, to the best of our knowledge. Hence, setting up the initial condition with strong magnetic fields overimposed on the slowly rotating transonic Bondi flow is a simple but working approach that we make here.

  The physical conditions in the flow, namely its density distribution and temperature are governed by the global parameters, such as the mass of the accreting cloud and accretion rate, black hole mass and spin, as well as the sonic radius,
  and the magnetic field normalization. The density normalized to the mass scale of the cloud enclosed in our computational domain is chosen to match the densest parts of the accretion torus, as can be found in the gamma ray burst central engines ($\rho \lesssim 5 \times 10^{10}$g cm$^{-3}$). Also, the chosen value of the sonic radius sets the temperature of the flow.
  At the beginning of the simulations, we have the temperatures range  above $3.76 \times 10^{10}\textrm{K}$, consistently with the physical conditions in the neutrino-cooled torus formed inside the collapsar, as modeled e.g. by \citep{Lopez2009}.
  We note that presupernovea models from \cite{2005Heger} and \cite{Heger00} show values  of the temperatures in the inner parts of the star around $8 \times 10^{9}\textrm{K}$, while the densities of these pre-supernova stars are $\rho \sim 8\times 10^{9}$ g cm$^{-3}$.

The position of the shock front and its expansion velocity may serve to estimate the energy dissipation rate in the shocks.
In case of simulations without magnetic field we detect four shock fronts which move outwards and we can determine their velocities. Two shock fronts were found in the model with $A_0=0.5$ $S=1.0$ and their velocities are $0.014c$ and $0.022c$. For the models with sub critical envelope rotation we obtained higher velocities of the shock fronts: $0.04c$ and $0.044c$ for $A_0=0.5$ and $A_0=0.85$, respectively.This trend seems opposite in comparison to \citet{Murguia2020}, where the shock front velocities were growing for the increasing angular momentum in the envelope. However, their simulations had always higher angular momentum assumed (corresponding to $S=1.5$, $S=1.75$ and $S=2.0$). Moreover, the fundamental difference in the physics of the code was that the which mass and spin of the black hole, hence Kerr metric coefficients, did not change. Such an approach is justified if only the black hole mass is much larger than the adopted envelope's mass, and if the simulation timescale is short.
In our model, such an assumption is no longer valid. We note that the shock velocities obtained here are on average one order of magnitude smaller than the escape velocities corresponding to positions of the shocks, which means that they cannot significantly halt the collapse.

  Interestingly, we find that the evolution pattern in most of models is similar, albeit one particular initial parameter is crucial and affects the final outcome. For the non magnetized envelopes, the evolution mostly depends on stellar rotation rate ($S$ value).
In particular, the large amplitude oscillations of the accretion rate appear only for super-critical rotation $S=1.4$. These models also result in the smallest final black hole masses, and largest spins, regardless of the initial values of the spin.

In the case of magnetized envelope, things change and evolution is not only determined by the rotation rate. 
For sub-critical rotation, the magnetic fields seem to not affect much the collapsars evolution. 
The distributions of density and Mach number are similar to the respectively modeled non-magnetized cases.
The accretion rate oscillations appear however in both critical and super-critical rotation models, in $beta=1$ case, while the amplitude of the peaks in lower than in thermally-dominated models without magnetic fields.
In contrast, in the case of $\beta=100$ the accretion rate in  simulations with $S=1.0$ behaves more smoothly, and is very similar to the sub-critical case. This is due to 
mostly almost spherically symmetric distribution of density and only slight flattening when a mini-disk forms at the end of collapse.

The grid resolution of our simulations is not enough to probe the magneto-rotational instability in great detail. Nevertheless, several results reveal the important coupling between the magnetic fields and rotation of the collapsing matter.
For the super-critical rotation, the magnetic fields affect the collapse process in strongest way.
In the $S=1.4$ models, the spikes and oscillations at accretion rate are present around $0.4 \textrm{s}$ for every initial spin value, $A_0$.

Among all the magnetized models, one specific combination of $A_0=0.85$ and $S=1.4$ presents a somewhat different evolution pattern. For $\beta=100$, the model results in the maximum spin of the black hole, which reaches $A=1$ temporarily (however it drops to $A\sim 0.65$ at the end of simulation). 
In other words, the rotation of the envelope influences the collapse depending on the specific $\beta$ value.

\begin{figure*}
	\begin{tabular}{cccc}
		\includegraphics[width=0.30\textwidth]{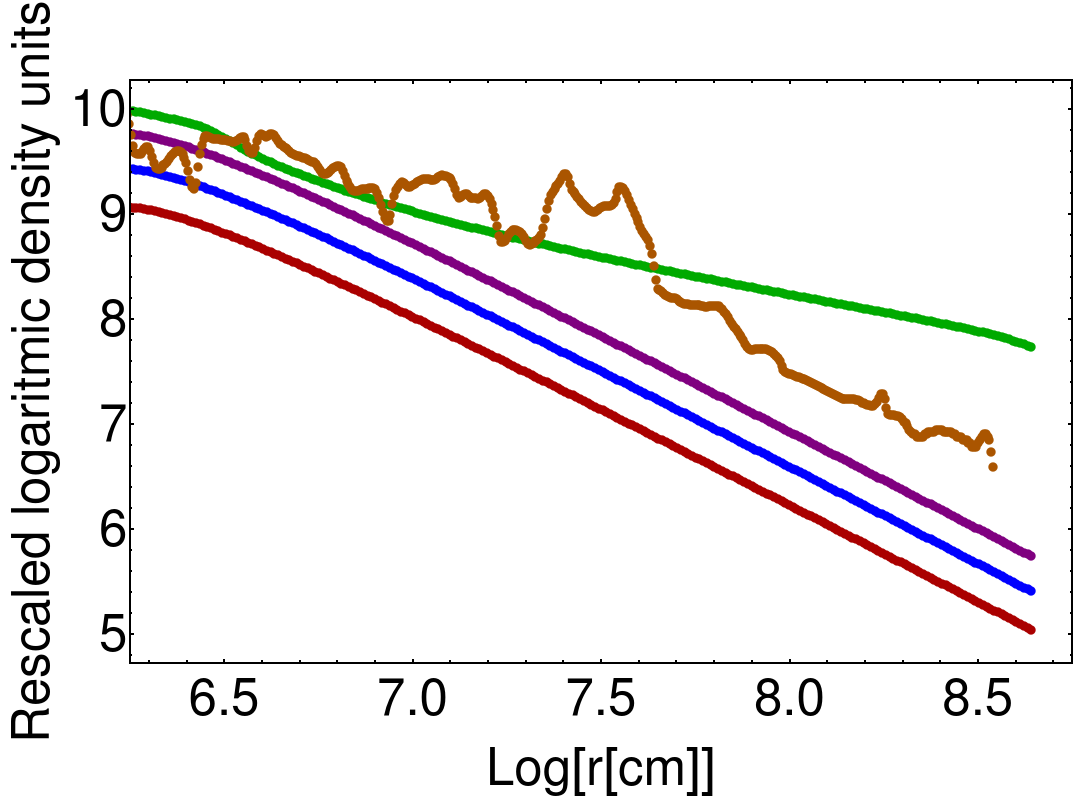}
		\includegraphics[width=0.30\textwidth]{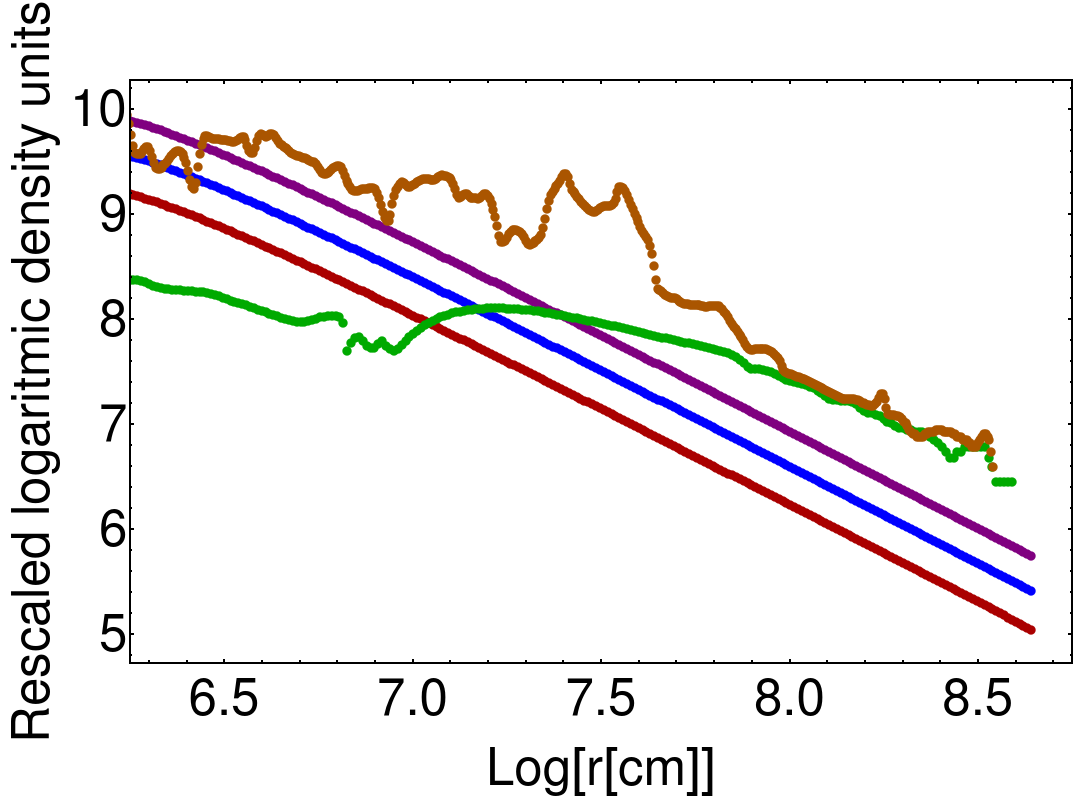} &
		\includegraphics[width=0.30\textwidth]{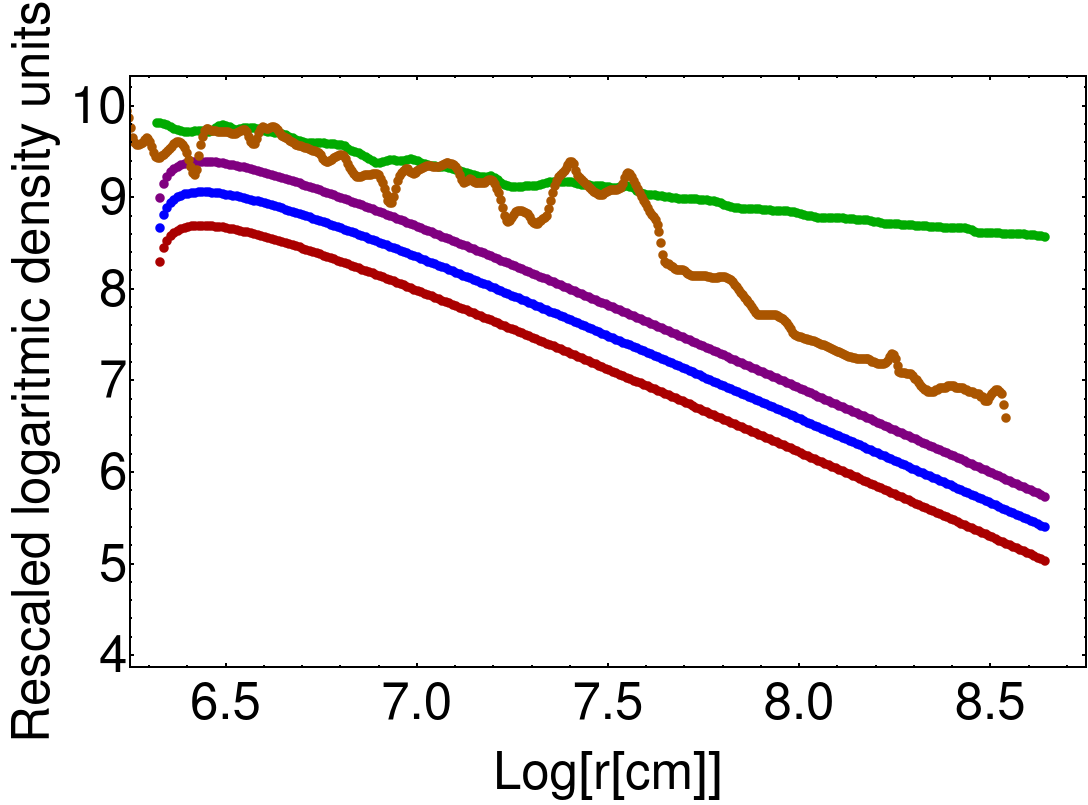} &
	\end{tabular}
	\caption{Comparison of the density profiles at the equator obtained during our simulations (green) with density profiles of the Shakura-Sunayev disk with $\alpha=0.1$ (red), $\alpha=0.3$ (blue), and $\alpha=1.0$ (purple). Additionally, we plotted density of the evolved magnetized torus model, started from the Chakrabarti solution, taken at time time $t=1500 t_{\rm g}$ (brown lines). Density units are rescaled.
          Left panel shows our simulation without magnetic field with $A_0=0.3$ and $S=1.4$, middle panel shows $A_0=0.85$ and $S=1.0$, and in the right panel we show simulation with $\beta=1$, $A_0=0.5$ and $S=1.0$.
          }
	\label{fig:profiles}
\end{figure*}

 In some models we obtained disk-like structures at the equatorial plane, which sustained for a period of time during magneto-hydrodynamic simulations. We can compare their structure, i.e., equatorial density profiles, with the standard solutions known from accretion disk theory \citet{1973A&A....24..337S} (SS73). Here we present the standard, stationary accretion disk model with $\alpha=0.3,0.5$ and $1$. It is shown with three different lines in the Fig. \ref{fig:profiles}.
 Our simulated disk-like profiles are shown for three chosen simulations: (i) small mini-disk which is formed around $t=1.5~\textrm{s}$ in the simulation with $A_0=0.3$ $S=1.4$, shown in the left panel (ii)  $A_0=0.85$ $S=1.0$, shown in the middle panel; here our simulated disk appears around $t=1.5~\textrm{s}$ and remains for few seconds and (iii) disk structure which is formed in simulation with $\beta=1$, $A_0=0.5$, $S=1.0$, presented in the right panel. The simulation results are marked with green solid line.
             Disk (i) presents snapshot taken at time $t=1.48$, cf. Fig. \ref{fig:models_lowA0_s14sim}, disk (ii) shows snapshot taken at $t= 2.04$, cf. Fig. \ref{fig:a085s10_2}, and disk (iii) shows snapshot taken at $t=0.163$, cf. Fig. \ref{fig:A05S10Beta1_profile}.
   
  We notice that the disk forming from the non-magnetized collapsar model with super-critical rotation speed, agrees with SS73 disk only in the inner parts of the flow. Then the density profile flattens in the outer parts. In this simulation, the angular momentum is not transported out by viscous forces, while the rotation of the envelope is sub-Keplerian. Also, the same holds for the non-magnetized simulation of the mini-disk with critical rotation.
  Also, the model with magnetic field and critical rotation diverges from the SS73 disk structure at the outer parts of the flow. This model however is best
  matched by another 'standard' accretion model.
  We compared it with density profile of evolved \cite{Chakrabarti1985} torus structure, taken from the GR MHD simulation, and evolved at late time, $t=1500M$ (see \cite{2021arXiv210111880J}).
    Profiles shown in the Fig. \ref{fig:profiles} have arbitrarily normalized density units.

Finally, let us speculate about the relevance of our findings to the observed black hole systems and their parameter estimates.
In Figure \ref{fig:maps_hydro} we presented the parameter space of the non-magnetized models, depicting the results depending on the initial black hole spin and rotation strength.
As the results show, the final black hole mass is anti-correlated with the final black hole spin, which is in a good agreement with the results from gravitational wave observations \citep{Safarzadeh2020}. They show that most heavy (stellar-mass) black holes tend to be spinless, and their spins do not exceed $a\sim 0.4$, as given by the probability distribution analysis \citep{Roulet2019}. 
It should be noted however, that our results consider only black holes formed in a collapsing star, without the interaction with any companion in a binary system. 
The parameter space used in our analysis (see Fig. \ref{fig:maps_hydro}) covers different values of an initial black hole spin a the beginning of collapse, and the angular momentum content in the envelope.
In general, these values might represent the state of collapsar as reached at some evolved state in a binary system, when the envelope and/or stellar core has been spun up by the companion. This state however refers to only a very close contact, or a common-envelope phase, rather than a detached system with long time accretion.

In case of the black holes accreting in the 
Low and High Mass X-ray Binaries (LMXBs and HMXBs), even in the wind fed HMXBs the accreted material has enough angular momentum from the orbit so that it forms a disk around the BH before it is accreted (only at larger scales the accretion flow will be Bondi-like). 
Typical black holes in HMXBs are expected to accrete only up to a few tenths or at most $1 M_{\odot}$ over their lifetimes, and the common belief is that no spin up is possible via accretion on this systems. 

On the other hand, the measured spins of black holes in LMXBs cover the whole range of spin parameters \citep{fragos2015}.
Our simulations cover the parameter space in which the specific angular momentum accreted by the BH is not much larger than that of the ISCO.
Nevertheless, the maximum black hole spin achieved temporarily in all the non-magnetized runs, is very high, if only rotation is larger than critical. Therefore, the constraint that we find on the angular
momentum content (or the magnetic fields in case of the magnetized flow),
might be relevant for the evolutionary history of some particular black
hole binary systems, such as GRS 1915+105 
\citep{McClintock2006}.

In systems like GRS 1915+105, the spin is 'directly' measured from the spectral fitting 
(in this microquasar we have a=0.95 estimate).
In addition, the black hole mass for this microquasar is largest among known sources of this type in our Galaxy (over 12 Solar mass).
In others, like
GX 339-4, the indirect method for the BH spin estimate, based on the estimated jet power is used (for this system it results in a=0.25).
For several further LMXBs there is no observational constraint for the BH spin, while the estimates of the black hole mass give values from 7.5 (for XTE J1118+480) up to
9.0 Solar mass (for GS 2023+338).

\section*{Acknowledgments}
We thank Tassos Fragos, Petra Sukova, Lukasz Stawarz, and Ishika Palit for helpful discussions.
This work was supported in part by the grant 
no. DEC-2016/23/B/ST9/03114, from the Polish National Science Center. D.K. was supported by Polish NSC grant 2016/22/E/ST9/00061.
We also acknowledge support from the Interdisciplinary Center for Mathematical Modeling of the Warsaw University, through the computational grant Gb79-9.

\bibliography{paper_spin_v1}{}
\bibliographystyle{aasjournal}

\end{document}